\documentclass[a4paper,11pt, final]{article}
\usepackage{test,graphicx,subcaption,booktabs,paralist,multirow,framed}
\usepackage[bottom]{footmisc} % makes sure the footnote of first page is at the borrom
\usepackage[authoryear,round]{natbib}
\usepackage[usenames,dvipsnames]{xcolor}
\usepackage[colorlinks=true,citecolor=Mahogany,linkcolor=Mahogany,urlcolor=Mahogany]{hyperref}
\usepackage{doi}
\usepackage{mathtools}
\usepackage{bm}
\usepackage{ushort}
\usepackage{xspace} 
\usepackage{mathpazo} 
\usepackage[margin=1.in]{geometry}
\usepackage{setspace}
\usepackage{pifont} 
\usepackage{nicefrac} 
\usepackage{stmaryrd}
\usepackage[ruled,vlined]{algorithm2e}

\spacing{1.25} 
\theoremstyle{plain}

\newtheorem{assumption}{Assumption} 
\newtheorem{definition}{Definition} 

\usepackage{float}
\usepackage[obeyDraft]{todonotes}

\definecolor{c1}{HTML}{4477AA}
\definecolor{c2}{HTML}{CC6677}

\usepackage{datetime}
\newdateformat{monthyeardate}{\monthname[\THEMONTH] \THEYEAR}
\usepackage{tabularx}

\newcolumntype{L}{>{\centering\arraybackslash}X}
\usepackage{siunitx}
\begin{document}

\title{\textbf{Structural Reinforcement Learning\\ for Heterogeneous Agent Macroeconomics}}
\author{
Yucheng Yang$^{*,1}$ 
\and Chiyuan Wang$^{*,2}$ 
\and Andreas Schaab$^3$ 
\and Benjamin Moll$^4$
}
\date{Preliminary \\[2ex]
First version: December 2025	\\
This version: \monthyeardate\today
	\\
\href{https://benjaminmoll.com/SRL/}{[latest version]}
}

% --- Acknowledgement as first, unnumbered footnote ---
\renewcommand\thefootnote{}% no printed marker
\footnotetext{We thank Jes{\'u}s Fern{\'a}ndez-Villaverde, Jakob Foerster, Felix K\"{u}bler, Dima Mukhin, Richard Rogerson, Tom Sargent, Bo Li, Sebastian Towers, Gianluca Violante, Clarisse Wibault, Tiphaine Wibault, Zhuoran Yang, and many seminar participants for helpful comments. Yang acknowledges support from the Swiss NSF (\#10003091), Wang acknowledges support from the NSFC (\#72450002), Moll acknowledges support from the Leverhulme Trust and the European Research Council (\#101200645). We will release codes implementing our computational experiments as soon as possible.}

\setcounter{footnote}{0}% so the next footnote (author affils) is numbered 1

% 2) Equal-contribution star footnote
\renewcommand\thefootnote{\fnsymbol{footnote}}% use *, †, ‡, ...
\setcounter{footnote}{1}% 1 -> *
\footnotetext{Equal contribution.}

% 3) Numeric affiliation footnotes
\renewcommand\thefootnote{\arabic{footnote}}

\footnotetext[1]{University of Zurich and SFI. Email: \href{mailto:yucheng.yang@uzh.ch}{yucheng.yang@uzh.ch}}
\footnotetext[2]{Peking University, School of Computer Science and BIGAI. Email: \href{mailto:wang2021@stu.pku.edu.cn}{wang2021@stu.pku.edu.cn}}
\footnotetext[3]{UC Berkeley. Email: \href{mailto:schaab@berkeley.edu}{schaab@berkeley.edu}}
\footnotetext[4]{London School of Economics, corresponding author. Email: \href{mailto:b.moll@lse.ac.uk}{b.moll@lse.ac.uk}}

\maketitle

\begin{abstract}
We present a new approach to formulating and solving heterogeneous agent models with aggregate risk. We replace the cross-sectional distribution with low-dimensional prices as state variables and let agents learn equilibrium price dynamics directly from simulated paths. To do so, we introduce a \emph{structural reinforcement learning} (SRL) method which treats prices via simulation while exploiting agents’ structural knowledge of their own individual dynamics. Our SRL method yields a general and highly efficient global solution method for heterogeneous agent models that sidesteps the Master equation and handles models traditional methods struggle with, like those with nontrivial market-clearing conditions. We illustrate the approach in the Krusell-Smith model, the Huggett model with aggregate shocks, and a HANK model with a forward-looking Phillips curve, all of which we solve globally within minutes.
\end{abstract}

\thispagestyle{empty}
\setcounter{page}{0}

\pagebreak

%%%%%%%%%%%%%%%%%%%%%%%%%%%%%%%%%%%%%%%%%%%%%%%%%%%%%%%%%%%%
%%%%%%%%%%%%%%%%%%%%%%%%%%%%%%%%%%%%%%%%%%%%%%%%%%%%%%%%%%%%
%%%%%%%%%%%%%%%%%%%%%%%%%%%%%%%%%%%%%%%%%%%%%%%%%%%%%%%%%%%%
\section{Introduction}
Many of the most important questions in macroeconomics call for studying models with heterogeneous agents and aggregate risk. A well-known difficulty is that, in standard recursive rational-expectations formulations of such models, the cross-sectional distribution of agents becomes a state variable in the Bellman equation characterizing agents' decision problems -- the ``Master equation."\footnote{See, e.g., \citet{denhaan}, \citet{krusell-smith}, \citet{schaab}, and \citet{bilal}. The name ``Master equation" comes from the mathematics literature on Mean Field Games \citep{cardaliaguet-delarue-lasry-lions}.} This difficulty arises even though agents do not directly ``care about'' the distribution, i.e. it does not enter their objective functions; instead, as is standard in competitive equilibrium models, they only care about prices. Intuitively, low-dimensional equilibrium prices do not follow a Markov process but the extremely high-dimensional distribution does. Therefore, agents with rational expectations forecast prices by forecasting distributions.

While the recent literature has made some impressive advances, the extreme curse of dimensionality inherent in the Master equation remains a central computational bottleneck for \emph{global} solutions to heterogeneous agent models with aggregate risk.\footnote{There is, of course, also the global solution method of \citet{krusell-smith} and \citet{denhaan} which assumes that agents forecast prices by forecasting \emph{moments} of cross-sectional distributions rather than the distributions themselves. We discuss similarities and differences to our approach further below.} Even seemingly simple model environments like a \citet{huggett} model with aggregate risk (in which there is a non-trivial market clearing condition) or a one-asset HANK model with a forward-looking Phillips curve lead to exceedingly difficult computational problems. This lack of a general and efficient global solution method limits the applicability of heterogeneous-agent macroeconomics, in particular to questions in which aggregate non-linearities play a key role.\footnote{While we do not consider such models in the present draft, we think that one particularly promising application of our global solution method will be modeling infrequent but large boom-bust cycles like financial crises.} 

The contribution of this paper is to develop an approach that sidesteps the Master equation by using ideas from reinforcement learning (RL). RL means learning value or policy functions of incompletely-known Markov decision processes via some form of Monte Carlo simulation \citep{sutton-barto}. We apply this idea to equilibrium prices and let agents compute price expectations directly from simulated paths. However, our approach differs from standard RL in that we assume that agents have \emph{structural knowledge} about the dynamics of their own individual states (e.g. their budget constraint and idiosyncratic income process). We term this hybrid approach \emph{structural reinforcement learning} (SRL) and our specific algorithm a \emph{structural policy gradient} (SPG) algorithm.\footnote{The ``structural" in SRL is analogous to that in structural vector autoregressions (SVARs).} By imposing that policy functions depend only on current prices (or a short price history) we keep the state space low-dimensional so that we can work with a grid-based (tabular) approach rather than deep neural networks. Finally, we provide an efficient implementation in JAX \citep{JAX,quantecon-JAX} that can be run on Google Colab.\footnote{Google Colab is a cloud computing platform that is easily accessible to all researchers.}

SRL delivers a new and highly efficient global solution method for heterogeneous agent models with aggregate risk. Importantly, it solves problems traditional methods struggle with. We demonstrate this with two example applications. First, a model environment with a nontrivial market-clearing conditions, namely a \citet{huggett} model with aggregate risk. Second, a HANK model with a forward-looking Phillips curve. Both model environments have proven notoriously difficult for traditional methods.\footnote{\label{foot:JEDC}There is an interesting historical note regarding the computational difficulty of the Huggett model with aggregate risk. According to \citet{maliar2020deep}, in the influential JEDC special issue on numerical solution methods for HA models with aggregate risk \citep{denhaan-judd-juillard}, all participants were initially asked to solve two benchmark models globally -- the \citet{krusell-smith} model and the \cite{huggett} model with aggregate shocks (Maliar and Maliar call this ``the HANC model with savings through bonds"). However, as they report, no single team was able to successfully solve the latter model, and it was ultimately dropped from the JEDC project.} We instead solve the Huggett model in around one minute and the HANK model in around three minutes.\footnote{All experiments are implemented in JAX and are run on a single NVIDIA A100 GPU on Google Colab. We will discuss the precise convergence criteria in Section \ref{sec:implementation}.} For completeness, we also solve the easier \citet{krusell-smith} model in around 55 seconds.

Almost all existing global solution methods for heterogeneous agent models use dynamic programming. They either use dynamic programming to directly tackle the high-dimensional Master equation with the distribution as state variable; or, as in \citet{krusell-smith}, they use model-generated data to estimate a low-dimensional Markovian ``perceived law of motion" (PLM) for moments of the distribution and then apply dynamic programming to this lower-dimensional approximate problem.\footnote{The PLM is simply an approximate Markov process for the low-dimensional vector of moments. The PLM may have a simple parametric (e.g. linear) functional form as in \citet{krusell-smith} or it may be a general, non-linear function, e.g. parameterized by a neural network as in \citet{FV-hurtado-nuno}.} Our SPG algorithm instead works directly with the sequential formulation of the problem and never attempts to force it into the standard Markovian structure required for applying dynamic programming. While we reduce the dimensionality of the state space in a fashion reminiscent of moment-based methods, we never estimate a PLM. Instead, we use the simple idea at the core of all RL methods that value functions are expected values -- here, expected discounted lifetime utilities -- and can therefore be approximated by averaging across simulated trajectories. We then find the low-dimensional policy function that maximizes expected lifetime utility using stochastic gradient ascent.

The key elements of our SPG method delivering fast computations even in challenging model environments are as follows. Most importantly, as already mentioned, we replace the cross-sectional distribution with low-dimensional prices as state variables. We do this in two steps. The first step is to impose that agents observe the history of equilibrium prices but not the cross-sectional distribution. By the Wold representation theorem, this assumption does not, by itself, imply any departure from full information rational expectations. In the second step, we restrict the agents' state space: we assume that their policy functions depend only on current prices or, perhaps, a short price history. This second assumption means that we do not solve for the model's rational-expectations equilibrium; instead, we solve for a restricted perceptions equilibrium (RPE) in the sense of \citet{sargent-RPE}, \citet{evans-honkapohja}, or \citet{branch}: while agents' expectations are restricted, they are nevertheless statistically consistent with actual equilibrium outcomes, thereby fulfilling one of the desiderata of rational expectations. 

The next key element delivering fast computations is the defining assumption of our SPG approach that agents have structural knowledge about the dynamics of their own individual states. In contrast to standard policy gradient methods which estimate approximate policy gradients, this assumption allows us to compute exact policy gradients by differentiating through these individual dynamics. The only part of the environment that is treated as unknown is the process for general‑equilibrium prices and aggregate shocks, which is learned from simulated data. Specifically, our SPG method discretizes the individual state space so that agents' individual states evolve according to a known transition matrix $\mathbf{A}_{\bm \pi}$ where $\bm \pi$ denotes the vector of individual policies. When computing policy gradients of lifetime utilities with respect to $\bm \pi$, we exploit this structural knowledge and differentiate through $\mathbf{A}_{\bm \pi}$ while using simulation only for prices and aggregate shocks.

Finally, our low-dimensional grid- and price-based policy functions allow us to (globally) simulate the economy forward in time in a very efficient manner. For any particular trajectory of aggregate shocks, we update the distribution using the ``histogram method" of \citet{young}. The fact that policy functions depend on current prices allows us to efficiently handle non-trivial market clearing conditions like in \citet{huggett}. Intuitively, \emph{policy functions double as individual supply schedules}. Integrating across the distribution to obtain the corresponding aggregate supply schedule, it is then straightforward to solve for equilibrium prices at each point in time along a simulation path. Our treatment of market clearing mirrors the RL literature's distinction between agents and environment: agents can interact with their environment under any given policy including suboptimal ones; finding optimal policies is conceptually separate. In line with this dichotomy, we treat market clearing as part of the environment and find equilibrium prices also for suboptimal policies. This approach differs from standard practice in macroeconomics which first finds optimal policies given prices in an inner loop and then finds equilibrium prices in an outer loop.

In summary, SRL enables efficient \emph{global} solutions of heterogeneous-agent models with full cross-sectional distributions. Importantly, while our restricted state space (and use of an RPE) simplifies agents' decision problems \emph{inside} the model, it does \emph{not} diminish the rich dynamics of the \emph{economy}  which still evolves stochastically and non-linearly, driven by the policy functions of forward-looking heterogeneous agents.

We illustrate our method in three benchmark environments: a \cite{huggett} economy, the classic \citet{krusell-smith} model, and a one-account heterogeneous agent New Keynesian (HANK) model with sticky prices. In all three cases, our price-based SRL algorithm converges quickly on modern hardware: solving the Krusell-Smith model takes 55 seconds
%in line with alternative global solution methods
while the Huggett and HANK models -- typically viewed as more challenging because of their non-trivial market-clearing conditions and, in the HANK case, a forward-looking Phillips curve -- take only modestly longer (about 75 seconds and roughly 3 minutes, respectively). We present extensive tests to verify the accuracy of our solutions. For the Krusell-Smith model, our method produces equilibrium dynamics that closely match the rational expectations obtained using deep-learning-based methods. As a complementary exercise, we extend the information set to allow agents to condition on lagged prices and show that this richer price history does not meaningfully change the solution, suggesting that current prices already contain most of the information that matters for behavior. Finally, in the HANK application we show how to use our approach to solve the household and firm problems jointly, using the same SPG algorithm not only for consumption-saving decisions but also for the forward-looking price-setting problem of firms.

%RELATED LITERATURE
\paragraph{Relation to Economics Literature.} 
A huge theoretical and quantitative literature studies environments in which heterogeneous households are subject to uninsurable idiosyncratic shocks. See  \citet{krusell-smith-survey}, \citet{heathcote-storesletten-violante}, \citet{quadrini-rios-rull}, \citet{krueger-mitman-perri}, \citet{sargent-HAOK-HANK}, and \citet{auclert-rognlie-straub-AR} for surveys.

Within this literature, a sizable subliterature is concerned with the \emph{global} solution of heterogeneous agent models with aggregate risk. A first strand of the literature tackles the Master equation directly, typically using deep neural networks -- see e.g. \citet{han-yang-e}, \citet{maliar-maliar-winant}, \citet{azinovic-gaegauf-scheidegger}, \citet{kahou-fv-perla-sood}, \citet{duarte-duarte-silva}, \citet{huang}, \citet{kase-melosi-rottner}, \citet{gu-lauriere-merkel-payne}, \citet{payne-rebei-yang}, and \citet{gopalakrishna-gu-payne}. Our approach differs from all of these papers in that we \emph{sidestep} the Master equation rather than attempting to ``tame the curse of dimensionality" inherent in it. Among these papers, \citet{han-yang-e} is most related to and influential for our work: like them, we learn value and policy functions of heterogeneous agents via simulation; also like them, we take advantage of agents' structural knowledge of their own individual dynamics -- what we have termed a ``structural" RL approach. The key difference is that \citet{han-yang-e} include the high-dimensional cross-sectional distribution in the agents' state space whereas we replace this distribution with low-dimensional prices as state variables.

A second strand of the literature estimates a low-dimensional Markovian PLM for moments of the distribution and then applies dynamic programming as in \citet{krusell-smith} and \citet{denhaan}. Similar to us, this approach sidesteps the Master equation by working with a low-dimensional state space that does not include the distribution. One variant of this approach formulates PLMs directly in terms of equilibrium prices so that -- like in our approach -- this low-dimensional state space includes prices.\footnote{See for example \citet{lee-wolpin}, \citet{storesletten-telmer-yaron}, \citet{gomes-michaelides}, \citet{favilukis-ludvigson-vanN}, \citet{llull}, \citet{kaplan-mitman-violante}, and \citet{FV-marbet-nuno-rachedi}.} However, this approach introduces an additional fixed-point loop in which the perceived law of motion is repeatedly updated and the associated dynamic programming problem re-solved, a procedure that is computationally slow in many economically interesting environments, particularly those with non-trivial market-clearing conditions. By contrast, we work directly with the problem’s sequential formulation and never estimate a PLM.

Our approach is also related to the adaptive learning literature \citep[e.g.][]{bray,marcet-sargent,evans-honkapohja}. \citet{jacobson} and \citet{giusto} apply adaptive learning in heterogeneous agent models with aggregate risk. RL and adaptive learning are linked because both are special cases of a more general set of stochastic approximation methods \citep{robbins-monro}. Similar to the learning literature, our approach computes an equilibrium that is ``self-confirming" \citep{sargent-conquest,cho-sargent}.\footnote{Agents form price expectations from data generated by the economy in which they live. Their expectations are therefore statistically consistent with actual equilibrium outcomes but may be incorrect for events that are infrequently observed (e.g. events off the equilibrium path).} But because we restrict the individual state space to not include the distribution, our self-confirming equilibrium is an RPE \citep{sargent-RPE,branch}. In the language of \citet{guarda}, our RPE features expectations that are both ``narrow" and ``short": narrow because they do not condition on the distribution and short because they do not condition on a long price history.\footnote{Our notion of equilibrium is also closely related to Guarda's nonparametric restricted perceptions equilibrium (NRPE). The difference is that Guarda defines his NRPE recursively and uses dynamic programming whereas our formulation is sequential.}

Our focus on a low-dimensional vector of equilibrium variables links SRL to the ``sequence space" approach of \citet{SSJ} and \citet{auclert-etal-2nd-order}. The difference is that SRL handles stochastic price sequences outside steady-state neighborhoods. By using Monte Carlo simulation, it yields a global rather than local solution method in sequence space.\footnote{Recent work has explored global solution methods in sequence space using high-dimensional approximation techniques, e.g., \citet{azinovic-zemlicka-deep-SSJ}.}
%\footnote{In future versions of this paper we plan to evaluate how well the first- and second-order sequence-space solutions approximate our global solution.}

To be clear, we do not view SRL as an empirically realistic model of human learning.\footnote{\citet{moll-challenge} proposed three criteria for alternatives to rational expectation in heterogeneous agent models:  (1) computational tractability, (2) consistency with empirical evidence, and (3) (some) immunity to the Lucas critique. Our SRL approach delivers on (1) and (3) but not (2).} Instead, it is an efficient computational method for finding RPEs in heterogeneous agent models with aggregate risk. Nevertheless, as we note in the conclusion, the core idea of our approach -- agents forming expectations about equilibrium prices by sampling -- could, in principle, serve as a building block for an empirical theory of expectations formation in macroeconomics.

\paragraph{Relation to RL Literature.} Our approach connects to several central ideas in the RL literature. As already mentioned, in RL, agents learn optimal policies in Markov decision processes using sampled trajectories \citep{sutton-barto}. RL is at the core of some impressive advances in artificial intelligence, e.g. RL agents learning to play Go and Atari games better than humans \citep{deepmind-atari,deepmind-go,deepmind-go-zero}  or post-training of ``reasoning" large language models \citep[LLMs][]{openai-o1,deepseek-r1}.\footnote{RL ideas have also been applied in economics. Besides \citet{han-yang-e}, see for example \citet{barberis-jin}, \citet{chen-etal-RL-monetary}, \citet{barrera-desilva}, \citet{calvano2020artificial}, \citet{dou-goldstein-ji}, and the references in \citet{FV-nuno-perla} for applications in finance and macroeconomics. RL ideas have also been applied in game theory \citep[e.g.][]{roth-erev-95,erev-roth,fudenberg-levine} but using a different formulation that does not work with value functions and instead directly reinforces the ``propensities" of choosing strategies.}

Conceptually, our SRL method adopts the RL principle of optimizing policies using Monte Carlo estimates of expected returns. But it exploits full structural knowledge of each agent's reward function and individual transition dynamics to compute exact end-to-end policy gradients. Only equilibrium objects are treated as unknown parts of the environment. In this sense, our SRL approach sits between model-based and model-free RL: we differentiate exactly through the known micro-level transition probabilities while learning the induced macro environment from simulation. Also see \citet{han-yang-e}.

While classical RL demonstrates considerable power in solving complex environments, its general-purpose nature implies suboptimal performance in specialized domains. For instance, when applied to the computationally intensive post-training phase of LLMs, classical RL algorithms introduce substantial overhead. To address this, \citet{rafailov-etal-DPO} reformulated RL from human feedback (RLHF) as a deep learning problem and proposed the Direct Preference Optimization (DPO) algorithm, which significantly reduces computational cost and enhances training stability.
%In addition to reformulating problem structures for better performance, researchers have explored information reuse in specialized domains.
Similarly, \citet{hu-etal-difftaichi} and \citet{freeman-etal-brax} introduced differentiable physics engines -- DiffTaichi and Brax (the latter also implemented using JAX) -- that enable gradient computation directly from built-in physical equations.\footnote{\label{foot:PINN}More generally, our approach of exploiting the problem's economic structure is akin to the use of physics-informed neural networks  \citep{raissi-perdikaris-karniadakis-PINN}. In this sense, SRL is ``economics-informed."} This allows policy updates via exact gradient methods, eliminating the need for sampling-based approximations and thereby improving both accuracy and efficiency. In Brax, this method is termed ``analytical policy gradient," which shares conceptual foundations with our SRL framework.

Heterogeneous-agent models in macroeconomics can be viewed as a special case of the more general Mean Field Games (MFGs) studied in the mathematics literature \citep[e.g.][]{lasry-lions,cardaliaguet-delarue-lasry-lions}. In the RL literature, the work closest to ours is therefore the line of research that applies RL methods to solve MFGs \citep[e.g.][]{yang-etal-MFMARL,lauriere-etal-RL,lauriere-etal-survey,xu-etal-RL,wu-lauriere-etal-common-noise}. With the exception of \citet{wu-lauriere-etal-common-noise}, none of these papers study the case with aggregate risk (or ``common noise" in MFG terminology) considered here. More importantly, none of the papers applying RL to MFGs exploit the structural knowledge of agents' individual dynamics. Specifically, they do not exploit the transition matrix (or infinitesimal generator) for individual states $\mathbf{A}_{\bm \pi}$  and its known dependence on the policy $\bm \pi$, even though this matrix (operator) is typically available and even used for updating the distribution.\footnote{\citet{gabriele-glielmo-taboga} uses various multi-agent RL (MARL) algorithms to solve the \citet{krusell-smith} model. These algorithms also do not exploit this structural knowledge. Another difference is that \citet{gabriele-glielmo-taboga} consider a relatively small finite number (e.g. $529$) of agents  rather than the MFG continuum limit.} 

In order to sidestep the Master equation and keep the state space manageable, we assume a form of \emph{partial observability}: agents do not observe the high-dimensional distribution which is the system's underlying Markov state and instead observe low-dimensional prices. That being said, a similar SRL approach can also be applied to fully observable MFGs as in \citet{han-yang-e}.

\paragraph{Roadmap.} Section \ref{sec:setup} starts by describing the setup our SRL method applies to starting with a simple example. Section \ref{sec:approach} describes the SRL method and Section \ref{sec:applications} reports our computational experiments. Section \ref{sec:conclusion} concludes.

%%%%%%%%%%%%%%%%%%%%%%%%%%%%%%%%%%%%%%%%%%%%%%%%%%%%%%%%%%%%
%%%%%%%%%%%%%%%%%%%%%%%%%%%%%%%%%%%%%%%%%%%%%%%%%%%%%%%%%%%%
%%%%%%%%%%%%%%%%%%%%%%%%%%%%%%%%%%%%%%%%%%%%%%%%%%%%%%%%%%%%
\section{Setup}\label{sec:setup}

To explain the logic of our approach in the simplest possible fashion, we present it in a context that is very familiar to most economists: a general equilibrium model with incomplete markets and uninsured idiosyncratic labor income risk. We first do this in an economy in which individuals save in unproductive bonds that are in zero net supply as in \citet{huggett} but with aggregate risk. This economy is ideal for illustrating our method because it features a non-trivial market clearing condition. We later consider different ways of closing the model, for example a \citet{krusell-smith} economy in which individuals save in productive capital.

%%%%%%%%%%%%%%%%%%%%%%%%%%%%%%%%%%%%%%%%%%%%%%%%%%%%%%%%%%%%
%%%%%%%%%%%%%%%%%%%%%%%%%%%%%%%%%%%%%%%%%%%%%%%%%%%%%%%%%%%%
\subsection{A Huggett Model with Aggregate Risk}\label{sec:setup_huggett}

\paragraph{Individuals.}
There is a continuum of individuals that are indexed by $i$. Each individual has an income which evolves stochastically over time. Specifically, income $y_{i,t}z_t$ is an endowment of the economy's final good and consists of an idiosyncratic component $y_{i,t}$ (idiosyncratic risk) and an aggregate component $z_t$ (aggregate risk). Individuals are heterogeneous in the idiosyncratic income component $y_{i,t}$ and their wealth $b_{i,t}$. The states of the economy are (i) the joint distribution of income and wealth which we denote by $G_t(b,y)$ and (ii) the aggregate shock $z_t$.

Individuals have standard preferences over utility from future consumption $c_t$:
\begin{equation}\label{eq:prefs}
	\mathbb E_0 \sum_{t=0}^\infty \beta^t u(c_t).
\end{equation}
The wealth of an individual takes the form of bonds and evolves according to
\begin{equation}\label{eq:bc}
	c_{i,t} + b_{i,t+1} = (1+r_t) b_{i,t} +  y_{i,t} z_t,
\end{equation}
where $r_t$ is the interest rate. Individuals also face a borrowing limit
\begin{equation}\label{eq:borr}
	b_{i,t} \geq \ushort{b},
\end{equation}
with $-\infty<\ushort{b} \leq 0$ and which we assume to be tighter than the ``natural borrowing constraint" \citep{aiyagari}. The two income components follow Markov processes:
\begin{equation}\label{eq:risk}
	y_{i,t+1} \sim \mathcal{T}_{y}(\cdot|y_{i,t}) \quad \mbox{and} \quad z_{t+1} \sim \mathcal{T}_{z}(\cdot|z_{t}),
\end{equation}
where $\mathcal{T}_{y}$ and $\mathcal{T}_z$ summarize the respective transition probabilities. We typically assume that the idiosyncratic component $y_{i,t}$ lives on a finite grid $\{y_1,...,y_{J_y}\}$ whereas the aggregate component $z_t$ is continuous, for example, the logarithm of $z_t$ may follow an AR(1) process.

Individuals maximize \eqref{eq:prefs} subject to \eqref{eq:bc}, \eqref{eq:borr} and \eqref{eq:risk}, taking as given the evolution of the equilibrium interest rate $r_t$ for $t\geq0$. The individuals' optimization problem gives rise to consumption and saving policy functions which denote by $c_t(b,y,z)$ and $b_t'(b,y,z)$ where the prime superscript indexes the next period, i.e. $b_{i,t+1} = b_t'(b_{i,t},y_{i,t},z_t)$. For future reference, we denote the collection of both policy functions by
\begin{equation}\label{eq:policy}
	\pi_t(b,y,z) = \{c_t(b,y,z),b_t'(b,y,z) \}.
\end{equation}

\paragraph{Equilibrium.} 
The economy can be closed in a variety of ways. We here follow \citet{huggett} and assume that the interest rate $r_t$ (which is the only price in the economy) is determined by the requirement that, in equilibrium, bonds must be in zero net supply:
\begin{equation}\label{eq:eq}
	\int b'_t(b,y,z_t) \ dG_t(b,y) = 0, 
	\quad \mbox{all} \ t \geq 0,
\end{equation}
where $b_t'(b,y,z)$ is saving of an individual with states $(b,y,z)$ as just discussed. 
%and where $0 \leq B < \infty$. If $B=0$, bonds are in zero net supply. Alternatively, $B$ can be positive. For instance, a government could issue debt and sell it to individuals or there could be saving opportunities abroad. 
We later consider alternative ways of closing the economy. For instance Section \ref{sec:applications_krusell_smith} assumes that wealth takes the form of productive capital hired by a representative firm so that the interest rate equals the aggregate marginal product of capital as in \citet{krusell-smith}.

A competitive equilibrium is defined in the usual way: quantities and prices $\{r_t\}_{t=0}^\infty$ such that
\begin{enumerate}
\item Individuals maximize \eqref{eq:prefs} subject to \eqref{eq:bc}, \eqref{eq:borr} and \eqref{eq:risk}, taking as given $\{r_t\}_{t=0}^\infty$.

\item Markets clear: \eqref{eq:eq} holds for all $t \geq 0$.
\end{enumerate}
Importantly, in this competitive equilibrium, individuals do \emph{not} ``care about" the cross-sectional distribution $G_t$, i.e. it does not enter their objective functions. Instead they only care about \emph{prices}, here the interest rate $r_t$.

\paragraph{Compact Notation.} 
To ease notation going forward and also with an eye toward other,  more complex heterogeneous agent models, we introduce some additional notation. First, we denote the vector of individual states by
$$s = (b,y).$$
and summarize the budget constraint \eqref{eq:bc} and $y_{i,t}$-process in \eqref{eq:risk} in terms of transition probabilities for this vector $s$:
\begin{equation}\label{eq:transitions}
	s_{i,t+1} \sim \mathcal T_s(\cdot|s_{i,t},c_{i,t},z_t,p_t).
\end{equation}
Second, we denote the vector of prices by $p_t$. Of course, in the Huggett model above, there is only one price $p_t=r_t$. But this notation will be useful in other applications, e.g. in \cite{krusell-smith} there is a wage $w_t$ in addition to the interest rate $r_t$ so that $p_t=(r_t,w_t)$.

Finally, in equilibrium, prices depend on the economy's state variables, the distribution $G_t(s)$ and the aggregate shock $z_t$. We therefore denote by
\begin{equation}\label{eq:price_functional}
	p_t = P^*(G_t,z_t)
\end{equation}
the ``equilibrium price functional" that summarizes this dependence. In the Huggett model above, this equilibrium price functional is implicitly determined by the market clearing condition \eqref{eq:eq}.

\paragraph{Generalizations.} 
The tools developed in this paper apply to a large class of heterogeneous agent models. We consider a few of these in our computational experiments in Section \ref{sec:applications}. The most general setup is as follows: agents solve
\begin{align*}
v_{i,0} &= \max_{\{a_{i,t}\}} \  \mathbb{E}_0\sum_{t=0}^\infty \beta^t R(s_{i,t},a_{i,t},z_{t},p_t)  \quad \mbox{subject to} \quad s_{i,t+1} \sim \mathcal{T}_s(\cdot|s_{i,t},a_{i,t},z_t,p_t),
\end{align*}
where $R(s,a,z,p)$ is a reward function that depends on individual states $s \in \mathcal{S} \subseteq \mathbb{R}^{n_s}$, actions $a \in \mathcal{A} \subseteq \mathbb{R}^{n_a}$, aggregate states $z\in \mathcal{Z} \subseteq  \mathbb{R}^{n_z}$, and prices $p \in \mathcal{P} \subseteq  \mathbb{R}^{n_p}$, and where $\mathcal{T}_s$ summarizes the transitions of individual states $s$. Prices $p_t$ are determined in equilibrium from a set of market clearing conditions which results in a mapping $p_t = P^*(G_t,z_t)$ just like in \eqref{eq:price_functional}. While our computational method applies to such general setups, going forward, we will explain it in terms of the simple Huggett model with two-dimensional individual state $s$, one-dimensional aggregate state $z$, and one-dimensional price $p$ for concreteness.

%%%%%%%%%%%%%%%%%%%%%%%%%%%%%%%%%%%%%%%%%%%%%%%%%%%%%%%%%%%%
%%%%%%%%%%%%%%%%%%%%%%%%%%%%%%%%%%%%%%%%%%%%%%%%%%%%%%%%%%%%
\subsection{Discretized Representation}\label{sec:setup_discretized}

To illustrate and implement our method it is convenient to discretize the individual state space. While idiosyncratic income realizations $y$ are already on a finite grid, wealth $b$ is continuous. We therefore place the individual state $s = (b, y)$ on a finite grid $s \in \{s_1, \ldots, s_J\}$ with $J = J_y \times J_b$. Objects such as the value function or cross-sectional distribution become $J$-dimensional vectors:
\begin{equation*}
	\bm v_t = 
	\left[\begin{matrix}
		v_t(s_1) \\
		\vdots \\
		v_t(s_J)
	\end{matrix}\right]
	\qquad \text{ and } \qquad
	\bm g_t = 
	\left[\begin{matrix}
		g_t(s_1) \\
		\vdots \\
		g_t(s_J)
	\end{matrix}\right],
\end{equation*}
where we use boldfaced notation to denote vectors. Note that the vector $\bm g_t$ is simply the ``histogram" which collects the fraction of agents at each point of the state space.

Similarly, the consumption-saving policy $\pi_t(s,z)$ becomes a vector $\bm \pi_t(z)$ defined on the $J$-dimensional grid.
Given a  policy $\pi_t(\cdot)$, the induced one-step transitions for $s$ can be collected in a $J \times J$ transition matrix,
\begin{equation*}
	\mathbf{A}_{\bm\pi_t(z_t)}
	\qquad \text{with entries} \qquad
	\text{Pr}(s_{i,t+1} = s_{j'} \, | \, s_{i,t} = s_j) = \mathcal T_s(s_{j'} \, | \, s_j, \pi_t(s_j, z_t), p_t) .
\end{equation*}
Entry $jj'$ of this matrix represents the probability that an individual in state $j$ transitions to state $j'$ next period, with rows summing to $1$.
These probabilities are encoded in $\mathcal T_s$ and depend on the policy $\pi_t(s, z)$ as well as the period-$t$ realization of prices $p_t$.

The cross-sectional distribution then evolves according to the discrete-time Chapman-Kolmogorov equation 
\begin{equation}\label{eq:CKE}
	\bm g_{t+1} = \mathbf{A}_{\bm\pi_t(z_t)}^{\rm T} \bm g_t ,
\end{equation}
where $\mathbf{A}_{\bm\pi_t(z_t)}^{\rm T}$ denotes the transpose of the transition matrix. 
Intuitively, think of probability mass at each grid point $s_j$ \textit{flowing out} across the row $j$ of $\mathbf{A}_{\bm\pi_t(z_t)}$. The transpose in \eqref{eq:CKE} simply accumulates these inflows at destinations $s_{j'}$. The use of the ``histogram" $\bm g_t$ on the discretized state space to track the cross-sectional distribution is as in \citet{young} and \citet{AHLLM}.

%%%%%%%%%%%%%%%%%%%%%%%%%%%%%%%%%%%%%%%%%%%%%%%%%%%%%%%%%%%%
%%%%%%%%%%%%%%%%%%%%%%%%%%%%%%%%%%%%%%%%%%%%%%%%%%%%%%%%%%%%
\subsection{Key Difficulty: Equilibrium Prices Are Not Markov}

Two immediate implications follow. First, the high-dimensional aggregate state $(\bm g_t, z_t)$ is Markov by construction. The transition probabilities for $(\bm g_{t+1}, z_{t+1})$ depend on the current realizations of $(\bm g_t, z_t)$ only. 
Second, equilibrium prices are not Markov on the other hand \citep{moll-challenge}. 
They are determined as a function of the aggregate state by the price functional $p_t = P^*(\bm g_t, z_t)$. The transition probabilities for $p_{t+1}$ therefore depend on the high-dimensional state $(\bm g_t, z_t)$. Concretely, taking as given a policy function $\pi_t(s, z)$, the law of motion of prices in equilibrium is given by the equations 
\begin{equation}\label{eq:price_process}
\begin{split}
	p_t &= P^*(\bm g_t, z_t) \\
	\bm g_{t+1} &= \mathbf{A}_{\bm\pi_t(z_t)}^{\rm T} \bm g_t \\
	z_{t+1} &\sim \mathcal T_z(\cdot \, | \, z_t).
\end{split}
\end{equation}
The transition probabilities for $p_{t+1}$ cannot be determined as a function of the current price realization $p_t$ alone. We have $p_{t+1} = P^*(\bm g_{t+1}, z_{t+1})$ so the conditional distribution of $p_{t+1}$ depends on $(\bm g_t, z_t)$ and not just on $p_t$. Thus, $(\bm g_t, z_t)$ is a Markov state but $p_t$ is not.

\paragraph{The Master equation.}
Dynamic programming requires Markov state variables. Since $p_t$ alone is not Markov, the Bellman equation cannot be written only in terms of $p_t$. The key idea of dynamic programming --- splitting the sequence problem into a current flow payoff and a continuation value term --- fails when transitions are not Markov, since the current state does not provide sufficient information to determine the continuation value. The recursive formulation must instead include the true aggregate state $(\bm g, z)$:
\begin{align}\label{eq:master_equation}
	V(s, \bm g, z) = &\max_c \; \; u(c) + \beta \mathbb E [V(s', \bm g', z') \, | \, s, \bm g, z] \\
	&\text{ s.t. } \hspace{4mm}  \nonumber
	s' \sim \mathcal T_s(\cdot \, | \, s, c, z, p) \\  \nonumber
	&\hspace{10.95mm} p = P^*(\bm g, z) \\  \nonumber
	&\hspace{10.95mm} \bm g' = \mathbf{A}_{\bm\pi}^{\rm T} \bm g  \nonumber
\end{align}
Equation \eqref{eq:master_equation} is often referred to as the ``Master equation" --- a Bellman equation on a state space that includes the cross-sectional distribution $\bm g$ \citep{cardaliaguet-delarue-lasry-lions,macro-annual,schaab,bilal,gu-lauriere-merkel-payne}.

\paragraph{Why this matters.}
Households care directly about prices because $p$ enters their budget sets; they do not care directly about $\bm g$. Yet under rational expectations the only way to forecast $p_{t+1}$ is to forecast $(\bm g_{t+1}, z_{t+1})$, so the high-dimensional $\bm g$ becomes a state variable and the associated Bellman equation inherits an extreme curse of dimensionality.
But what if there was a way to approximate value and policy functions directly in terms of the current prices $p_t$, for which there are no Markov transition probabilities? This is the approach we develop in the next section.

%%%%%%%%%%%%%%%%%%%%%%%%%%%%%%%%%%%%%%%%%%%%%%%%%%%%%%%%%%%%
%%%%%%%%%%%%%%%%%%%%%%%%%%%%%%%%%%%%%%%%%%%%%%%%%%%%%%%%%%%%
%%%%%%%%%%%%%%%%%%%%%%%%%%%%%%%%%%%%%%%%%%%%%%%%%%%%%%%%%%%%
\section{Sidestepping the Master Equation via Structural Reinforcement Learning}\label{sec:approach}

This section describes our SRL method. Rather than solving the Master equation \eqref{eq:master_equation} with the full cross-sectional distribution as a state variable, we work with a low-dimensional state consisting of prices $p_t$ (and the aggregate shock $z_t$). We adapt RL ideas to let agents learn optimal behavior from simulated equilibrium data, taking $(s_t, z_t, p_t)$ as their state. In contrast to standard RL, our SRL method exploits agents' structural knowledge of their own individual dynamics.

%%%%%%%%%%%%%%%%%%%%%%%%%%%%%%%%%%%%%%%%%%%%%%%%%%%%%%%%%%%%
%%%%%%%%%%%%%%%%%%%%%%%%%%%%%%%%%%%%%%%%%%%%%%%%%%%%%%%%%%%%
\subsection{The Key Idea of Reinforcement Learning: Monte Carlo instead of Bellman Equations}\label{sec:approach_monte_carlo}
Before proceeding, we briefly summarize the basic ideas of RL.\footnote{See \citet{sutton-barto} and \citet{zhao} for brilliant textbook treatments. Also see \citet{murphy} and \citet{silver-RL-course}.} RL means learning value or policy functions of incompletely-known Markov decision processes via some form of Monte Carlo simulation. The key problem addressed by RL is: what to do in dynamic optimization problems in which the agent does not know the exact environment she is operating in, specifically the stochastic process for the underlying state variables? The key insight of RL is that, in such environments, one can still approximate optimal value and policy functions \emph{as long as one can simulate}.

A simple analogy is how to compute the expected value $\mathbb{E}[x]$ of a random variable $x$. The standard way is to compute $\mathbb{E}[x] = \int x f(x)dx$ for a known probability distribution $f(x)$. But what if $f$ is unknown? In such cases, one can often still sample from $f$ and approximate the expected value $\mathbb{E}[x]$ with the sample mean $\bar{x} = \frac{1}{N} \sum_{n=1}^N x_n$.

Building on this intuition, consider the question of how to calculate the following value function:
$$v_{0} = \mathbb{E} \left[\sum_{t=0}^\infty \beta^t u(p_t)\right],$$
where $u$ is a utility function and $p_t$ is some exogenous stochastic process. The standard approach is to use dynamic programming: assume that $p_t$ is Markov with known transition probabilities $f(p'|p)$; then write and solve the Bellman equation
$$v(p) = u(p) + \beta \int v(p')f(p'|p)dp'.$$
An alternative approach is to use Monte Carlo simulation: simply sample $N$ trajectories $\left\{p_t^{n}\right\}_{t=0}^{\rm T}$ for $n=1,...,N$ and approximate the expected value $v_0$ as
$$v_0 \approx \widehat{v}_0 = \frac{1}{N}\sum_{n=1}^N \sum_{t=0}^{T} \beta^t u(p_t^{n}).$$
This basic idea -- to compute expected values via simulation -- lies at the heart of all RL algorithms. Crucially this simulation-based approach does not require knowledge of the transition probabilities $f$. It also works directly with the sequential formulation of the problem. In particular, it is unnecessary to force the problem into the standard Markovian structure required for applying dynamic programming, e.g. by estimating a perceived law of motion for prices $p_t$. As we explain next, our SRL approach to heterogeneous agent macroeconomics uses this same approach to compute expectations about equilibrium prices.

%%%%%%%%%%%%%%%%%%%%%%%%%%%%%%%%%%%%%%%%%%%%%%%%%%%%%%%%%%%%
%%%%%%%%%%%%%%%%%%%%%%%%%%%%%%%%%%%%%%%%%%%%%%%%%%%%%%%%%%%%
\subsection{Revisiting the Agents' Decision Problem}\label{sec:approach_intro}

Recall from Section \ref{sec:setup} that agents choose $\{c_{i, t}\}$ to solve
\begin{equation}\label{eq:objective_general}
    v_{i, 0} 
    = \max_{\{c_{i, t}\}} \mathbb E \bigg[ \sum_{t=0}^\infty \beta^t u(c_{i, t}) \bigg]
    \quad \text{s.t.} \quad
    s_{i, t+1} \sim \mathcal T_s(\cdot \, | \, s_{i,t}, c_{i,t}, z_t, p_t),
    \quad
    p_t = P^*(G_t, z_t) .
\end{equation}
We begin by specifying what individuals observe and on what they condition their decisions.

\begin{assumption}[Information]\label{ass:info}
	At date $t$, an individual observes the entire history of aggregate prices $\{p_t, p_{t-1}, \ldots\}$, but not the cross-sectional distribution $\bm g_t$.
\end{assumption}

\noindent
Assumption~\ref{ass:info} rules out direct conditioning on the distribution but does not, by itself, imply any departure from full information rational expectations. Under standard regularity conditions, a stationary price process $\{p_t\}$ admits a Wold representation and can be written as an infinite-order vector moving average process
\begin{equation*}
	p_t = \sum_{j=0}^{\infty} \kappa_j \varepsilon_{t-j},
\end{equation*}
for some coefficients $\{\kappa_j\}$ and white-noise innovations $\varepsilon_t$. Under additional restrictions (invertibility), the stationary price process also has an equivalent VAR($\infty$) representation,
\begin{equation*}
	p_{t+1} \sim \mathcal T_p \big(\, \cdot \, \big| \, p_t, p_{t-1}, p_{t-2}, \ldots \big),
\end{equation*}
so that the infinite history of prices is sufficient for forecasting under rational expectations. In this sense, the price history is rich enough to recover all information that is relevant to the agents, even though they never observe the cross-sectional distribution directly. Note that the Wold representation theorem effectively converts the recursive formulation for the price process \eqref{eq:price_process} into a sequential stochastic process. It is also worth pointing out that, in the language of the RL literature, Assumption \ref{ass:info} is a ``partial observability" assumption.

Of course, working with the infinite price history is infeasible in practice. Instead, starting from the rational expectations benchmark in Assumption \ref{ass:info}, we restrict attention to low-dimensional summaries of this history. We begin with the most restrictive case, in which agents condition only on current prices.

\begin{assumption}[Restricted state space]\label{ass:policy}
	Agents' decision rules (policy functions) take the form
\begin{equation*}
	\pi(s_t, z_t, p_t),
\end{equation*}
so that policies do not depend on lagged prices.\footnote{
    Whether $z_t$ is payoff-relevant depends on the particular application. If, conditional on $(s_t, p_t)$, $z_t$ affects neither current payoffs nor individual transitions, then it can be dropped from the state vector for the individual's decision problem. For example, $z_t$ is directly payoff-relevant in the Huggett model of Section \ref{sec:setup_huggett} because it scales current income $y_{i, t} z_t$. By contrast, in our Krusell-Smith and HANK applications in Sections \ref{sec:applications_krusell_smith} and \ref{sec:applications_HANK}, $z_t$ matters for individuals only through equilibrium prices, so policies could equivalently be written as $\pi(s_t, p_t)$. For notational uniformity we keep $z_t$ in $\pi(s_t, z_t, p_t)$, but in these cases it is redundant from the household's point of view.
}
\end{assumption}

\noindent
Assumption~\ref{ass:policy} imposes a particular restriction on perceptions: agents treat the current price vector as a sufficient statistic for decision making, even though in the true equilibrium the price process is not Markov in $p_t$ alone. Within this class of policies we then solve problem \eqref{eq:objective_general} by optimizing over many simulated equilibrium paths.

In Section \ref{sec:applications_huggett} we relax Assumption \ref{ass:policy} and allow policies to depend on a short history of past prices. More generally, one could let the policy depend on the hidden state of a recurrent neural network that summarizes the price history; this fits directly into the same RL-based framework \citep{hausknecht-stone}.

Given any (possibly suboptimal) policy $\pi(s, z, p)$, the discretization of individual states on a finite grid $s \in \{s_1,\dots,s_J\}$ allows us to write the policy in vector form
\begin{equation*}
	\bm \pi(z, p) = 
	\left[\begin{matrix}
		\pi(s_1, z, p) \\
		\vdots \\
		\pi(s_J, z, p)
	\end{matrix}\right] .
\end{equation*}
As in Section \ref{sec:setup_discretized}, we represent the cross-sectional distribution as a vector $\bm g_t$, and the individual transitions induced by the policy are encoded in a sparse transition matrix $\mathbf{A}_{\bm\pi(z, p)}$.

%%%%%%%%%%%%%%%%%%%%%%%%%%%%%%%%%%%%%%%%%%%%%%%%%%%%%%%%%%%%
%%%%%%%%%%%%%%%%%%%%%%%%%%%%%%%%%%%%%%%%%%%%%%%%%%%%%%%%%%%%
\subsection{Sequential Restricted Perceptions Equilibrium}

Under Assumptions \ref{ass:info} and \ref{ass:policy}, an individual's relevant state is simply $(s_t, z_t, p_t)$. Given this restricted state space, we define equilibrium as a variant of a restricted perceptions equilibrium (RPE) in the sense of \citet{sargent-RPE} and \citet{branch}. Because we work directly with the sequential formulation of the agents' problem, we term it a ``sequential RPE".

\begin{definition}[Sequential restricted perceptions equilibrium]\label{def:RPE}
A sequential restricted perceptions equilibrium consists of a pair of mappings $(\pi^*(s,z,p), P^*(\bm g, z))$ such that:
\begin{enumerate}
\item \textbf{Optimality.} 
	Given a stochastic process $\{z_t\}_{t \geq 0}$ as well as a price process $\{p_t\}_{t \geq 0}$ generated by $p_t = P^*(\bm g_t, z_t)$, the policy $\pi^* = \{c^*, b^{\prime \ast}\} \in \Pi$ solves 
\begin{equation*}
	\max_{\pi \in \Pi}
	\mathbb E \bigg[ \sum_{t=0}^\infty \beta^t u(c(s_t, z_t, p_t)) \bigg]
	\quad \text{s.t.} \quad
	s_{t+1} \sim \mathcal T_s(\cdot \, | \, s_t, \pi(s_t, z_t, p_t), z_t, p_t),
\end{equation*}
where $c(s, z, p)$ is the consumption component of $\pi(s, z, p)$. Here $\Pi$ denotes the set of measurable policies $\pi: \mathcal{S} \times \mathcal{Z} \times \mathcal{P} \to \mathcal{C} \times \mathcal{B}$ that satisfy the budget and borrowing constraints.

\item \textbf{Market clearing.}  
For every $t$, the market clearing conditions hold: in the Huggett model
\begin{equation}\label{eq:RPE_mkt_clearing}
	\int b'(s, z_t, p_t) \, dG_t(s) = 0,
\end{equation}
where $b'(s, z, p)$ is the saving component of $\pi(s, z, p)$. The solution is a mapping $p_t = P^*(\bm g_t, z_t)$.

\item \textbf{Consistency.}  
When all agents follow $\pi^*$, the cross-sectional distribution evolves according to
\begin{equation*}
	\bm g_{t+1} = \mathbf{A}_{\bm\pi^*(z_t, p_t)}^{\rm T} \bm g_t,
\end{equation*}
where $\mathbf{A}_{\bm\pi(z, p)}$ is the transition matrix induced by $\mathcal T_s$ and $\pi$ and prices are given by $p_t = P^*(\bm g_t, z_t)$.
\end{enumerate}
\end{definition}

\noindent
This equilibrium notion has three features that are important for our purposes. First, it is a \emph{sequential} equilibrium: We work directly with time paths of states and prices rather than with a recursive formulation in terms of a Markov state. Second, it is \emph{self-confirming} in the sense that, given their information and policy, agents' beliefs about price dynamics are consistent with the equilibrium price process along the realized paths. Third, it features \emph{restricted perceptions} because agents condition only on prices, not on the full distribution. This both reflects a realistic informational environment and greatly reduces the dimensionality of the state space. As mentioned in the introduction, in the language of \citet{guarda}, agents' beliefs are both ``narrow" and ``short".

These features make it natural to use RL. In practice, we parameterize the low-dimensional policy function $\pi(s, z, p)$ on the discretized state space and approximate the maximization problem \eqref{eq:objective_general} by evaluating it along many simulated equilibrium paths. In the next subsection, we describe how to simulate these trajectories efficiently under a given candidate policy.

%%%%%%%%%%%%%%%%%%%%%%%%%%%%%%%%%%%%%%%%%%%%%%%%%%%%%%%%%%%%
%%%%%%%%%%%%%%%%%%%%%%%%%%%%%%%%%%%%%%%%%%%%%%%%%%%%%%%%%%%%
\subsection{Simulating the Economy for Given Policy Functions}

Starting from an initial pair $(\bm g_0, z_0)$, the simulated economy evolves according to
\begin{align*}
	z_{t+1} &\sim \mathcal T_z(\cdot \, | \, z_t) \\
	\bm  g_{t+1} &= \mathbf{A}_{\bm\pi(z_t, p_t)}^{\rm T} \bm g_t \\
	p_t &= P^*(\bm g_t, z_t).
\end{align*}
%Updating the aggregate shock $z_{t+1}$ and the distribution $\bm g_{t+1}$ is straightforward: we draw the aggregate shock according to its transition kernel and {\color{red}update the distribution via a sparse matrix–vector multiplication similar to} \citet{young} 
Updating the aggregate shock $z_{t+1}$ is straightforward. For the distribution $\bm g_{t+1}$, we adopt \citet{young}'s non-stochastic simulation method and extend it to a full matrix formulation: the policy function induces a sparse transition matrix over the grid, and the distribution evolves deterministically via matrix-vector multiplication. 
The main computational difficulty lies in the last step. In some models, such as \cite{krusell-smith}, the price functional $P^*(\bm g_t, z_t)$ is available in closed form. In others, such as the Huggett model in Section \ref{sec:setup_huggett}, prices are defined only implicitly by market-clearing conditions. Computing such implicit prices represents a significant challenge. %existing numerical solution methods for heterogeneous agent models with aggregate risk. 
Most existing numerical methods solve a non-linear root-finding problem in every period of the simulation, typically making this step the slowest part of the algorithm; see for example \cite{krusell1997income}, \cite{schaab}, as well as the historical note in footnote \ref{foot:JEDC} in the introduction.
% More broadly, there are three main differences between our approach and the method in \cite{krusell1997income}. First, as is obvious now, \cite{krusell1997income} involves a fixed-point iteration over the perceived law of motion for the bond price and the household policy function, which could be numerically fragile. Our algorithm sidesteps with this outer loop entirely. Second, if prices do not appear in the state space as is typical the case in the Krusell–Smith approach, each iteration requires solving an auxiliary policy function $b'(b,y,Z,K,q)$ given the continuation value function $V(b,y,Z,K)$—see page 5 of my note “Nontrivial Market Clearing Conditions in HA Models” on Dropbox; Gianluca’s notes also discuss this issue. (Note that in our implementation of KS method for the Huggett model in Appendix \ref{supp:KS4Huggett}, we do have price in state space, so there is no such a problem.) Third, our procedure for finding the market-clearing price is different. Along each simulated trajectory, we compute the entire vector of aggregate savings $S(p,z)$ across all grid values of $p$ and $z$ (a step that is computationally cheap due to the vectorized JAX implementation). Given this precomputed aggregate saving schedule, the market-clearing price $p_t$ in each period and trajectory can be found rapidly using a linear interpolation method, rather than a nonlinear root solver as in Krusell–Smith or in Andreas’s job-market paper.
Our method delivers an efficient way to compute these implicit prices along simulated paths. We show next how this works in the Huggett economy of Section \ref{sec:setup_huggett}.

\paragraph{Efficient handling of non-trivial market clearing conditions.} 
In the Huggett model, the equilibrium interest rate $p_t=r_t$ is pinned down by the requirement that bonds must be in zero net supply, see \eqref{eq:eq}. The key insight that allows us to find equilibrium prices efficiently is that, due to Assumption \ref{ass:info}, individual policy functions $\pi(s,z,p)$ depend on current prices $p$ rather than the cross-sectional distribution $G_t(s)$. In addition to the much lower dimensionality of $p$ as compared to $G_t(s)$, this has another key payoff: \emph{policy functions double as individual supply schedules} which can easily be aggregated to obtain \emph{aggregate} supply curves at each point in time which also depend on the current price $p$. Given an aggregate supply curve as a function of $p$, it is then straightforward to solve for the equilibrium price $p_t = P^*(G_t,z_t)$.

To see this in more detail, consider the market clearing condition in our restricted perceptions equilibrium \eqref{eq:RPE_mkt_clearing}. Equivalently,
\begin{equation*}
S_t(p_t,z_t)=0 \quad \mbox{where} \quad S_t(p, z) = \int b'(s, z, p) dG_t(s)
\end{equation*}
is aggregate savings implied by the individual saving policy function $b'(s, z, p)$. The key observation is that the individual saving policy function depends on the interest rate $p$. This means that varying $p$ traces out an entire individual supply \emph{schedule} $p \mapsto b'(s, z, p)$. Aggregating yields the analogous aggregate supply schedule $p \mapsto S_t(p,z)$. The equilibrium price can then be computed time-period by time-period along each simulated path. Our approach of including current prices in the state space to clear markets at each point along a simulation is similar to \citet[][Section 3.5]{krusell-smith-survey}.

\begin{figure}[ht!]
	\centering
	\includegraphics[width=0.6\textwidth]{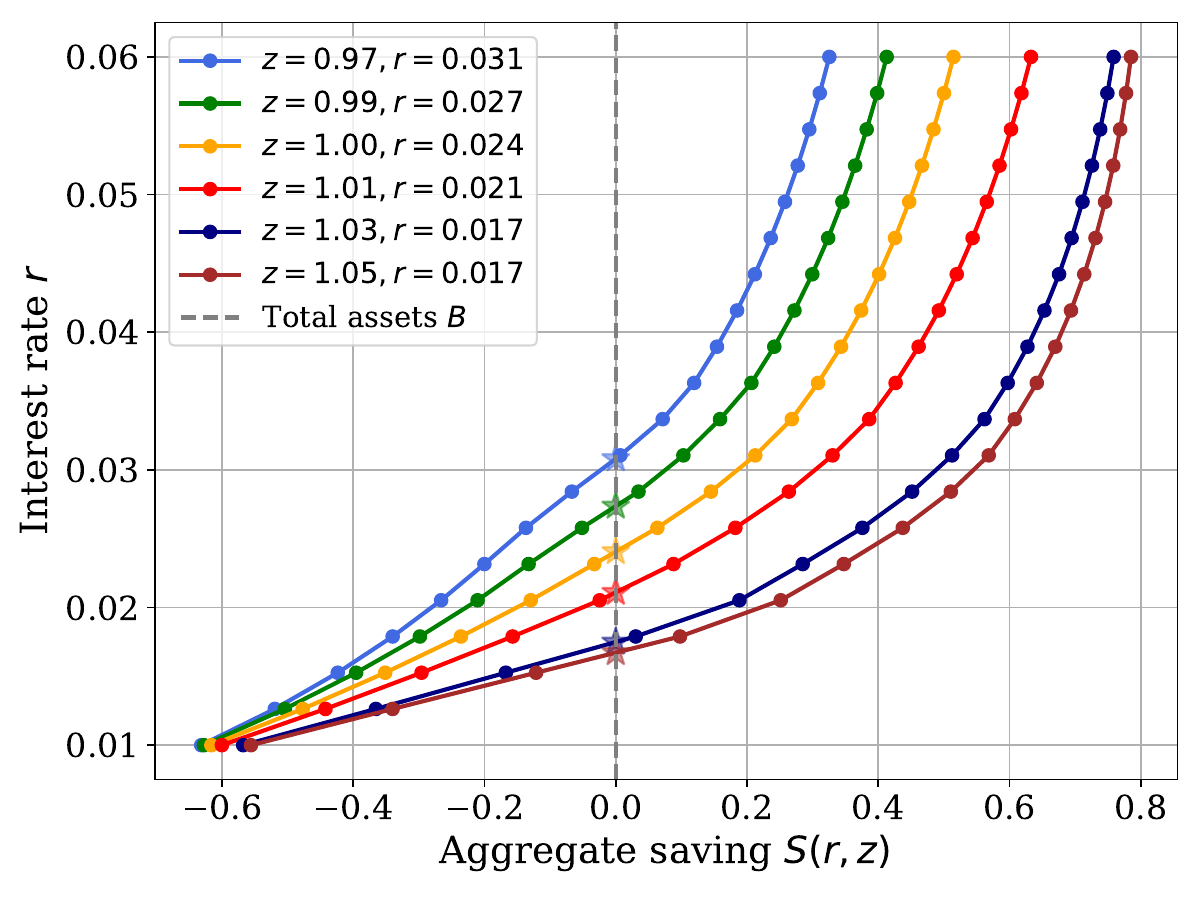}
	\caption{Aggregate saving function $S(r, z)$ in the Huggett economy}
	\label{fig:nontrivial_simulation}
\end{figure}

Finding the equilibrium interest rate can be numerically implemented in a variety of ways. In our experiments, we found that the simplest way of doing this is to use the discretized representation. Once we know the vectors $\bm g_t$ and $\bm b'(z, p)$ for the discretized cross-sectional distribution and saving policy function, we can compute the entire saving schedule at time $t$ for all grid values of $(z, p)$ as
\begin{equation}\label{eq:St_discrete}
	S_t(p, z) = \bm b'(z, p)^{\rm T} \bm g_t.
\end{equation}
Figure \ref{fig:nontrivial_simulation} plots the aggregate saving function $p \mapsto S(p, z)$ in the Huggett economy at some date $t$, with each line corresponding to a grid value of $z$. Given the realized aggregate state $z_t$, we select the corresponding supply curve $S_t(\cdot,z_t)$ and determine the interest rate $p_t$ so that $S_t(p_t, z_t) = 0$. In our numerical experiments the function $S_t(p,z)$ is weakly increasing in $p$ under the optimal policy $\bm b'(z, p)$ and in a neighborhood of the market-clearing price, so the solution is unique.\footnote{
	On the bounded price grid $\{p_k\}_{k=1}^K$ we compute $\{S_t(p_k,z_t)\}_{k=1}^K$ and find the two adjacent points that bracket the root. If
\begin{equation*}
S_t(p_k,z_t) \le 0 \le S_t(p_{k+1},z_t),
\end{equation*}
we set the market-clearing price to the linear interpolant
\begin{equation*}
p_t 
= p_k - \frac{S_t(p_k,z_t)}{S_t(p_{k+1},z_t) - S_t(p_k,z_t)} \big( p_{k+1} - p_k \big).
\end{equation*}
}
However, market clearing is part of the agents' environment and needs to hold for all candidate policy functions, not just at the optimal policy. As we explain in Section \ref{sec:implementation} below, we solve for optimal policies $\pi(s,z,p) = \{c(s,z,p),b'(s,p,z)\}$ iteratively starting from some initial guess. This means that it is important to have a reasonable initial guess for $b'(s,z,p)$ that delivers an upward-sloping aggregate saving supply under this initial guess.\footnote{\label{footnote:guess}
	For example, in our numerical experiments with power utility $u'(c)=c^{-\sigma}$, we have found that the initial guess $c^0(b,y,z,r) = \Bigl(1+r - (\beta (1+r))^{1/\sigma}\Bigr)\Bigl(b + \frac{yz}{r}\Bigr)$ and $b^{\prime 0}(b,y,z,r) = (\beta (1+r))^{1/\sigma}\left(b + \frac{yz}{r}\right) - \frac{yz}{r}$ works well. This would be the optimal policy function if all of $(y,z,r)$ were fixed over time rather than varying stochastically. One can also use this same initial guess but with a parameter $\sigma_0<\sigma$ in place of $\sigma$ which makes saving more responsive to the interest rate $r$ and thus the aggregate saving supply better behaved.
}

Both the dot product in \eqref{eq:St_discrete} and the interpolation step are simple vector operations and are very fast when implemented with JAX on GPUs. It is therefore computationally cheap to compute the entire vector of aggregate savings $S(p, z)$ along each simulated trajectory. Given this precomputed saving schedule, we use linear interpolation to find the market clearing price $p_t$ in each period and trajectory instead of calling a separate non-linear solver inside an outer fixed-point iteration as is common in existing methods \citep{krusell1997income, schaab}.

%%%%%%%%%%%%%%%%%%%%%%%%%%%%%%%%%%%%%%%%%%%%%%%%%%%%%%%%%%%%
%%%%%%%%%%%%%%%%%%%%%%%%%%%%%%%%%%%%%%%%%%%%%%%%%%%%%%%%%%%%
\subsection{Exploiting Agents' Structural Knowledge of their Individual Dynamics}

Standard RL applications, such as board games or Atari environments, typically treat both rewards and state transitions as unknown and learn them entirely from simulated data. Economic models are different. Agents know their preferences and they know how their choices affect their own individual state next period \citep{han-yang-e}. We make explicit use of this structure and refer to our approach as \emph{structural reinforcement learning} (SRL).

The value for an individual $i$ starting from state $(s, z, p)$ and following a given policy $\pi(s, z, p) = \{c(s, z, p), b'(s, z, p)\}$ is 
\begin{equation}\label{eq:value_given_policy}
	v_\pi(s,z,p) = 
	\mathbb E \bigg[ \sum_{t=0}^\infty \beta^t 
		% u(\pi(s_{i, t}, z_t, p_t)) 
		u(c(s_{i, t}, z_t, p_t)) 
		\, \bigg| \, s_{i,0} = s, \; z_0 = z, \; p_0 = p \bigg].
\end{equation}
Here $v_\pi(s, z, p)$ is the value associated with a \emph{fixed} policy $\pi$ and a given stochastic process for the aggregate variables $(p_t, z_t)$. The evolution of the individual state $s_{i, t}$ is governed by the known transition kernel $\mathcal T_s$ or, when individual states are discretized, by the known transition matrices $\mathbf{A}_{\bm\pi(z_t, p_t)}$. Agents know this mapping from current states and actions into next-period states because it is implied by their budget constraint and the exogenous process for idiosyncratic income, which we assume agents know, i.e. agents have rational expectations about their idiosyncratic income process. By contrast, agents \emph{do not} know the law of motion for the aggregate variables $(p_t, z_t)$ which are determined in general equilibrium from everyone's decisions and market clearing. Under rational expectations, one would require agents to know the stochastic process for $(p_t, z_t)$ exactly and to treat $(\bm g_t, z_t)$ as a Markov state so that prices become a function $p_t = P^*(\bm g_t, z_t)$. In our setup, agents instead take the process for $(p_t, z_t)$ as an object to be learned from simulated data.

SRL takes advantage of this separation between individual and aggregate dynamics by \emph{partitioning the state space} $(s,z,p)$ into individual states $s$ and aggregate states $(z,p)$ and treating these differently. We use the structural model to simulate individual transitions and payoffs exactly, i.e., conditional on a policy $\bm \pi(z,p)$, we treat the transition matrix $\mathbf{A}_{\bm\pi(z, p)}$ and hence the mapping $(s_t, z_t, p_t) \mapsto s_{t+1}$ as known by the agent and use this matrix to compute the value function the agent maximizes -- see \eqref{eq:value_vector_given_policy} below. We then use RL only to deal with the low-dimensional but non-Markov aggregate state $(z_t,p_t)$.

In particular, our algorithm updates the policy $\pi$ so as to increase $v_\pi(s, z, p)$ based on simulated histories of $(z_t,p_t)$ rather than using a Bellman equation on the high-dimensional state space $(\bm g_t, z_t)$. This keeps the learning problem low-dimensional while preserving the full heterogeneous-agent structure of the economy. While the simulations feature the full cross-sectional distribution which evolves stochastically and non-linearly over time, we never approximate that distribution nor any mapping from it. Instead only low-dimensional prices enters the agent's state.

%%%%%%%%%%%%%%%%%%%%%%%%%%%%%%%%%%%%%%%%%%%%%%%%%%%%%%%%%%%%
%%%%%%%%%%%%%%%%%%%%%%%%%%%%%%%%%%%%%%%%%%%%%%%%%%%%%%%%%%%%
\subsection{Problem To Be Solved}\label{sec:problem_summary}

The value $v_\pi(s, z, p)$ associated with a given policy $\pi(s, z, p)$ was defined in \eqref{eq:value_given_policy}. The agent's problem is to choose a policy $\pi=(c,b')$ to maximize her value $v_\pi$ given an initial state $(s, z, p)$. It is convenient to use discretized notation and rewrite the value function $v_\pi(s, z, p)$ in vector form as
\begin{equation}\label{eq:value_vector_given_policy}
	\bm v_\pi(z, p) = \mathbb E \bigg[
		\sum_{t=0}^\infty \beta^t \, \mathbf{A}_{\bm\pi, 0 \to t} \, \bm u(\bm c(z_t, p_t))
		\, \bigg | \, z_0 = z, \, p_0 = p \bigg] ,
\end{equation}
where $\bm \pi(z, p) = \{\bm c(z,p),\bm b'(z,p)\}$ is the discretized policy vector and 
\begin{equation}\label{eq:A_cumulative}
	\mathbf{A}_{\bm\pi, 0 \to t} = \mathbf{A}_{\bm\pi(z_0, p_0)} \times \cdots \times \mathbf{A}_{\bm\pi(z_{t-1}, p_{t-1})} 
\end{equation}
denotes the transition matrix of individual states between time $0$ and time $t$ under a particular trajectory $\{z_\tau,p_\tau\}_{\tau=0}^{t-1}$. Note again our partitioning of the state space into $s$ and $(z,p)$: the transition matrices $\mathbf{A}_{\bm\pi(z_t,p_t)}$ keep track of all $s$-transitions while the expectation in \eqref{eq:value_vector_given_policy} is taken only over $(z,p)$-trajectories.\footnote{Tracking value vectors using the transition matrix $\mathbf{A}_{\bm\pi}$ in this way is analogous to the use of such matrices in finite-difference methods for continuous-time HJB equations and HA models \citep{AHLLM}.} Importantly, the transition matrix $\mathbf{A}_{\bm\pi(z_t,p_t)}$ encodes all relevant structural knowledge about the dynamics of agents' own individual states $s$. The presence of this transition matrix in the optimization objective \eqref{eq:value_vector_given_policy} is why we refer to our approach as \emph{structural} RL.

When agents choose policies $\bm \pi$ they \emph{take as given} the evolution of equilibrium prices which evolve according to the true general-equilibrium dynamics,
\begin{equation}\label{eq:prices_as_given}
	p_t = P^*(\bm g_t, z_t) ,
	\qquad
	\bm g_{t+1} = \mathbf{A}_{\bm\pi(z_t, p_t)}^{\rm T} \, \bm g_t, 
	\qquad
	z_{t+1} \sim \mathcal T_z(\cdot \, | \, z_t),
\end{equation}
with $(\bm g_0, z_0)$ given. 

The agents' objective is to find a policy vector $\bm \pi(z, p)$ that maximizes $\bm v_\pi(z, p)$ for all initial values $(z, p)$ taking as given the evolution of equilibrium prices in \eqref{eq:prices_as_given}. Note that maximizing agents take into account the dependence of the transition matrix $\mathbf{A}_{\bm\pi,0\rightarrow t}$ in \eqref{eq:value_vector_given_policy} and \eqref{eq:A_cumulative} on the policy $\bm\pi$; in contrast, because they take prices as given, they \emph{do not} take into account how price dynamics depend on the policy $\bm \pi$ via the term $\mathbf{A}_{\bm\pi(z_t, p_t)}^{\rm T}$ in the Chapman-Kolmogorov equation for $\bm g_t$ in \eqref{eq:prices_as_given}.\footnote{As we explain below, in our SPG algorithm which maximizes the Monte Carlo counterpart to \eqref{eq:value_vector_given_policy} with respect to $\bm \pi$, we apply a stop-gradient to simulated prices to ensure that agents take prices as given.}

%This formulation highlights a simple but important distinction. 

An important feature of this problem is that, while the true state of the economy $(\bm g_t, z_t)$ is extremely high-dimensional (because $\bm g_t$ is a full cross-sectional distribution), the state that enters agents' policy and value functions $(s_t, z_t, p_t)$ is low-dimensional. The agent does not work with a perceived law of motion for prices. Instead, she treats the process for $(p_t, z_t)$ as given, observes realized sequences along simulated paths, and bases decisions on these realizations. As a result, there is no inner-outer fixed-point loop over perceived price laws of motion as in \cite{krusell-smith}. Our algorithm therefore operates directly on the low-dimensional state $(s_t, z_t, p_t)$ from the agent's perspective, while the high-dimensional state $(\bm g_t, z_t)$ only appears in the background through the law of motion for prices.

%%%%%%%%%%%%%%%%%%%%%%%%%%%%%%%%%%%%%%%%%%%%%%%%%%%%%%%%%%%%
%%%%%%%%%%%%%%%%%%%%%%%%%%%%%%%%%%%%%%%%%%%%%%%%%%%%%%%%%%%%
\subsection{Implementation: Structural Policy Gradient Algorithm}\label{sec:implementation}

To evaluate a candidate policy $\pi$, we approximate the value vector using Monte Carlo simulation as explained in Section \ref{sec:approach_monte_carlo}. We simulate $N$ trajectories of the economy under this policy and form the sample analog of the value vector
\begin{equation}\label{eq:vhat}
	\widehat{\bm v}_\pi = \frac{1}{N} \sum_{n=1}^N \bigg[ 
		\sum_{t=0}^{T} \beta^t \mathbf{A}_{\bm\pi, 0 \to t}^n
		u(\bm c(z_t^n, p_t^n)) \bigg],
\end{equation}
where $T$ is a large truncation horizon and the simulated paths are generated from
\begin{equation}\label{eq:prices_as_given_simulation}
	p_t^n = P^*(\bm g_t^n, z_t^n) ,
	\qquad
	\bm g_{t+1}^n = \mathbf{A}_{\bm\pi(z_t^n, p_t^n)}^{\rm T} \bm g_t^n ,
	\qquad
	z_{t+1}^n \sim \mathcal T_z(\cdot \, | \, z_t^n),
\end{equation}
starting from initial conditions $\bm g_0^n \sim \psi_g$ and $z_0^n \sim \psi_z$. Thus, $\widehat{\bm v}_\pi$ in \eqref{eq:vhat} is the sample analog of $\bm{v}_\pi(z,p)$ in \eqref{eq:value_vector_given_policy} averaged over the initial distribution of $(z,p)$ induced by the initial distribution of $(\bm g,z)$, i.e. $\widehat{\bm v}_\pi \approx \mathbb{E}_{z_0 \sim \psi_z,\bm g_0 \sim \psi_g}[\bm{v}_\pi(z_0,p_0)]$ with $p_0=P^*(\bm g_0, z_0)$.

In practice, we work with a scalar objective that averages over initial individual states. Let $\bm d_0$ denote the uniform distribution on the individual-state grid, so that $d_0(s_j) = 1/J$. We then maximize
\begin{equation}\label{eq:scalar}
	\mathcal L(\bm\theta) = \bm d_0^{\rm T} \widehat{\bm v}_\pi.
\end{equation}
Because our state space is low-dimensional, we can work with a grid-based (tabular) approach and we parameterize the policy as
\begin{equation*}
	\bm \theta = \left[\begin{matrix}
	\bm \pi(z_1, p_1)\\ \vdots\\ \bm \pi(z_K, p_L)
	\end{matrix}\right] = \left[\begin{matrix}
\pi(s_1,z_1, p_1)\\ \vdots\\ \pi(s_J,z_K, p_L)
	\end{matrix}\right] ,
\end{equation*}
where $J$, $K$, and $L$ are the numbers of grid points on the $s$, $z$, and $p$ grids. That is, the parameter vector $\bm \theta$ is simply the $J \times K \times L$-dimensional vector of values of the policy on the $(s,z, p)$ grid.\footnote{
	One can instead parameterize the policy as a neural network $\pi(s, z, p; 
	\bm\theta)$. For the low-dimensional policy functions in this paper, a grid-based parameterization is sufficient.
}

Furthermore, in practice it is costly to compute the high-dimensional matrix multiplications $\mathbf{A}_{\bm\pi, 0 \to t}^n = \mathbf{A}_{\bm\pi(z_0^n, p_0^n)} \times \cdots \times \mathbf{A}_{\bm\pi(z_{t-1}^n, p_{t-1}^n)} $ in \eqref{eq:vhat}. After substituting in \eqref{eq:scalar}, we therefore rewrite the optimization objective as
\begin{equation*}
\mathcal L(\bm\theta) = \frac{1}{N}\sum_{n=1}^N\sum_{t=0}^T \beta^t  (\bm{d}_{\bm \pi, t}^n)^{\rm T} u(\bm c(z_t^n, p_t^n)),
\end{equation*}
where $\bm{d}_{\bm \pi, t}^n = (\mathbf{A}_{\bm\pi,0 \rightarrow t}^n)^{\rm T} \bm{d}_0$ is the cross-sectional distribution of $s$ at time $t$ under policy $\bm \pi$ starting from the initial uniform distribution $\bm{d}_0$. We compute this distribution iteratively by solving the Chapman-Kolmogorov equation $\bm{d}_{\bm \pi, t+1}^n = (\mathbf{A}^n_{\bm \pi(z_t,p_t)})^{\rm T} \bm{d}_{\bm \pi,t}^n$ forward in time using the \citet{young} method.
\footnote{\citet{azinovic-gaegauf-scheidegger} use the \citet{young} method in a related fashion to construct an optimization objective involving a discretized cross-sectional distribution. However, they use neural networks to minimize Euler equation errors whereas we use a grid-based approach to maximize a value vector.}

\begin{figure}[ht!]
\centering
\includegraphics[width=\textwidth]{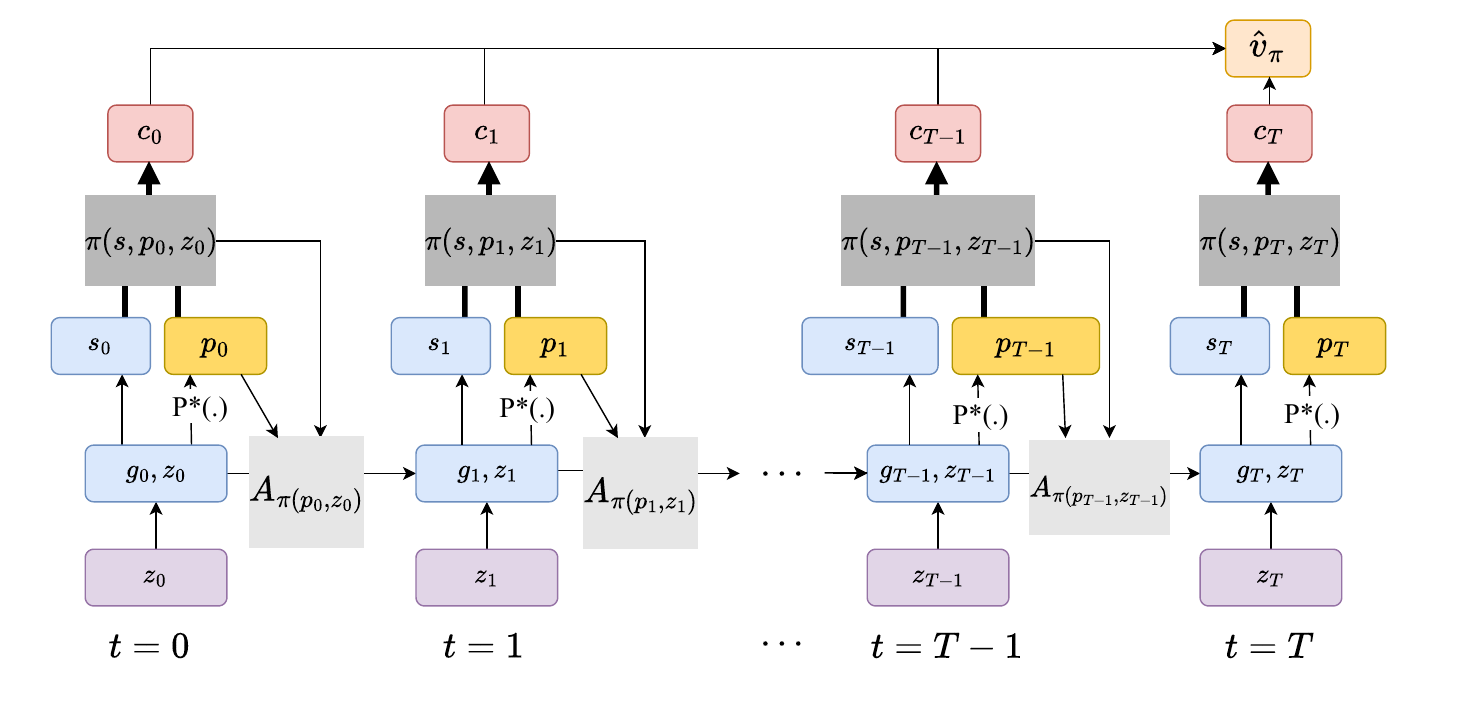}
\caption{Computational graph}
\label{fig:HARL_graph_p_z}
\end{figure}

Figure \ref{fig:HARL_graph_p_z} presents a computational graph that illustrates the algorithm. Since the mapping from parameters $\bm \theta$ to the objective $\mathcal L(\bm \theta)$ is differentiable, we use stochastic gradient ascent (or variants) to update
\begin{equation*}
	\bm\theta^{k+1} = \bm\theta^k + \eta_k \nabla_\theta \mathcal L(\bm\theta^k),
\end{equation*}
where $\eta_k$ is the learning rate at iteration $k$. A stop-gradient is applied to the price update $p_t^n = P^*(\bm g_t^n, z_t^n)$ in \eqref{eq:prices_as_given_simulation} so that, when computing gradients, prices are treated as exogenous. This is consistent with the standard notion of competitive equilibrium, in which agents take the process for $(p_t, z_t)$ as given and do not internalize how their behavior affects price formation.
Finally, we assess convergence of the algorithm by tracking the change in the policy vector $\bm \theta$ over the $(s, z, p)$ grid. Convergence is achieved when the $L^\infty$ (sup) norm of the policy update falls below a tolerance,
\begin{equation*}
	\| \bm \theta^{k+1} - \bm \theta^k \|_\infty < \epsilon_\text{converge}
	\qquad \text{ where } \qquad
	\| \bm x \|_\infty \equiv \max_{j, k, l} |x_{jkl}|.
\end{equation*}
We report our choice of convergence tolerance $\epsilon_\text{converge}$ for each application in Appendix \ref{app:applications}. Algorithm \ref{alg:SPG} summarizes the implementation.

\begin{algorithm}[ht!]
\SetAlgoLined
\SetKwInOut{Input}{Input}
\SetKwInOut{Output}{Output}

\Input{Initial policy parameters $\bm\theta^{0}$; step size sequence $\{\eta_k\}$;  
number of simulated trajectories $N$; horizon $T$.}
\Output{Approximate optimal policy parameters $\theta^{*}$.}

\begin{enumerate}
    \item Initialize parameters $\bm\theta^{0}$.
    \item \textbf{For each iteration $k=0,1,2,\dots$}:
    \begin{enumerate}
        \item Simulate $N$ trajectories
			\begin{equation*}
        \{(z_t^n,p_t^n,\mathbf{g}_t^n)\}_{t=0}^{T},
        \qquad n=1,\dots,N,
	\end{equation*}
        using policy $\bm\pi(\cdot;\bm\theta^{k})$ and market clearing conditions.
        \item Compute the sample objective
			\begin{equation*}
           \mathcal{L}(\bm\theta^{k})
              =  \bm d_0^{\rm T}\widehat{\bm v}_{\pi} \quad \mbox{where} \quad 
        \widehat{\bm v}_{\pi}
            = \frac{1}{N}\sum_{n=1}^{N}
                \Bigg[
                    \sum_{t=0}^{T}
                        \beta^{t}\,
                        \mathbf{A}_{\bm\pi, 0 \to t}\,
                        u(\bm c(z_t^n, p_t^n))
                \Bigg].
			\end{equation*}
        \item Update parameters by stochastic gradient ascent (or variants):
			\begin{equation*}
            \bm\theta^{k+1}
                = \bm\theta^{k}
                  + \eta_{k}\,\nabla_{\theta}\mathcal{L}(\bm\theta^{k}).
			  \end{equation*}
        \item Stop when convergence criteria are met.
    \end{enumerate}
\end{enumerate}
\caption{Structural Policy Gradient Algorithm}
\label{alg:SPG}
\end{algorithm}

%%%%%%%%%%%%%%%%%%%%%%%%%%%%%%%%%%%%%%%%%%%%%%%%%%%%%%%%%%%%
%%%%%%%%%%%%%%%%%%%%%%%%%%%%%%%%%%%%%%%%%%%%%%%%%%%%%%%%%%%%
%%%%%%%%%%%%%%%%%%%%%%%%%%%%%%%%%%%%%%%%%%%%%%%%%%%%%%%%%%%%
\section{Computational Experiments}\label{sec:applications}

In this section we report computational experiments for three benchmark economies: the Huggett model from Section \ref{sec:setup}, the classic \cite{krusell-smith} model, and a one-account heterogeneous agent New Keynesian (HANK) model with nominal rigidities.

We implement all three models in JAX and run them on a single NVIDIA A100 GPU on Google Colab. Table \ref{tab:Table_runtime_all} summarizes performance. 
Since our algorithm is stochastic and uses Monte Carlo simulations, we run our algorithm $10$ times for each specification and report averages across runs. The first column shows the average number of epochs until convergence, and the last column the corresponding average time for a single run.

\begin{table}[H]
	\centering
	\begin{tabular}{lcccc}
\hline\hline
Model & Average converge epoch & \# Runs & Average Runtime (sec) & \\
\hline
Krusell-Smith & 438.4 & 10 & 56.55 \\
Huggett with agg. shocks & 480.6 & 10 & 75.29 \\
HANK with agg. shocks & 496.5 & 10 & 199.53 \\
\hline
Partial equilibrium (Huggett) & 289.3 & 10 & 39.49 \\\hline\hline
\end{tabular}
	\caption{Runtimes}
	\label{tab:Table_runtime_all}
\end{table}

Solving the Krusell-Smith model takes about 55 seconds, in line with other global solution methods in the literature. By contrast, the Huggett and HANK models are typically viewed as more challenging because they feature non-trivial market clearing conditions: standard approaches nest an inner loop that repeatedly solves for prices until markets clear. Nevertheless, our method solves the Huggett model in 75 seconds and the HANK model in about 3 minutes. 

Finally, we also compare the cost of computing the model's general equilibrium (GE) to that of computing the corresponding partial equilibrium (PE) problem (see below). We find that, while computing the GE problem takes longer as expected, the difference in runtime is modest. In the Huggett model, for example, moving from partial to general equilibrium increases runtime from 39 seconds to 75 seconds. This is because we do not solve general equilibrium with a nested inner-outer loop that alternates between solving optimal policies and updating price functions or perceived laws of motion. Instead, prices are learned in an online fashion: along each simulated path we compute the market-clearing price implied by current policies, and the policy update uses these realized prices directly.

%%%%%%%%%%%%%%%%%%%%%%%%%%%%%%%%%%%%%%%%%%%%%%%%%%%%%%%%%%%%
%%%%%%%%%%%%%%%%%%%%%%%%%%%%%%%%%%%%%%%%%%%%%%%%%%%%%%%%%%%%
\subsection{Huggett Model with Aggregate Risk}\label{sec:applications_huggett}

\begin{figure}[ht!]
\centering
    \includegraphics[width=0.8\textwidth]{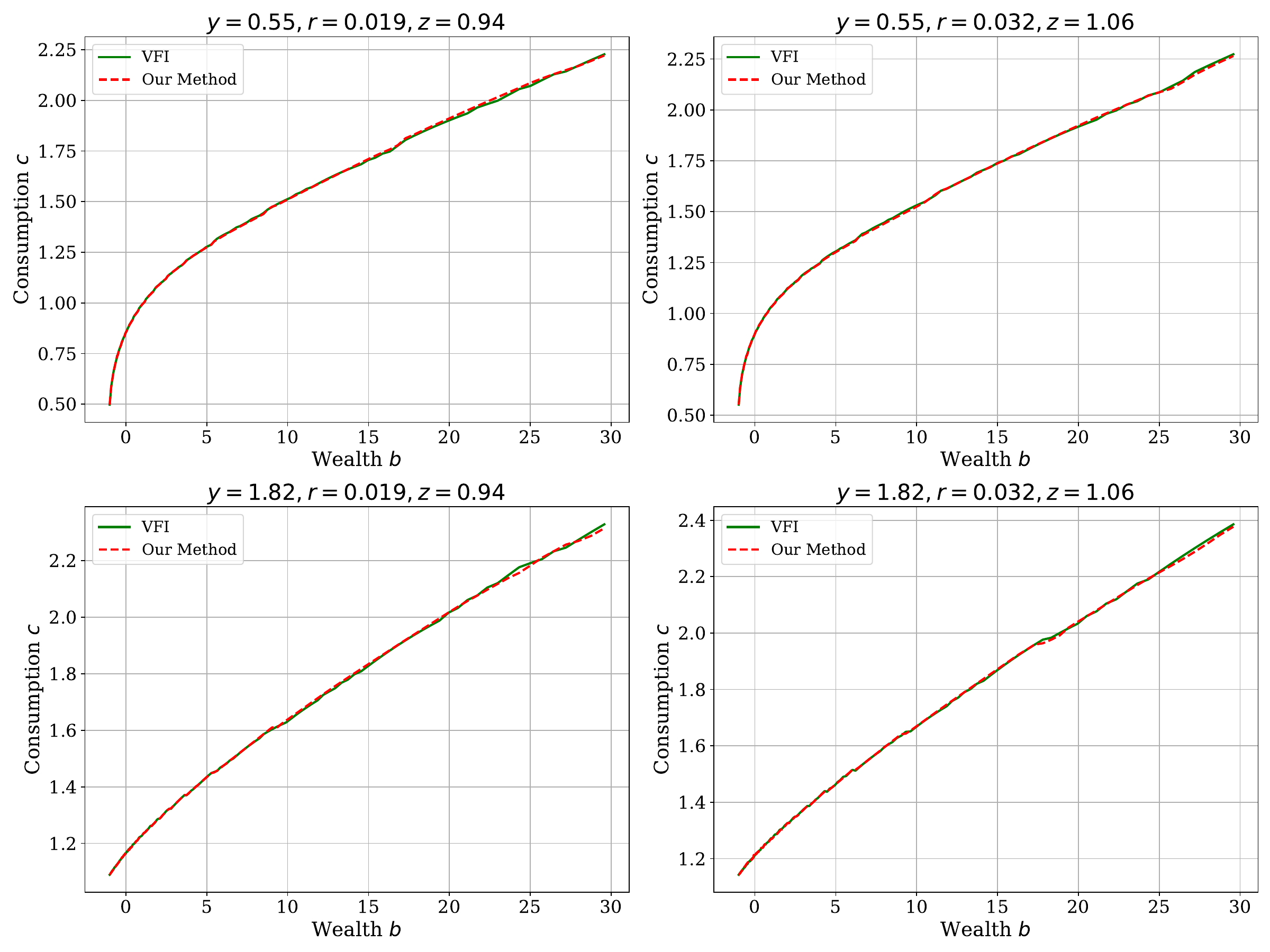}
    \caption{Solution comparison for the PE problem: SRL vs VFI}
    \label{fig:policy_compare_vfi_PE_Huggett}
\end{figure}

We start our computational experiments with the Huggett economy described in Section \ref{sec:setup}.\footnote{
	Our comparison of runtimes in Table \ref{tab:Table_runtime_all} and Figure \ref{fig:policy_compare_vfi_PE_Huggett} below make reference to the partial equilibrium problem of the Huggett economy. We present the details in Appendix \ref{app:applications_huggett}.
}

\paragraph{Calibration.}
We interpret one period as a year and set the household discount factor to $\beta = 0.96$. Preferences are isoelastic $u(c) = \frac{c^{1-\sigma}}{1-\sigma}$ and we set $\sigma = 2$.
In the Huggett model, both the idiosyncratic and the aggregate income components follow log AR(1) processes. We set the persistence parameters to $\rho_y = 0.6$ and $\rho_z = 0.9$ and the standard deviations of the innovations to $\nu_y = 0.2$ and $\nu_z = 0.02$. We discretize these processes on finite grids using a standard Tauchen procedure (details in the Appendix \ref{app:applications}). Finally, we impose a borrowing limit $\underline b = -1$ and fix aggregate bond supply at $B = 0$, so bonds are in zero net supply. The full calibration table is presented in Appendix \ref{app:applications}.

\paragraph{Hyperparameters.}
We discuss and report hyperparameter choices for all our experiments in the Appendix \ref{app:applications}.

\begin{figure}[ht!]
\centering
\begin{subfigure}[t]{.32\textwidth}
\centering
\includegraphics[width=\linewidth]{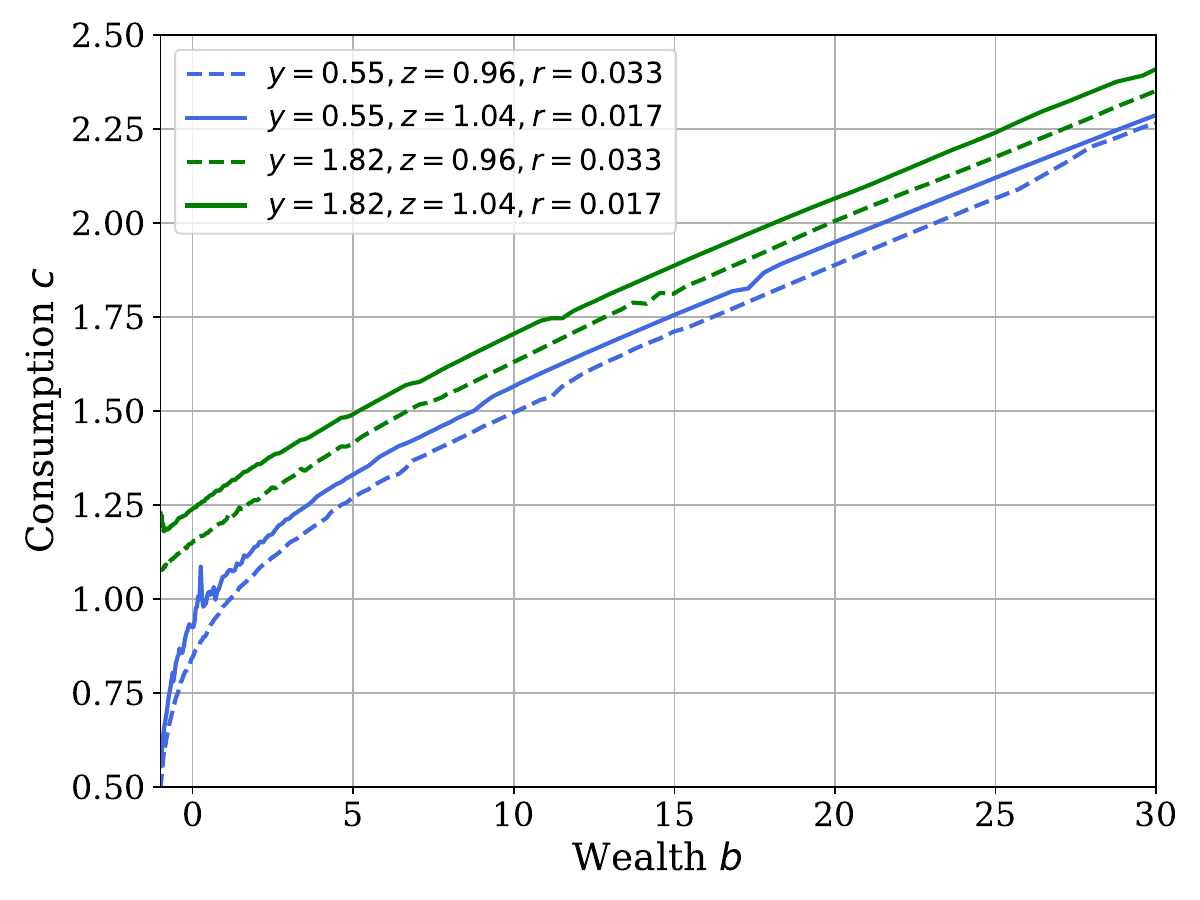}
\vspace{-7mm}
\caption{Optimal consumption policy}
\end{subfigure}
\begin{subfigure}[t]{.32\textwidth}
\centering
\includegraphics[width=\linewidth]{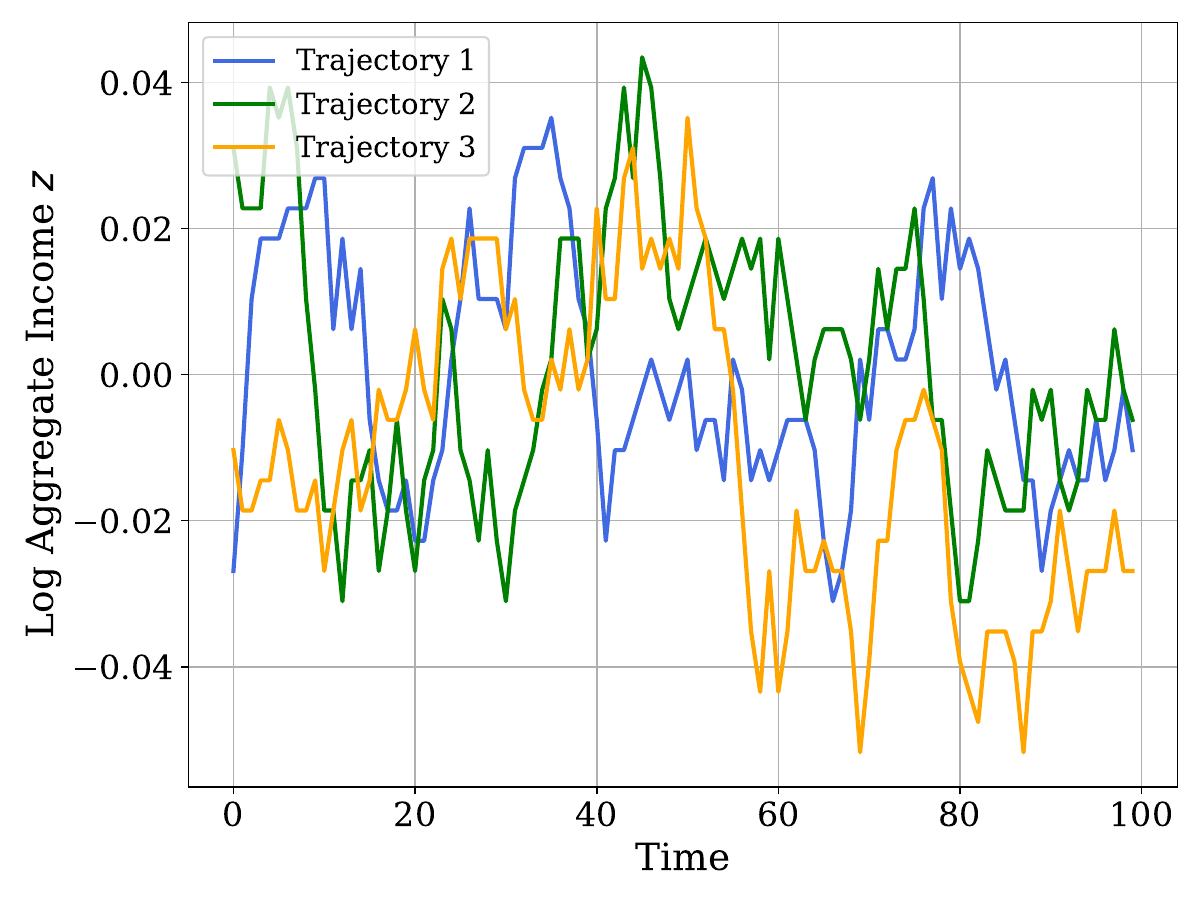}
\vspace{-7mm}
\caption{Aggregate state}
\end{subfigure}
\begin{subfigure}[t]{.32\textwidth}
\centering
\includegraphics[width=\linewidth]{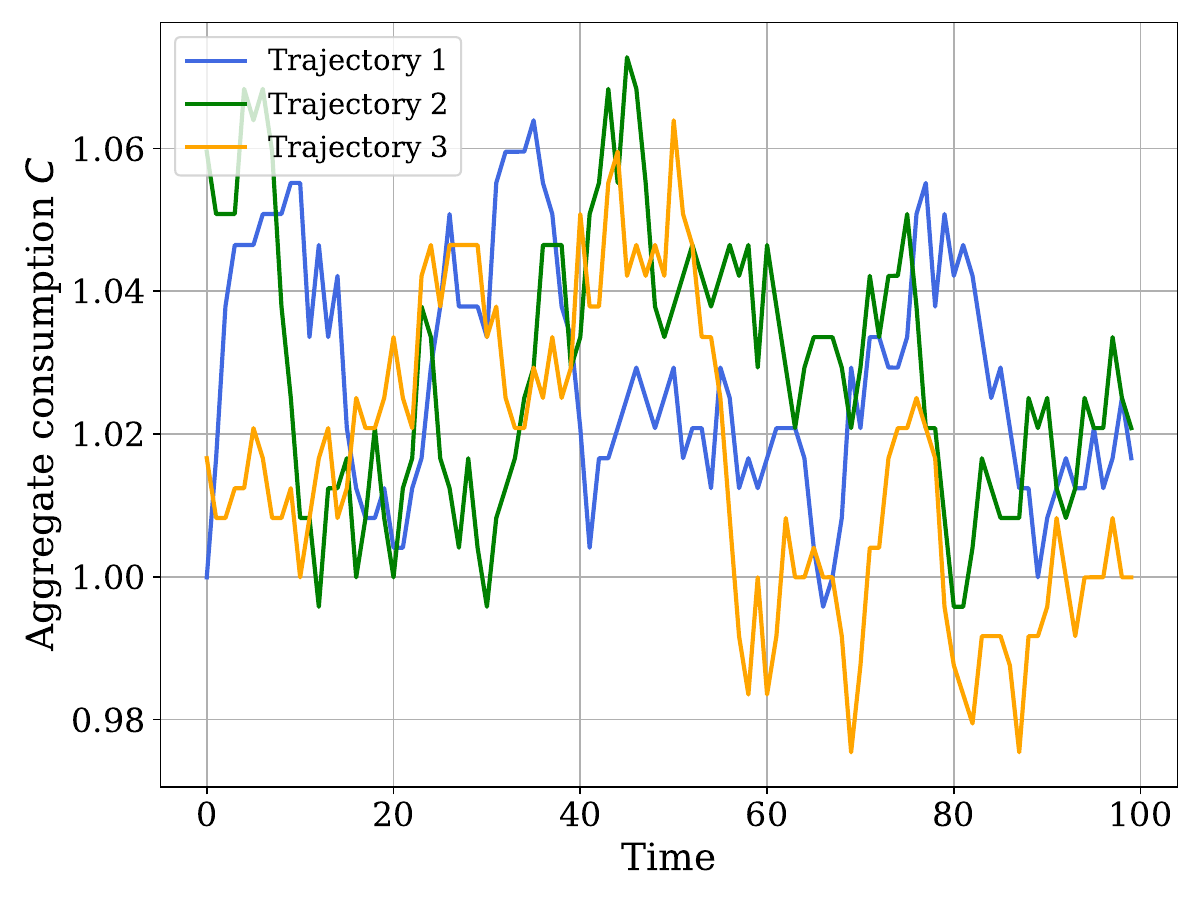}
\vspace{-7mm}
\caption{Aggregate consumption}
\end{subfigure}
\vspace{2mm}

\begin{subfigure}[t]{.32\textwidth}
\centering
\includegraphics[width=\linewidth]{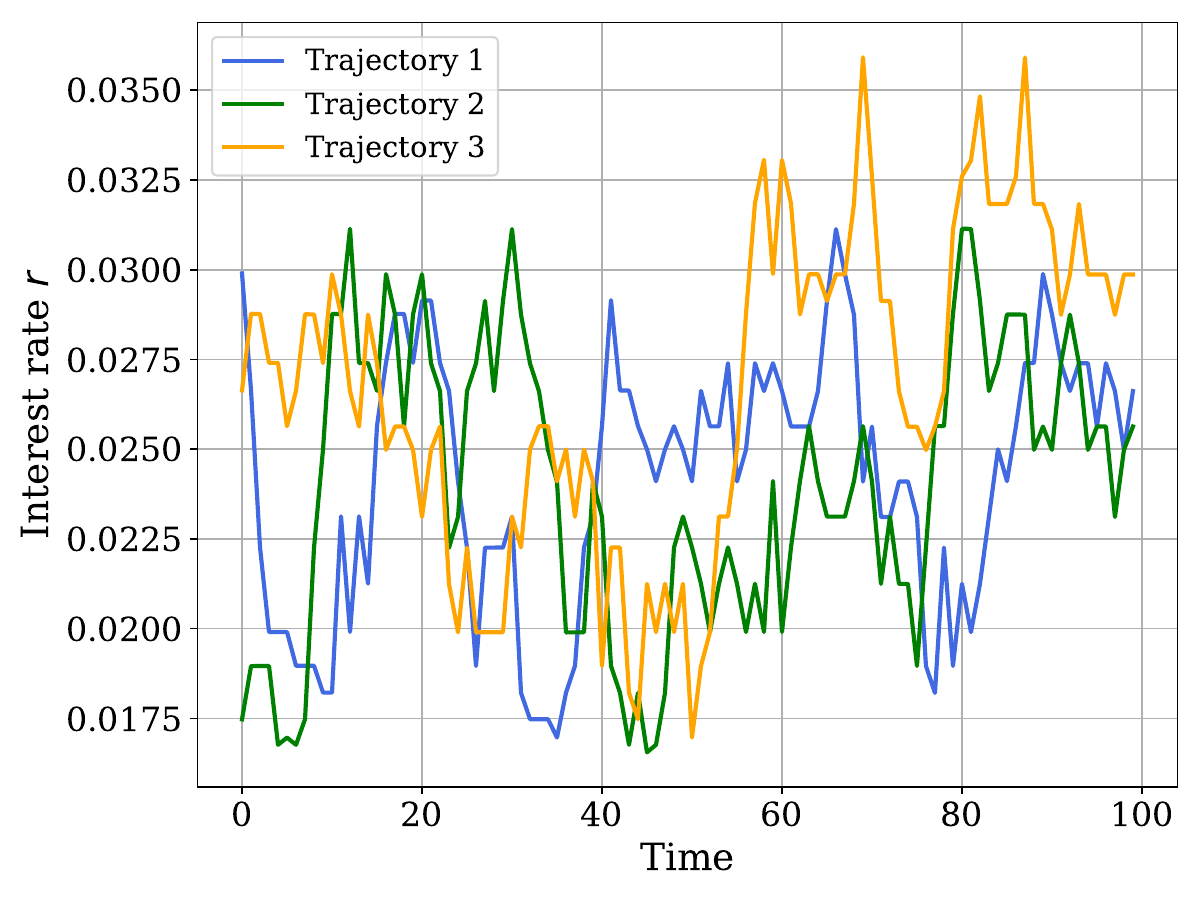}
\vspace{-7mm}
\caption{Interest rate}
\end{subfigure}
\begin{subfigure}[t]{.32\textwidth}
\centering
\includegraphics[width=\linewidth]{figures/HUGGETT/HUGGETT_S_p_z.pdf}
\vspace{-7mm}
\caption{Saving schedule}
\end{subfigure}
\begin{subfigure}[t]{.32\textwidth}
\centering
\includegraphics[width=\linewidth]{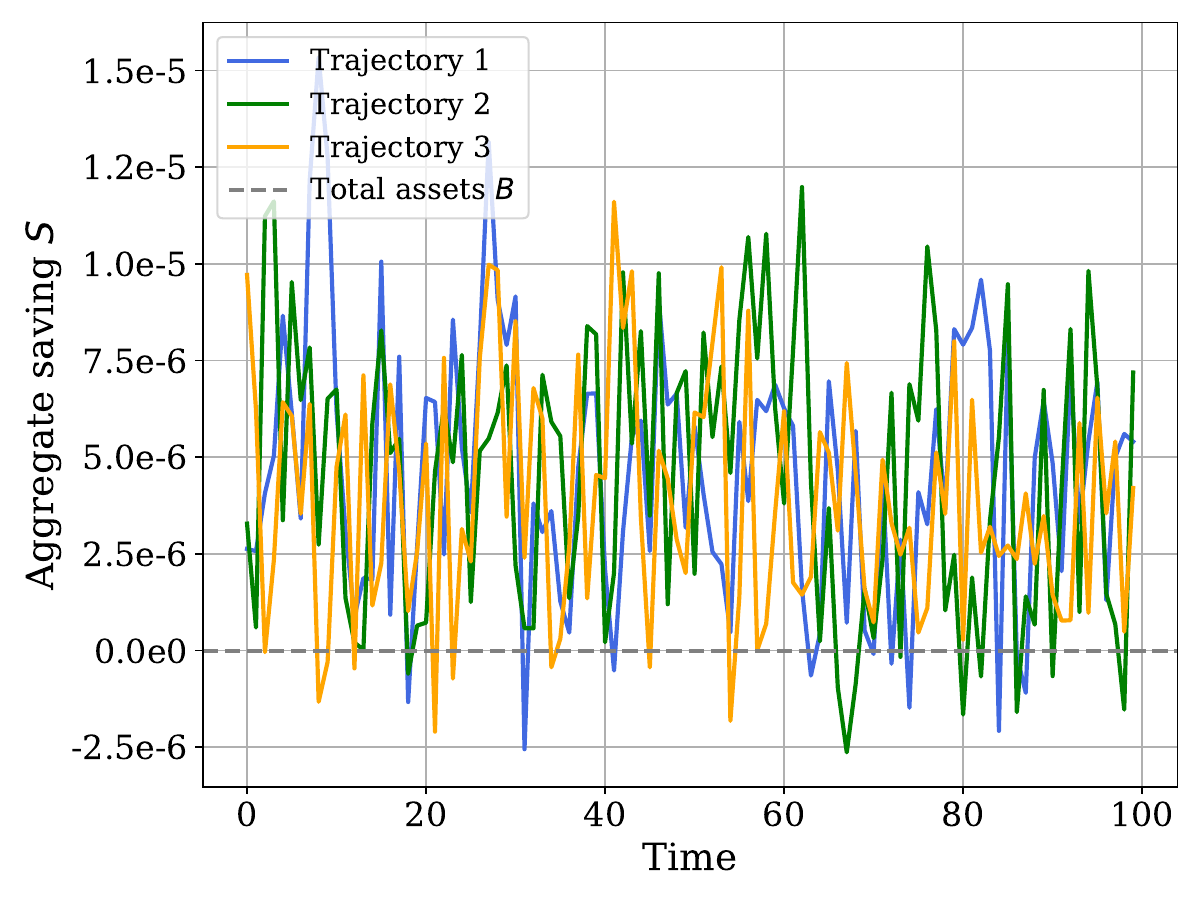}
\vspace{-7mm}
\caption{Aggregate saving}
\end{subfigure}
\caption{Simulation Results}
\label{fig:huggett_simulations}
\end{figure}

\paragraph{Numerical Results.}
Figure \ref{fig:huggett_simulations} reports the numerical solution and a simulation for the Huggett economy.

Panel (a) plots the optimal consumption policy as a function of wealth $b$ on the horizontal axis. Each line corresponds to one combination of individual income, aggregate income, and prices. We choose the realizations of $(y_t, z_t, p_t)$ that occur frequently in our simulations. The policy is monotonically increasing and concave in $b$, as expected from standard theory: richer households consume more, but at a decreasing marginal propensity.

Panels (b)-(d) display simulated time series for the aggregate state, aggregate consumption and equilibrium prices. Panel (b) shows the exogenous process for log aggregate income $z_t$. Panel (c) plots aggregate consumption $C_t$, which co-moves with $z_t$ but is somewhat smoother due to precautionary savings and imperfect risk sharing. Panel (d) shows the resulting interest rate $r_t$, which adjusts endogenously to clear the bond market in the presence of incomplete markets and zero net bond supply.

Panel (e) revisits the equilibrium bond demand schedule $S(p, z)$ discussed in Section 3.4, now evaluated at the trained policy. For different realizations of the aggregate state $z$, the figure shows how the aggregate demand for bonds varies with the interest rate. Market clearing corresponds to the intersection of $S(p, z)$ with zero.

Finally, Panel (f) plots the bond-market clearing residual along the simulated path, i.e. the difference between aggregate bond holdings implied by households' policies and the fixed supply of zero. The residual remains very close to zero throughout the simulation. The small deviations that do arise are due to numerical interpolation in prices rather than to a failure of the algorithm to enforce equilibrium. In practice, these deviations are negligible both in absolute terms. The average gap in bond market clearing for a single run is about $4.4\times 10^{-6}$.

\paragraph{Partial equilibrium problem.}
To gauge the accuracy of our SRL approach, we first consider a partial equilibrium (PE) version of the Huggett economy. In PE, households take as given an exogenous Markov process for interest rates and solve their individual dynamic program using either our SRL method or a standard value function iteration (VFI) algorithm. Because the PE environment dispenses with the fixed point over prices and distributions, it is a setting in which there is broad agreement on the correctness of conventional VFI solutions. This makes it a natural benchmark against which to compare the policies implied by our method. We describe the details of the PE specification and calibration in Appendix \ref{app:applications_huggett}. 

Figure \ref{fig:policy_compare_vfi_PE_Huggett} reports the comparison. Each panel plots the optimal consumption policy as a function of wealth $b$ for four combinations of individual income $y$ and interest rates $r$. The dashed line shows the policy obtained from our SRL algorithm, while the solid line shows the corresponding VFI solution in the PE environment. Across all four panels, the two sets of policy functions are almost indistinguishable. This comparison is a first reassuring test of the accuracy of our SRL method. It shows that, in a setting where a trusted VFI benchmark is available, our SRL approach replicates the rational expectation solution closely. 

\paragraph{Solutions with Lagged Price History.} 
A complementary way to assess the restrictiveness of conditioning only on $p_t$ is to enlarge the observable state with lagged prices. Conceptually, this moves us part of the way toward the full MA($\infty$) representation of the agent's problem, which in the limit would reproduce the rational expectations solution. Concretely, we now re-solve the Huggett model allowing households to keep track of one lagged price, so that $p_{t-1}$ becomes an additional state variable.

\begin{figure}[ht!]
\centering
\includegraphics[width=\textwidth]{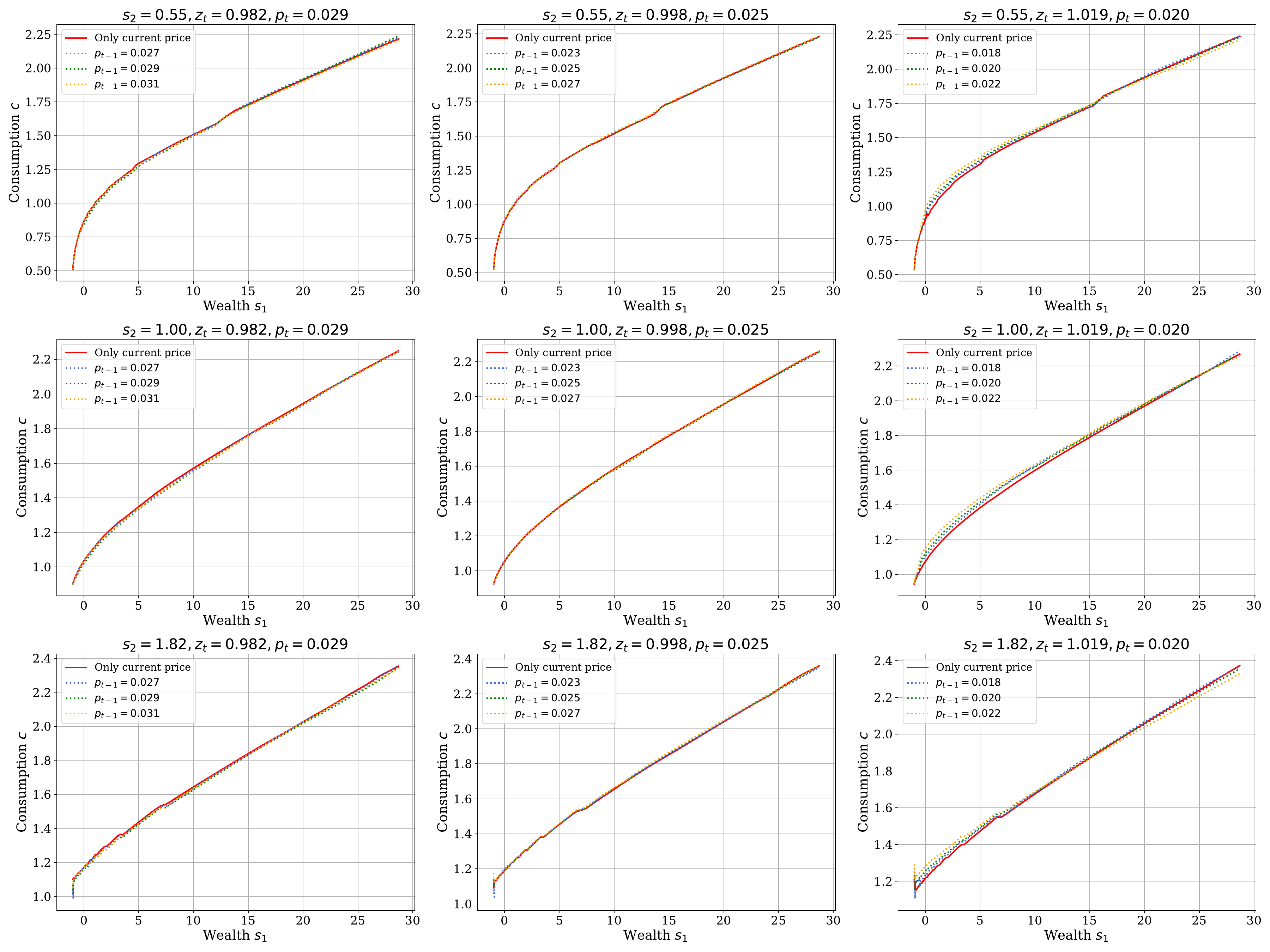}
\caption{Consumption Policy Function with Price Lags $p_{t-1}$ as State Variable}
\label{fig:huggett_price_lag}
\end{figure}

From a computational standpoint, this extension is straightforward: our method can accommodate a small number of lagged observables without reintroducing the curse of dimensionality. From an economic standpoint, it should, in principle, help agents forecast: because prices are not Markov in $p_t$ alone, a longer price history ought to contain incremental information about future prices.

Figure \ref{fig:huggett_price_lag} shows that, in practice, the effect of this additional information is minimal. The figure compares solutions to the Huggett model when agents condition (i) only on the current price $p_t$ (solid blue line) and (ii) on both $p_t$ and its lag $p_{t-1}$ (dashed lines). Each panel plots the consumption policy across wealth $b$ for fixed values of the individual income state $y$, the aggregate income state $z$, and the current interest rate $r$. Different dashed lines correspond to different realizations of \textit{past} prices.

Across all panels, the dashed lines lie almost on top of the solid line. That is, once we fix the current state $(y_t, z_t, p_t)$, optimal consumption is almost insensitive to the additional information contained in $p_{t-1}$. This suggests that, at least in the Huggett environment, current prices already summarize the relevant aspects of the history for household decisions, and that extending the observable state to include one lag has only a negligible impact on behavior.

\begin{figure}[ht!]
\centering
\begin{subfigure}[t]{.45\textwidth}
\centering
\includegraphics[width=\linewidth]{figures/HUGGETT/HUGGETT_policy_s1.pdf}
\vspace{-7mm}
\caption{Policy with $N_\text{sample} = 512$}
\end{subfigure}
\begin{subfigure}[t]{.45\textwidth}
\centering
\includegraphics[width=\linewidth]{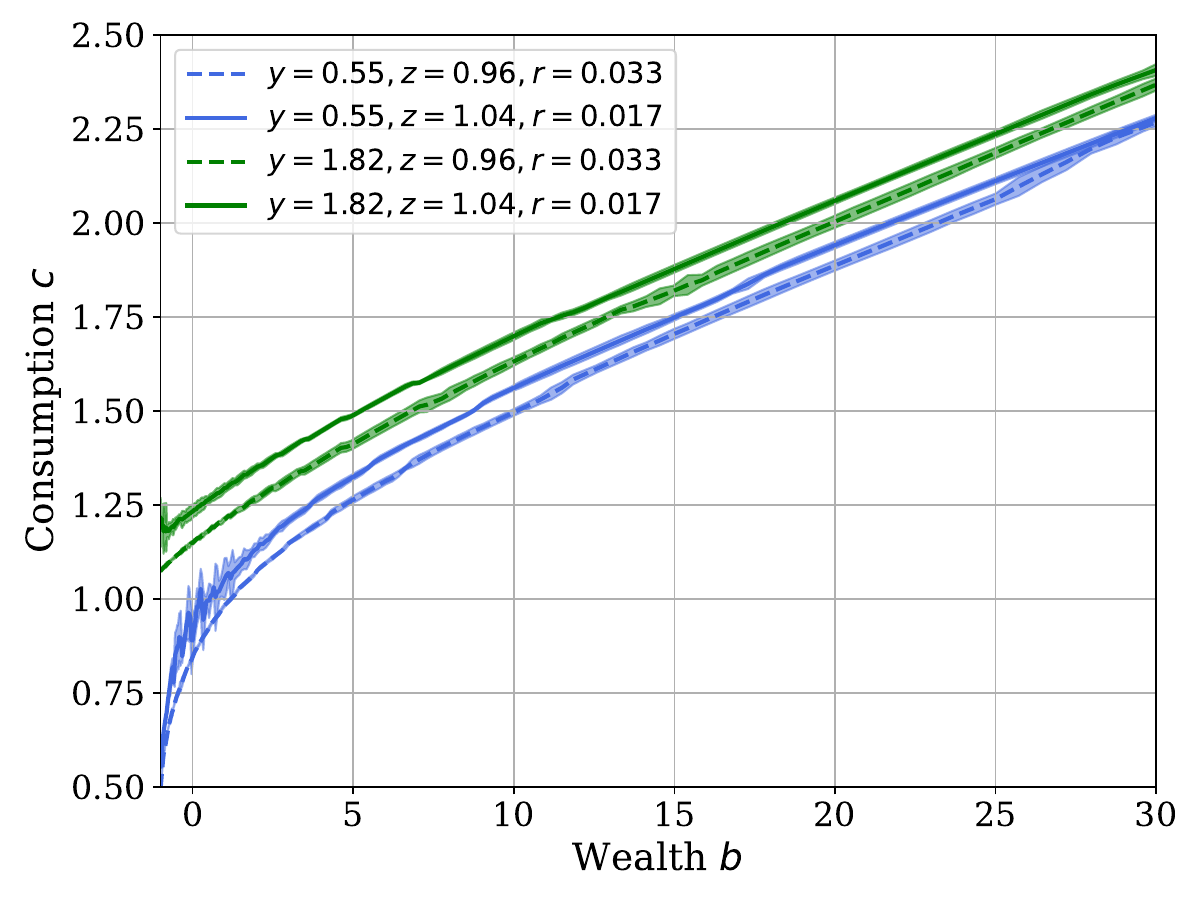}
\vspace{-7mm}
\caption{Policy CI with $N_\text{sample} = 512$}
\end{subfigure}
\vspace{2mm}

\begin{subfigure}[t]{.45\textwidth}
\centering
\includegraphics[width=\linewidth]{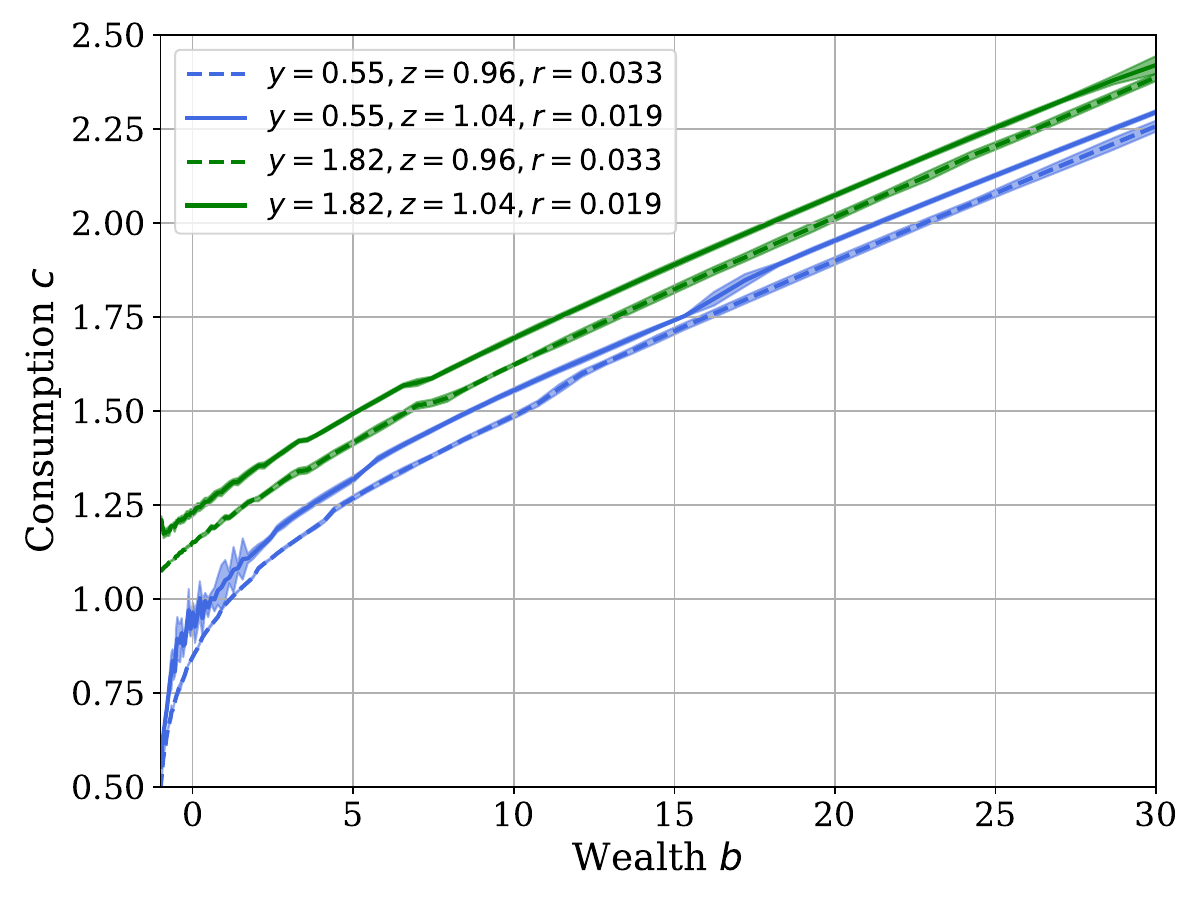}
\vspace{-7mm}
\caption{Policy CI with $N_\text{sample} = 2048$}
\end{subfigure}
\begin{subfigure}[t]{.45\textwidth}
\centering
\includegraphics[width=\linewidth]{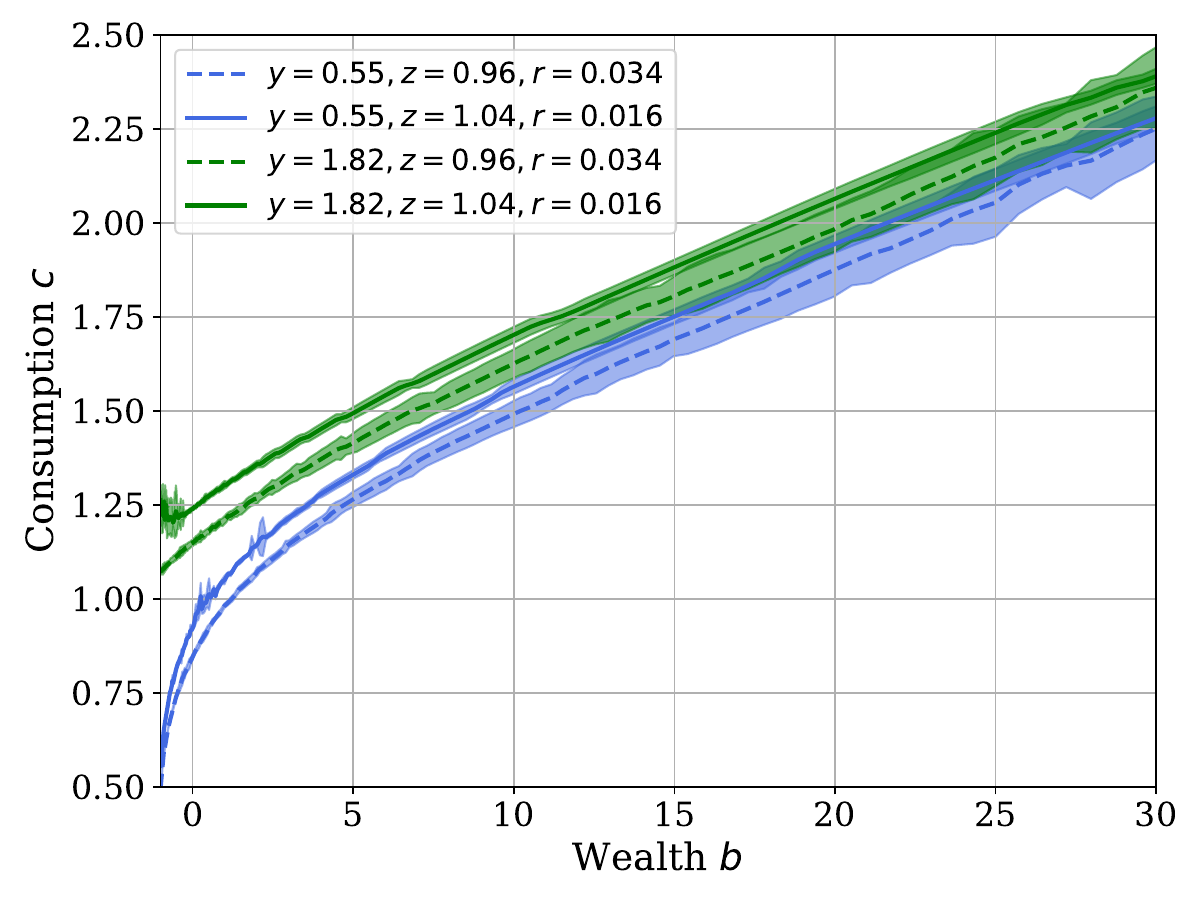}
\vspace{-7mm}
\caption{Policy CI with $N_\text{sample} = 24$}
\end{subfigure}
\caption{Dependence of Policies on Sample Size in the Huggett model}
\label{fig:Huggett_sample_size_comparisons}
\end{figure}

\paragraph{Dependence on the number of trajectories (sample size).} 
Next, we study how the quality and stability of the learned policy depend on the number of simulated trajectories used for training. Figure \ref{fig:Huggett_sample_size_comparisons} summarizes these results.

Panel (a) reports the consumption policy obtained from a single training run with 512 simulated trajectories. This number of trajectories is a key hyperparameter in our algorithm: it controls how many distinct state-price paths agents observe and learn from. The resulting policy is monotone and concave in wealth, as theory would suggest. % and visually very similar to the policies we documented in the Huggett experiment.

Panel (b) turns to sampling uncertainty. Here we compute pointwise 95\% confidence intervals (CIs) for the policy based on 10 independent training runs, each again using 512 trajectories. The figure shows that the confidence bands are quite tight across the wealth distribution. The bands widen modestly at higher wealth levels. This pattern is natural: states with high wealth are visited only rarely in the simulations --- indeed, beyond roughly $b > 10$ there is essentially no mass in the stationary distribution, so the algorithm has fewer observations from which to learn. Regions of the state space that are visited infrequently thus come with more sampling noise in the estimated policy.

Panels (c) and (d) vary the number of training trajectories to make this dependence more transparent. In Panel (c), we increase the number of trajectories. The point estimates of the policy change very little, but the confidence bands shrink, including in the high-wealth region. In other words, feeding the algorithm more data primarily reduces uncertainty; it does not systematically move the policy itself. Conversely, Panel (d) shows the case with fewer trajectories. Here the confidence intervals become much wider, especially at high wealth levels where visits are sparse. Put differently, when agents have learned from only a small number of simulations, agents who happen to find themselves in the same high-wealth state can end up taking quite different actions across independent runs.

These results suggest a useful way to think about our stochastic solution method. If we could associate to each point in the state space a simple measure of ``confidence'' --- for example, based on the width of the confidence band or on how often that state is visited in simulation --- it would reveal that agents are very certain about their behavior in frequently visited states, but much less certain in rare states. Our experiments indicate that, for economically relevant regions of the state space (low and medium wealth), the learned policies are both stable and precise, while the main residual uncertainty is confined to tails that households almost never reach in practice.

%%%%%%%%%%%%%%%%%%%%%%%%%%%%%%%%%%%%%%%%%%%%%%%%%%%%%%%%%%%%
%%%%%%%%%%%%%%%%%%%%%%%%%%%%%%%%%%%%%%%%%%%%%%%%%%%%%%%%%%%%
\subsection{Krusell-Smith Model}\label{sec:applications_krusell_smith}

\paragraph{Setup.}
The household side of the Krusell-Smith economy is as in the Huggett model of Section \ref{sec:setup}, except that financial wealth is now productive capital owned by households and rented to a representative firm. The firm uses capital and labor to produce according to
\begin{equation*}
	Y_t = z_t K_t^\alpha L_t^{1-\alpha}.
\end{equation*}
Under perfect competition, factor prices equal marginal products, $w_t = (1-\alpha) \frac{Y_t}{L_t}$ and $r_t^K = \alpha \frac{Y_t}{K_t}$, where $w_t$ is the real wage rate and $r_t^K$ the rental rate of capital. 
Since households own the capital and pay for depreciation, their net rate of return on capital is $r_t = r_t^K - \delta$. The market clearing condition for capital is 
\begin{equation*}
	\int b \ d G_t(b, y) = K_t ,
\end{equation*}
and labor market clearing condition is given by $L_t = 1$ because each household supplies one unit of labor inelastically.

\paragraph{Calibration.}
One period corresponds to a year. On the preference side, we set $\beta = 0.95$ and use CRRA preferences with coefficient of relative risk aversion $\sigma = 3$. 
On the production side, we set the capital share to $\alpha = 0.36$ and the depreciation rate to $\delta = 0.08$, standard in the quantitative macro literature.
For idiosyncratic income, we retain the same AR(1) specification and parameters as in the Huggett model, so that the cross-sectional heterogeneity is directly comparable across the two experiments. Aggregate productivity $z_t$ also follows a log AR(1) process with persistence $\rho_z = 0.9$ and innovation volatility $\nu_z = 0.03$. All log AR(1) processes are discretized on finite grids using a standard Tauchen procedure; the details of the grids are reported in Appendix \ref{app:applications}.

\begin{figure}[ht!]
\centering
\begin{subfigure}[t]{.32\textwidth}
\centering
\includegraphics[width=\linewidth]{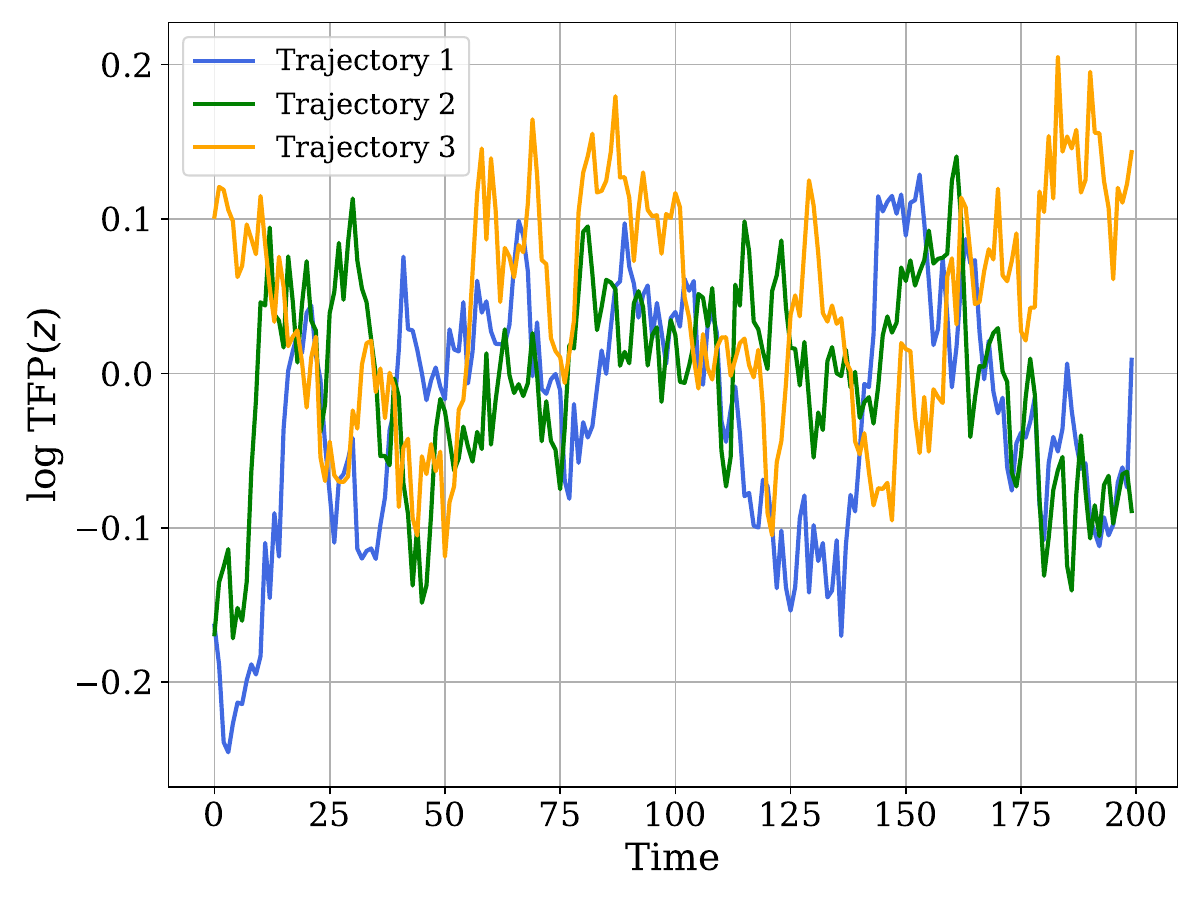}
\vspace{-7mm}
\caption{$\log$ TFP $\log(z)$}
\end{subfigure}
\begin{subfigure}[t]{.32\textwidth}
\centering
\includegraphics[width=\linewidth]{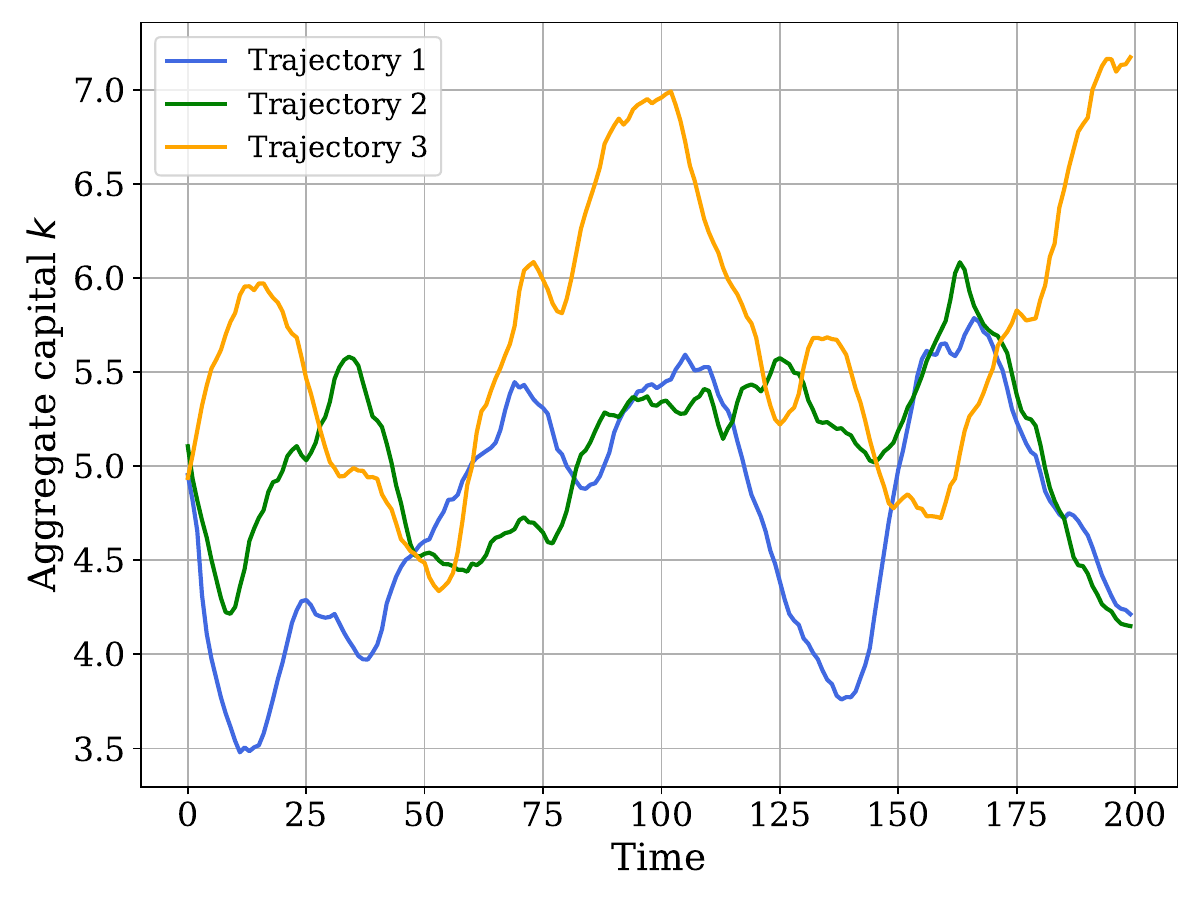}
\vspace{-7mm}
\caption{Aggregate capital $K$}
\end{subfigure}
\begin{subfigure}[t]{.32\textwidth}
\centering
\includegraphics[width=\linewidth]{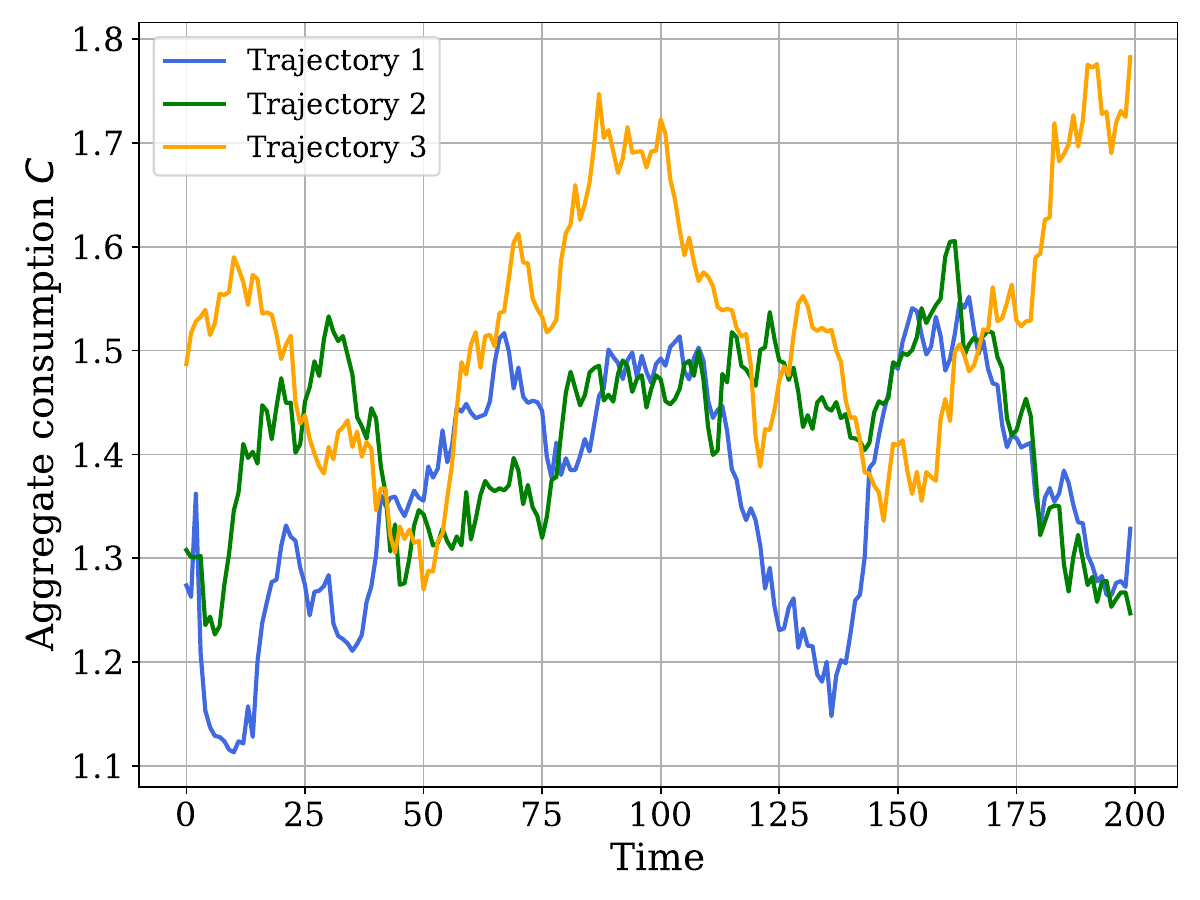}
\vspace{-7mm}
\caption{Aggregate consumption $C$}
\end{subfigure}
\vspace{2mm}

\begin{subfigure}[t]{.32\textwidth}
\centering
\includegraphics[width=\linewidth]{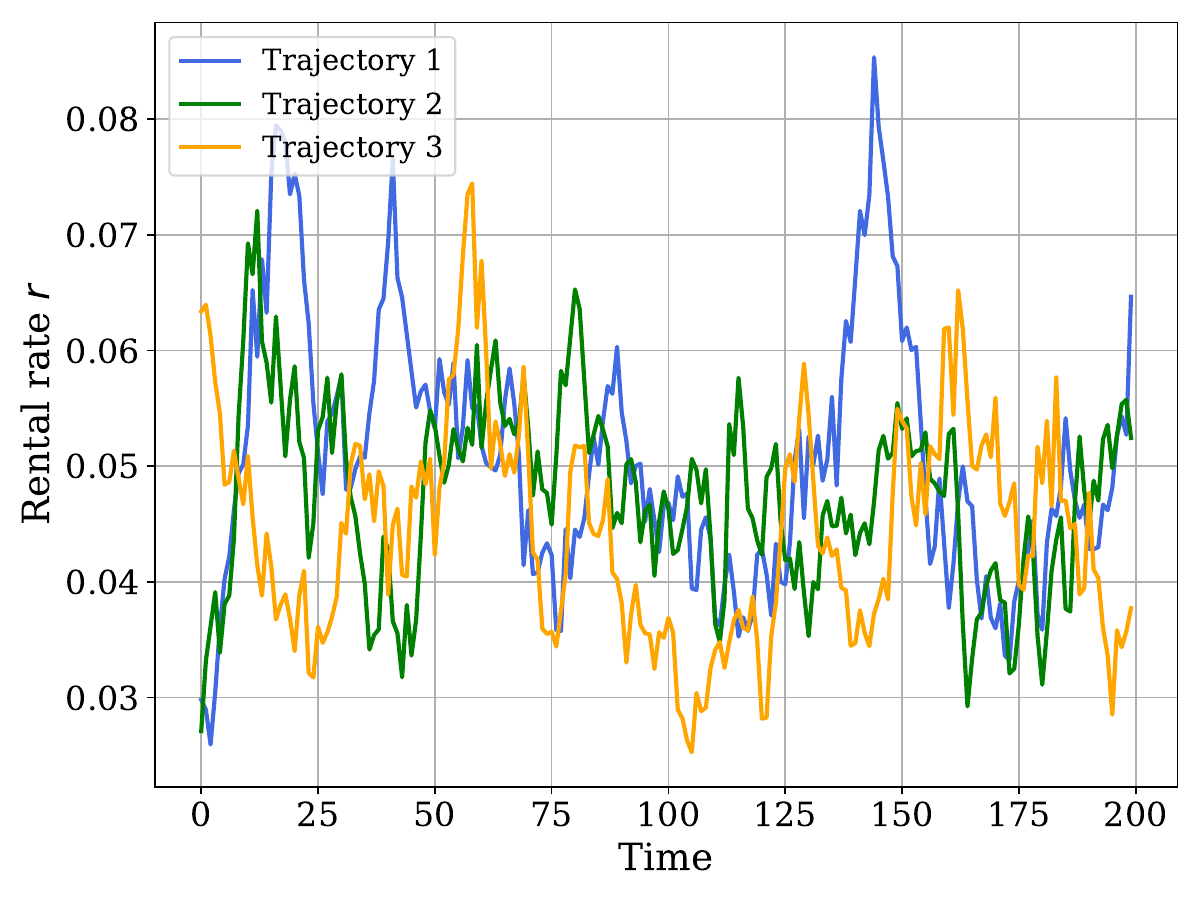}
\vspace{-7mm}
\caption{Rental rate $r$}
\end{subfigure}
\begin{subfigure}[t]{.32\textwidth}
\centering
\includegraphics[width=\linewidth]{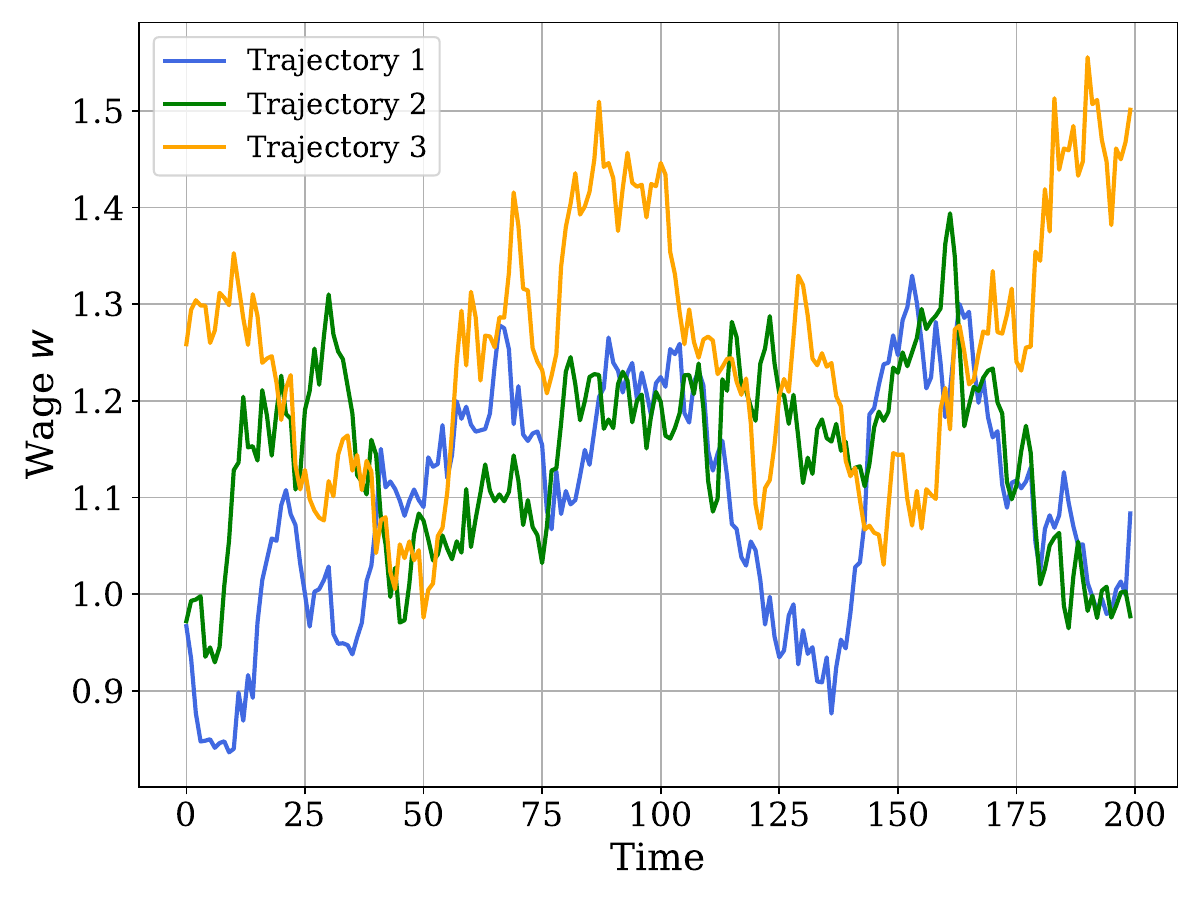}
\vspace{-7mm}
\caption{Wage $w$}
\end{subfigure}
\begin{subfigure}[t]{.32\textwidth}
\centering
\includegraphics[width=\linewidth]{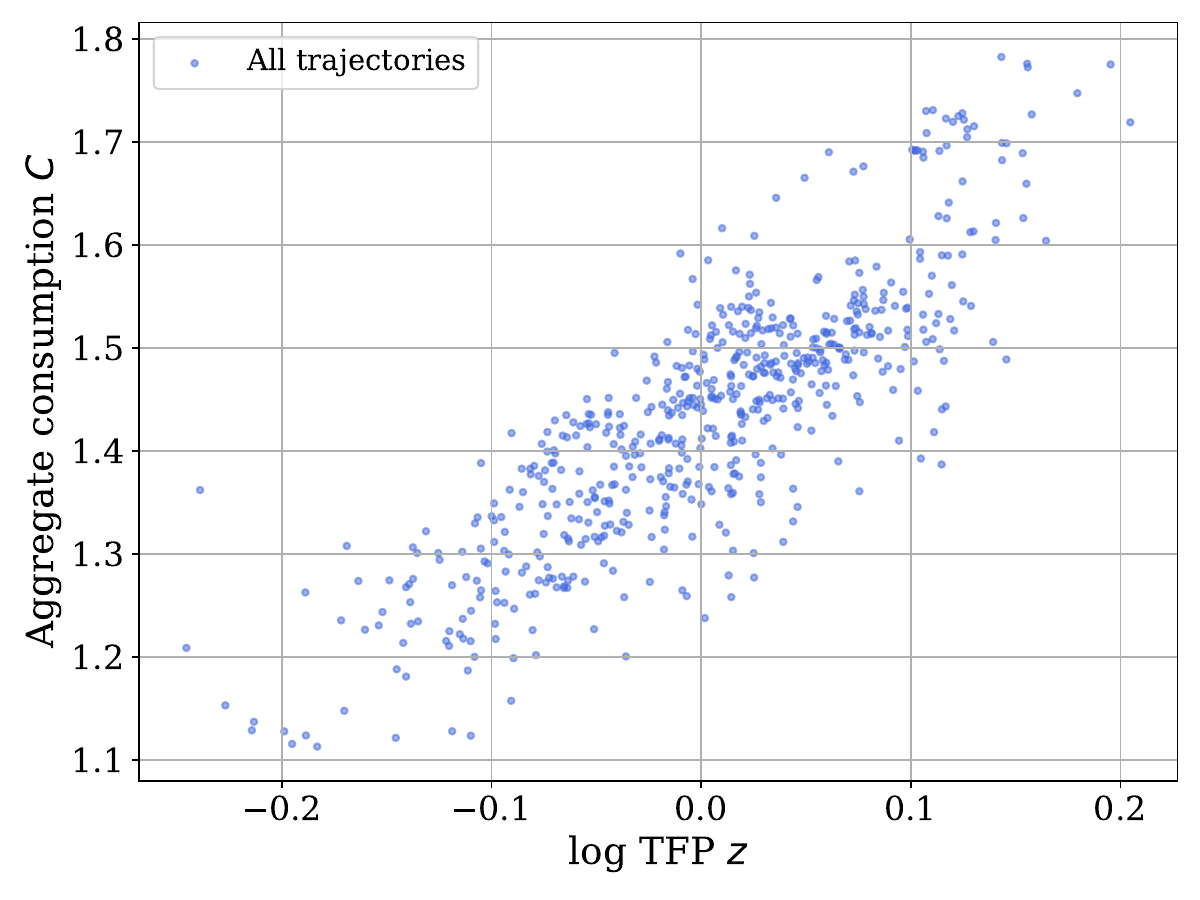}
\vspace{-7mm}
\caption{$C$ vs $\log$ TFP$(z)$}
\end{subfigure}
\caption{Simulation Results}
\label{fig:KS_simulations}
\end{figure}

\paragraph{Numerical Results.}
Figure \ref{fig:KS_simulations} plots simulation results for our solution of the Krusell-Smith model. We start in Panel (a) with the exogenous aggregate productivity process $z_t$, and plot in Panel (b) aggregate capital $K_t$, in Panel (c) aggregate consumption $C_t$, in Panel (d) the aggregate rental rate $r_t^K$ and in Panel (e) the aggregate wage $w_t$. As in standard neoclassical models, $K_t$ and $C_t$ comove strongly with $z_t$, while $r_t^K$ and $w_t$ move in opposite directions.

The final panel (f) presents a scatter plot of aggregate consumption $C_t$ against the exogenous productivity shock $z_t$. We find substantial \emph{vertical} dispersion: for a given realization of $z_t$, different periods in the simulation exhibit quite different levels of aggregate consumption. If all points lay on a single curve, then the same aggregate productivity level would always be associated with the same $C_t$. Instead, the cross-sectional wealth distribution shifts over time in ways that matter for aggregates, so the mapping $z_t \mapsto C_t$ is history-dependent. The blue dots in Panel (f) represent states actually visited along the simulated path, so this vertical spread measures the quantitative importance of distributional dynamics for aggregate outcomes. For values of $z_t$ very close to zero, simulated realizations of $C_t$ range roughly from 1.24 to 1.6, i.e. a difference on the order of 20\% of steady-state consumption. This illustrates that, even in this relatively simple heterogeneous-agent model, the cross-sectional distribution has a non-trivial impact on the aggregate response.

\begin{figure}[ht!]
\centering
\begin{subfigure}[t]{.45\textwidth}
\centering
\includegraphics[width=\linewidth]{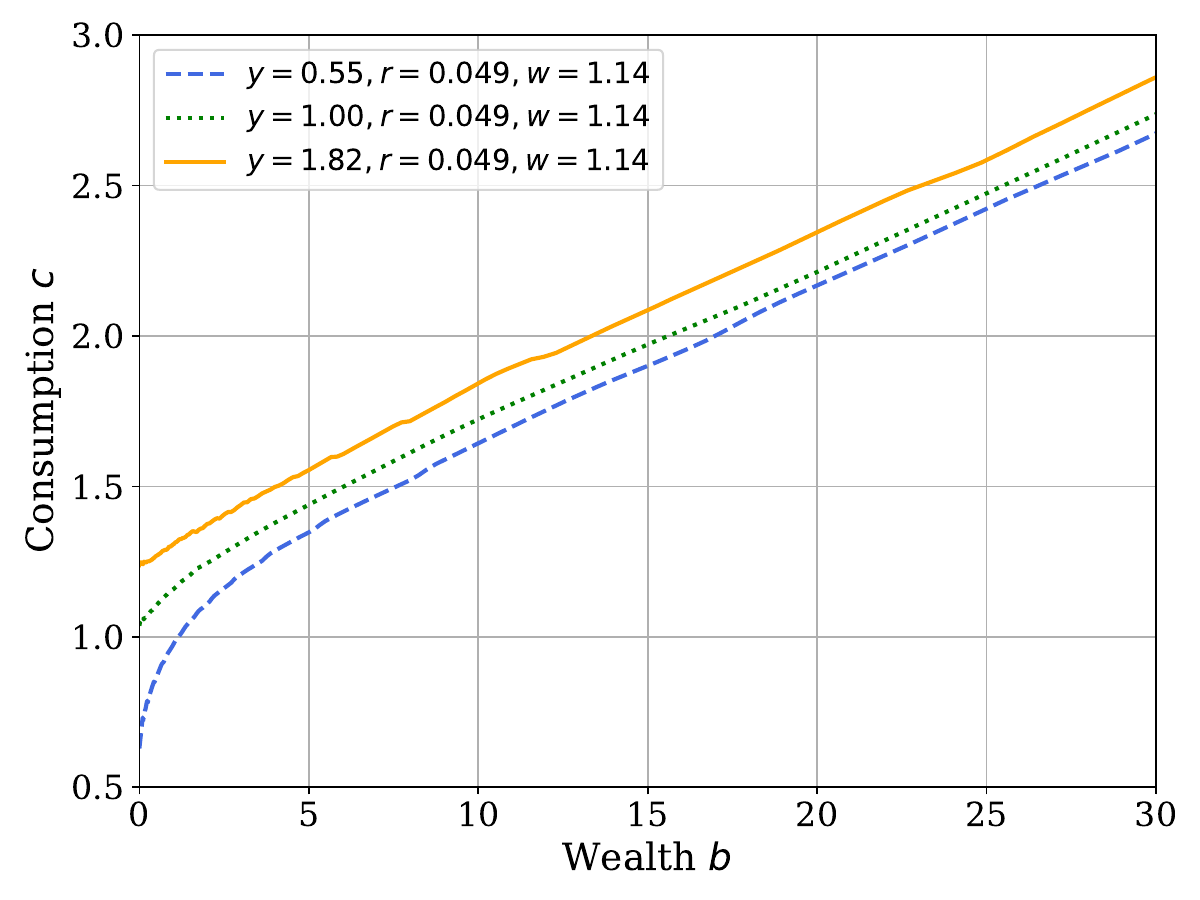}
\vspace{-7mm}
\caption{Policy with $N_\text{sample} = 512$}
\end{subfigure}
\begin{subfigure}[t]{.45\textwidth}
\centering
\includegraphics[width=\linewidth]{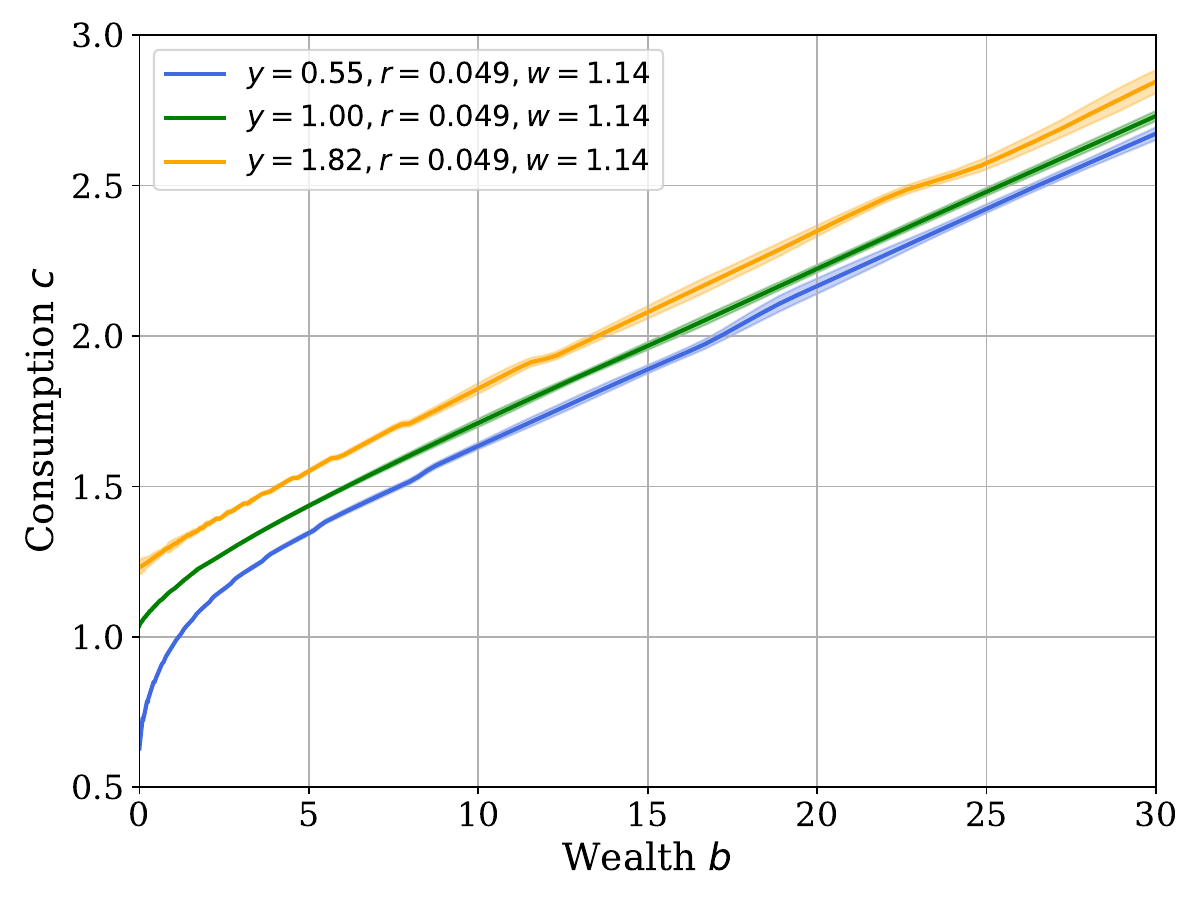}
\vspace{-7mm}
\caption{Policy CI with $N_\text{sample} = 512$}
\end{subfigure}

\begin{subfigure}[t]{.45\textwidth}
\centering
\includegraphics[width=\linewidth]{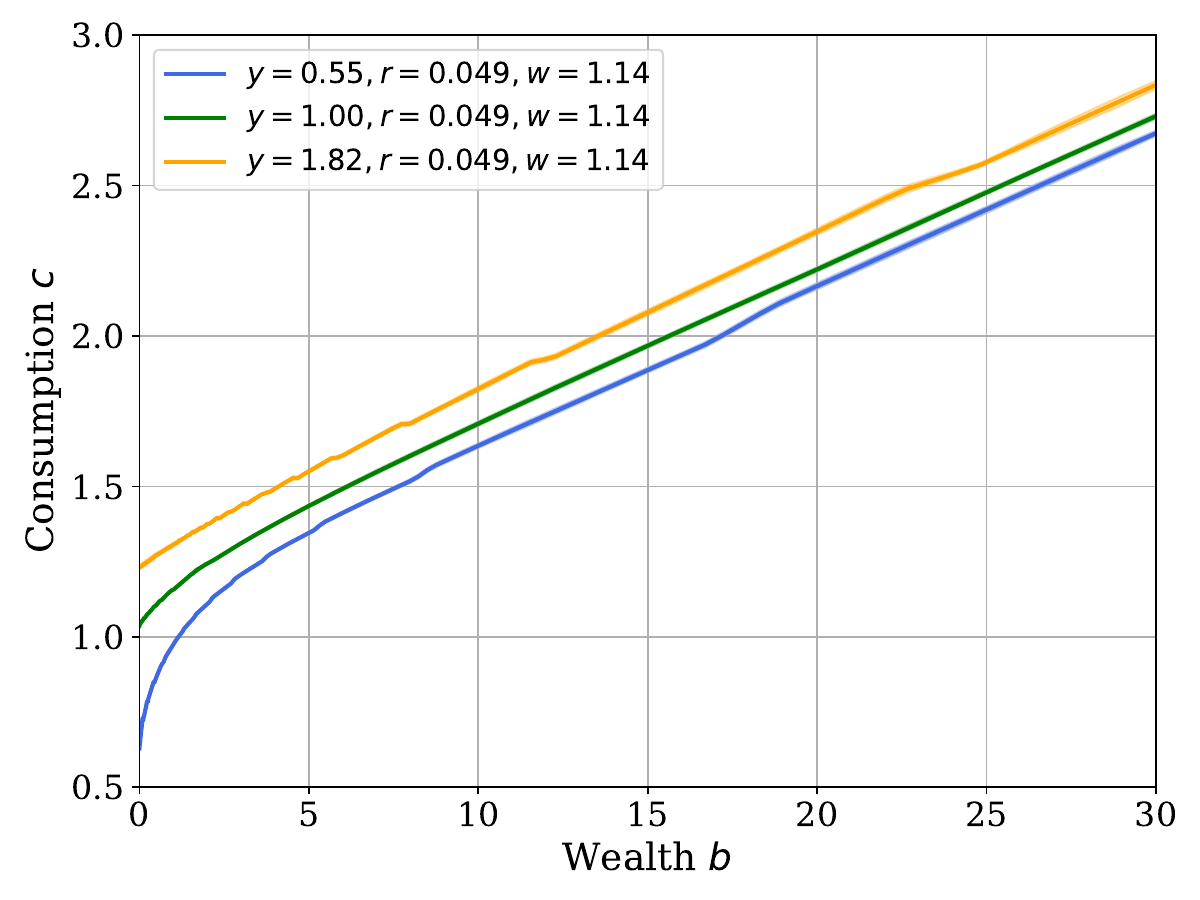}
\vspace{-7mm}
\caption{Policy CI with $N_\text{sample} = 2048$}
\end{subfigure}
\begin{subfigure}[t]{.45\textwidth}
\centering
\includegraphics[width=\linewidth]{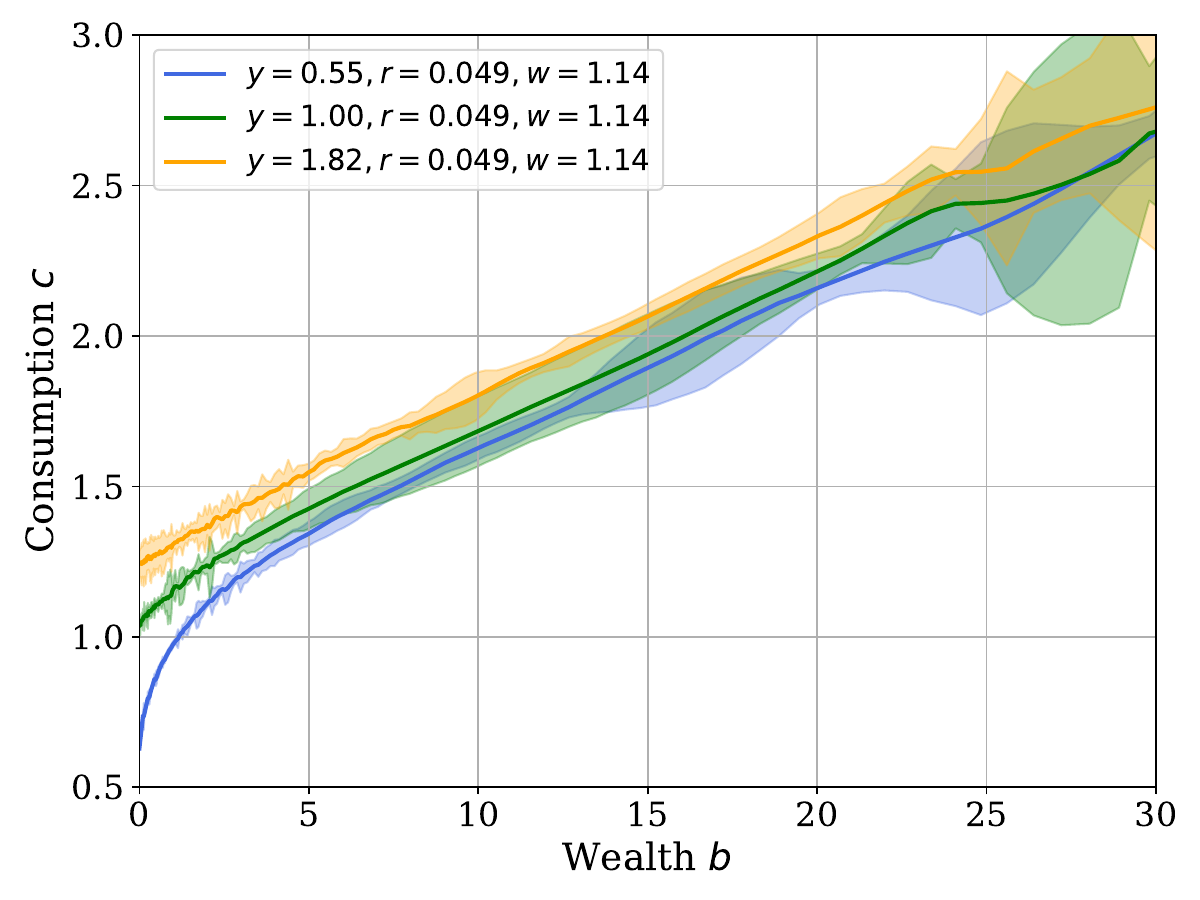}
\vspace{-7mm}
\caption{Policy CI with $N_\text{sample} = 24$}
\end{subfigure}
\caption{Dependence of Policies on Sample Size in the \citet{krusell-smith} model}
\label{fig:KS_policy_comparisons}
\end{figure}

\paragraph{Dependence on the number of trajectories (sample size).}
We now illustrate how the learned policy depends on the number of simulated trajectories used for training, similar to our discussion of Figure \ref{fig:Huggett_sample_size_comparisons} for the Huggett model. Panels (a) and (b) report the consumption policy from a single training run and the associated 95\% confidence intervals based on 10 independent training runs, both using 512 trajectories, respectively. The resulting policy is monotone and concave in wealth and visually very similar to the policies we documented in the Huggett experiment.There is only mild sampling uncertainty. Confidence bands are tight, especially for low wealth levels.  

Panel (c) increases the number of trajectories to 2048 and illustrates that confidence bands shrink even for the highest wealth levels. Panel (d) instead reduces the number of trajectories to 24 and illustrates that sampling uncertainty across runs increases substantially, especially at high wealth levels which are visited more rarely during the simulation.

\paragraph{Comparison to rational expectation solution.}
The partial equilibrium comparison we used in Section \ref{sec:applications_huggett} isolates the individual dynamic programming problem, where there is broad agreement on the accuracy of conventional VFI solutions. In general equilibrium, by contrast, obtaining the rational expectations (RE) solution requires treating the entire cross-sectional distribution as a state variable and solving the Master equation, a problem of much higher computational complexity. A growing literature proposes global solution methods for such RE equilibria. One recent example is the DeepHAM approach of \citet{han-yang-e}, which uses deep neural networks to approximate high-dimensional policy and value functions.

\begin{figure}[ht!]
\centering
\includegraphics[scale=0.3]{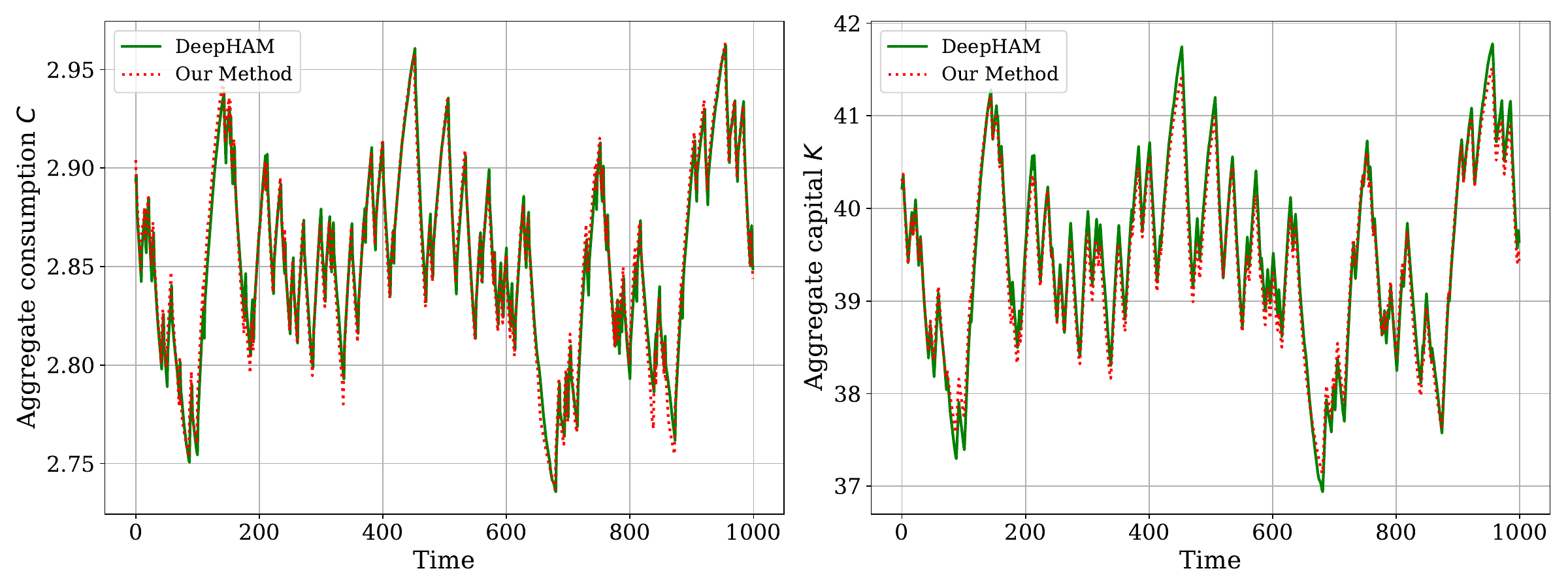}
\caption{Comparison to RE Solution with the DeepHAM method in \cite{han-yang-e}}
\label{fig:KS_comparison_RE}
\end{figure}

To benchmark our SRL method in this environment, we compare it directly to DeepHAM. Because the RE policy functions in DeepHAM conditions on the full cross-sectional distribution, whereas our approach conditions only on prices, the policy functions are not directly comparable. We therefore focus on equilibrium dynamics. Specifically, we initialize both economies from the same cross-sectional distribution and expose them to identical sequences of aggregate shocks. Figure \ref{fig:KS_comparison_RE} reports the resulting paths of aggregate consumption and capital. Across both panels, the two methods generate nearly indistinguishable aggregate dynamics.

This comparison indicates that, at least in the Krusell-Smith environment, our SRL approach can replicate the rational expectations solution while retaining the flexibility and scalability of a reinforcement-learning implementation.

%%%%%%%%%%%%%%%%%%%%%%%%%%%%%%%%%%%%%%%%%%%%%%%%%%%%%%%%%%%%
%%%%%%%%%%%%%%%%%%%%%%%%%%%%%%%%%%%%%%%%%%%%%%%%%%%%%%%%%%%%
\subsection{HANK Model}\label{sec:applications_HANK}

Our final benchmark economy is a one-account HANK model with sticky prices. 
This environment adds nominal rigidities and a richer firm block to the incomplete-markets structure from Section \ref{sec:setup}.

\paragraph{Setup.}
The problem of household $i$ is identical to that in Section \ref{sec:setup}, except that labor supply is now endogenous. In sequence form, 
\begin{align*}
	v_{i, 0} =& \max_{ \{c_{i, t}, n_{i, t}\} } 
	\mathbb E_0 \sum_{t=0}^\infty \beta^t u(c_{i, t}, n_{i, t}) \\
			  & \hspace{2.5mm} \text{s.t.} \hspace{5mm}
			  c_{i, t} + b_{i, t+1} = (1+r_t) b_{i, t} + w_t y_{i, t} n_{i, t} + d_t - T_t, 
			  \qquad 
			  b_{i, t+1} \geq 0,
\end{align*}
where $w_t$ is the real wage, $d_t$ denotes dividend payouts, and $T_t$ is a lump-sum tax. Households are the ultimate owners of firms but equity shares are not traded. The idiosyncratic income process $y_{i, t}$ is as in Section \ref{sec:setup}.

On the production side, a competitive final-good firm aggregates a continuum of intermediate inputs using a CES production technology with elasticity of substitution $\varepsilon$. Denoting by $Y_t$ aggregate output of the final good, cost minimization implies that demand for intermediate input $j$ is 
\begin{equation}\label{eq:firm_demand}
	y_{j, t} = \bigg(\frac{P_{j, t}}{P_t}\bigg)^{-\varepsilon} Y_t,
\end{equation}
where $P_{j, t}$ is the price of good $j$ and 
\begin{equation*}
	P_t = \bigg( \int_0^1 P_{j, t}^{1-\varepsilon} dj \bigg)^\frac{1}{1-\varepsilon}
\end{equation*}
is the aggregate price index.

Each intermediate good $j$ is produced by a monopolistically competitive firm with technology $y_{j, t} = z_t L_{j, t}$. The productivity term $z_t$ is common to all firms and follows a Markov process $z_{t+1} \sim \mathcal T_z(\cdot \, | \, z_t)$. Firm $j$ chooses its price $\{P_{j, t}\}$ to maximize the discounted value of profits subject to a quadratic adjustment cost as in \cite{Rotemberg1982}:
\begin{equation*}
	J_{j, 0} = \max_{\{P_{j, t}\}} 
	\mathbb E_0 \sum_{t=0}^\infty R_{0\to t}^{-1} \bigg[ \frac{P_{j, t}}{P_t} y_{j, t} - \frac{w_t}{z_t} y_{j, t} - \frac{\theta}{2} \bigg(\frac{P_{j, t} - P_{j, t-1}}{P_{j, t-1}}\bigg)^2 Y_t \bigg]
\end{equation*}
subject to the demand function \eqref{eq:firm_demand} and taking as given an initial price $P_{j, -1}$. Here $R_{0\to t} = R_1 \times \ldots \times R_t$ denotes the gross real interest rate between periods $0$ and $t$ with $R_t = 1+r_t$.

We focus on a symmetric distribution of initial prices, $P_{j, -1} = P_{j', -1}$, which implies symmetry ex post. In equilibrium we therefore have $P_{j, t} = P_t$ and $y_{j, t} = Y_t$ for all $j$. Denoting net inflation by 
\begin{equation*}
	\Pi_t = \frac{P_t - P_{t-1}}{P_{t-1}},
\end{equation*}
the firm's problem gives rise to the New Keynesian Phillips curve 
\begin{equation}\label{eq:NKPC}
	\Pi_t(1+\Pi_t) = \frac{\varepsilon}{\theta} \bigg(\frac{w_t}{z_t} - \frac{\varepsilon-1}{\varepsilon} \bigg) + \mathbb E_t \bigg[ R_{t+1}^{-1} \frac{Y_{t+1}}{Y_t} \Pi_{t+1}(1+\Pi_{t+1}) \bigg].
\end{equation}
The first term captures the gap between real marginal cost $\frac{w_t}{z_t}$ and the desired markup $\frac{\varepsilon-1}{\varepsilon}$, while the second term reflects expected future inflation, discounted by the real interest rate and scaled by output growth. Given inflation, firm dividend payments are
\begin{equation*}
	d_t = \bigg(1-\frac{w_t}{z_t}\bigg) Y_t - \frac{\theta}{2} \Pi_t^2 Y_t .
\end{equation*}

Monetary policy follows a Taylor rule 
\begin{equation*}
	1 + i_{t+1} = \bar R (1+\Pi_t)^\phi e^{\epsilon_t},
\end{equation*}
where $\phi > 1$ and the monetary policy shock $\epsilon_{t+1} \sim \mathcal T_\epsilon(\cdot \, | \, \epsilon_t)$ follows a Markov process. The real and nominal interest rates are linked by the Fisher equation
\begin{equation*}
	R_{t} = \frac{1+i_t}{1+\Pi_t}.
\end{equation*}
The government has a fixed supply of debt $B$ outstanding and finances interest payments using lump-sum taxes on households, $r_t B = T_t$.

Finally, three markets must clear in equilibrium. Goods market clearing requires
\begin{equation*}
	Y_t = \int c(b, y) dG(b, y) + \frac{\theta}{2} \Pi_t^2 Y_t.
\end{equation*}
so aggregate output is absorbed by consumption and price-adjustment costs. Labor market clearing implies
\begin{equation*}
	L_t = \int n(b, y) dG(b, y) ,
\end{equation*}
and the bond market clears when
\begin{equation*}
	B = \int b \ dG(b, y),
\end{equation*}
where we assume that bonds are in fixed positive supply $B$. The definition of competitive equilibrium is standard.

\paragraph{Firm Policy Gradient Method for the Phillips Curve.}
The price-setting block introduces an additional difficulty relative to the Huggett and Krusell-Smith models: Firm optimality gives rise to the forward-looking Phillips curve \eqref{eq:NKPC}. Standard approaches typically parameterize the conditional expectation term in Equation \eqref{eq:NKPC} and solve a non-trivial fixed point for the law of motion of inflation; see for example \citet{kase-melosi-rottner,FV-marbet-nuno-rachedi}. 

By contrast, we treat the firm problem exactly as we treat the household problem and solve it using the same SPG method. In practice, this means that we represent the firm's inflation policy as a function of individual firm states, aggregate shocks, and prices that are payoff relevant: $\Pi_t = \Pi(w_t, z_t)$. All policy functions are updated \emph{simultaneously}. Households and firms thus learn jointly from the \emph{same} simulated trajectories, and there is no separate fixed-point step for equilibrium expectations. 

A detailed description of the implementation is provided in Appendix \ref{app:applications_HANK}. Here we simply note that, in our experiments, this symmetric treatment of households and firms has very good convergence properties. As Table \ref{tab:Table_runtime_all} shows, solving the HANK model with the forward-looking Phillips curve is only mildly more costly than solving the baseline Huggett model, despite the added forward-looking structure in the firm block.

\paragraph{Calibration.}
A time period corresponds to a year, with $\beta = 0.95$. Preferences are CRRA and separable in consumption and labor, $u(c, n) = \frac{1}{1-\sigma} c^{1-\sigma} - \frac{1}{1+\eta} n^{1+\eta}$, with coefficient of relative risk aversion $\sigma = 1$ implying log utility over consumption, and inverse Frisch elasticity $\eta = 1$.

On the production side, we set the elasticity of substitution across intermediate goods to $\varepsilon = 10$ and the Rotemberg adjustment cost parameter to $\theta = 100$. We set the Taylor rule coefficient to $\phi = 1.5$, and we set the fixed government bond supply to $B = 5$.  

This economy features three shocks: one idiosyncratic and two aggregate. 
For households' idiosyncratic income process, we use the same AR(1) specification and parameterization as in the Huggett model so that cross-sectional heterogeneity is directly comparable across applications. Aggregate risk comprises the TFP process $z_t$ and the monetary policy shock $\epsilon_t$, both following AR(1) processes. We set their persistence to $\rho_z = \rho_\epsilon = 0.9$ and their innovation volatilities to $\nu_z = 0.07$ and $\nu_\epsilon = 0.002$, respectively. We use a standard Tauchen procedure to discretize all three processes on finite grids and report all remaining details in Appendix \ref{app:applications}.

\begin{figure}[ht!]
\centering
\begin{subfigure}[t]{.38\textwidth}
\centering
\includegraphics[width=\linewidth]{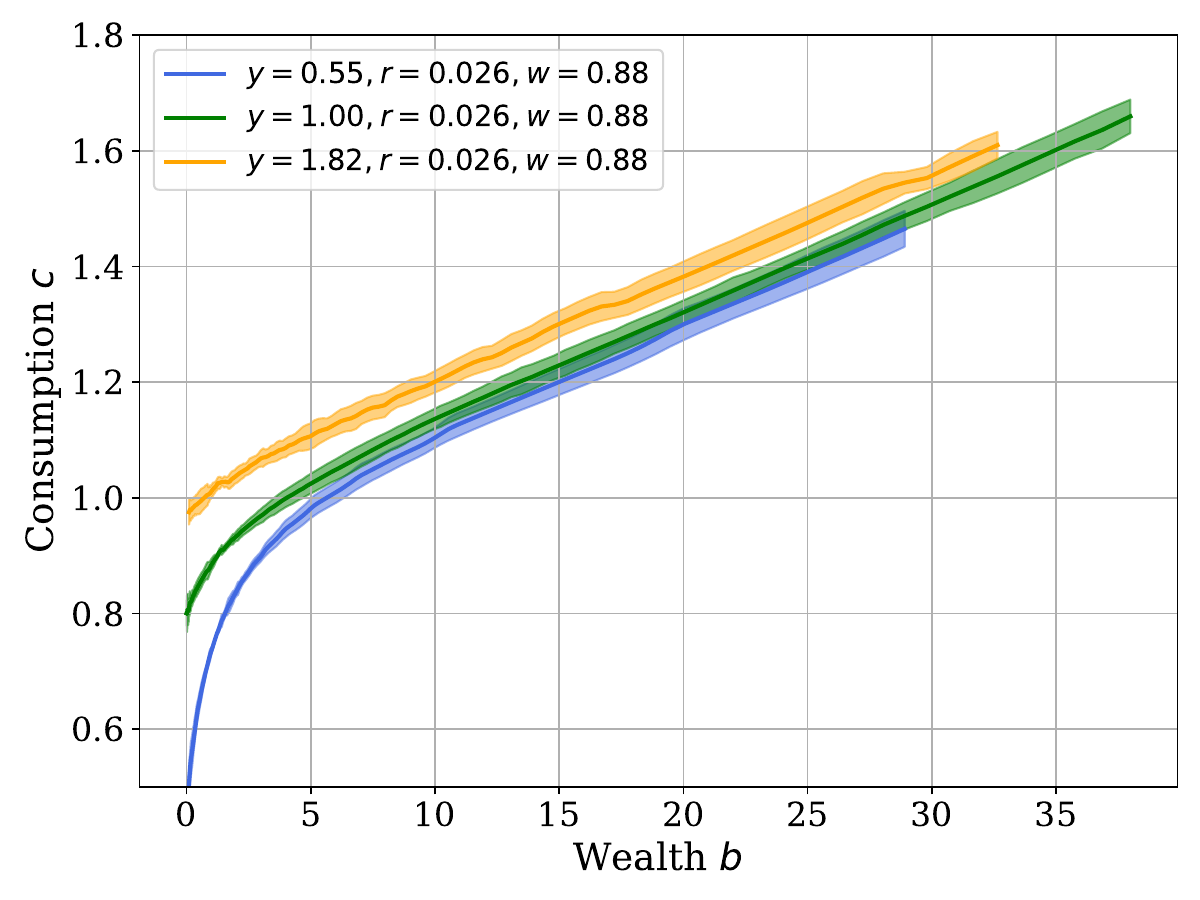}
\vspace{-7mm}
\caption{Consumption policy}
\end{subfigure}
\begin{subfigure}[t]{.38\textwidth}
\centering
\includegraphics[width=\linewidth]{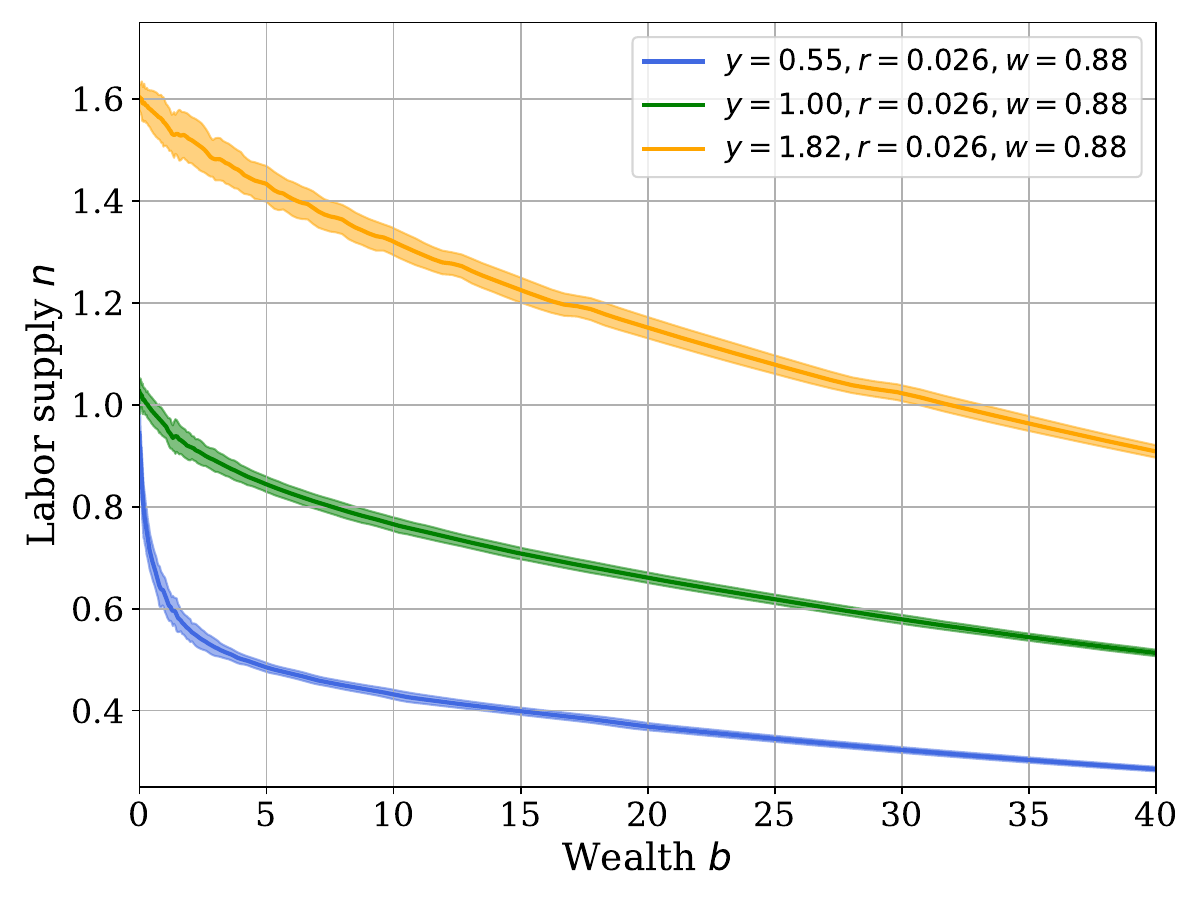}
\vspace{-7mm}
\caption{Labor supply policy}
\end{subfigure}
\vspace{3mm}

\begin{subfigure}[t]{.38\textwidth}
\centering
\includegraphics[width=\linewidth]{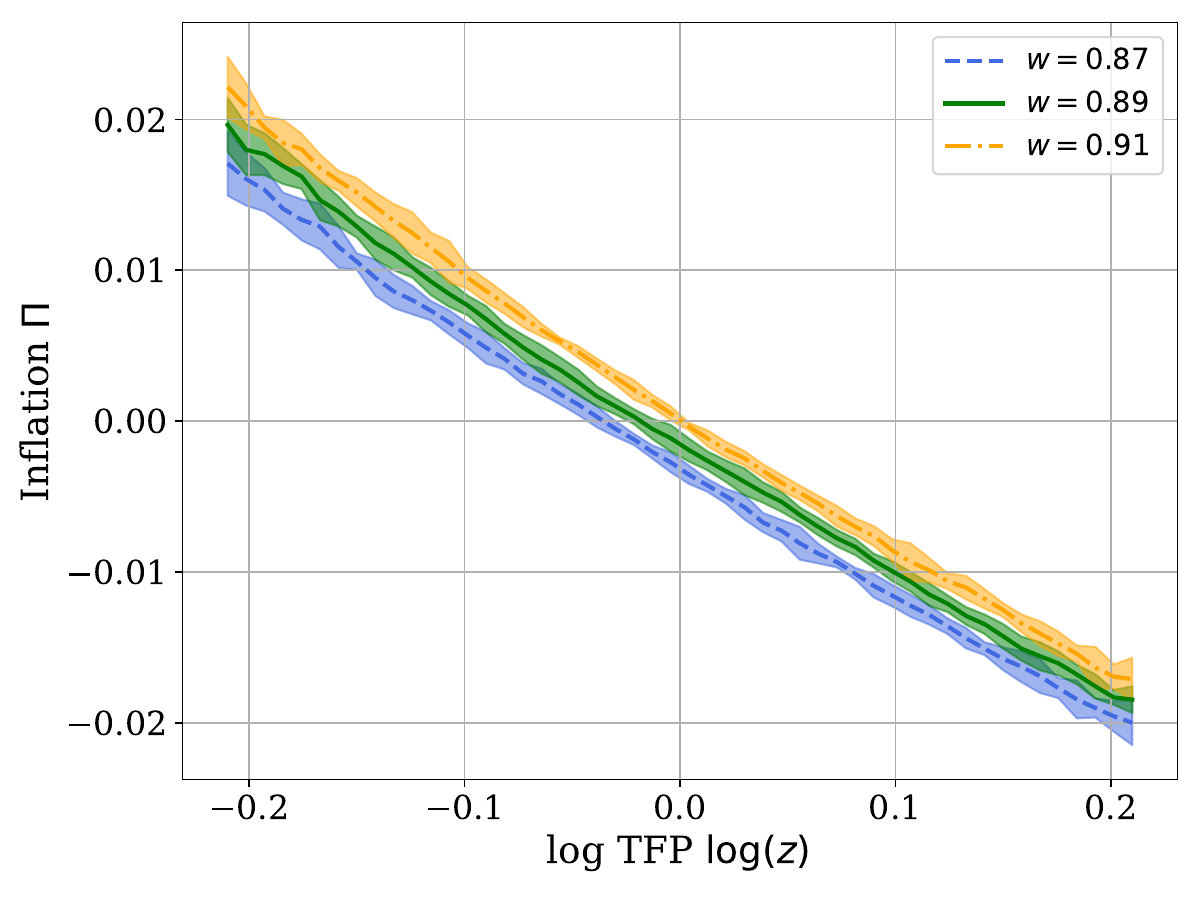}
\vspace{-7mm}
\caption{Inflation policy}
\end{subfigure}
\caption{Household and Firm Policy Functions in HANK}
\label{fig:HANK_policies}
\end{figure}

\paragraph{Numerical Results.} 
Figure \ref{fig:HANK_policies} reports the optimal policy functions for households and firms in the HANK model. Panels (a) and (b) show household consumption and labor supply policies, while Panel (c) displays the firm's inflation policy. For each policy, we plot the point estimate together with confidence bands obtained from 10 independent training runs with $N=512$ simulated trajectories each.

Panel (a) shows that consumption is increasing and concave in wealth and increasing in idiosyncratic labor productivity, as standard theory would suggest. Panel (b) shows the corresponding labor supply policy, which is decreasing and convex in wealth and increasing in labor productivity: richer households work less at the margin, while high-productivity households supply more labor.

Panel (c) displays the firm's inflation policy function. It is decreasing and almost linear in log aggregate productivity and increasing in marginal cost. In particular, positive supply (TFP) shocks lower marginal cost and induce lower inflation, consistent with the New Keynesian Phillips curve \eqref{eq:NKPC}.

Across all three panels, the confidence bands are tight over the bulk of the wealth distribution and widen somewhat only in the far tails, where states are rarely visited in simulation. The firm's policy is the most challenging to learn --- reflected in somewhat wider confidence bands --- but remains well behaved and economically sensible. Overall, Figure \ref{fig:HANK_policies} suggests that our SPG method recovers accurate policy functions in this richer HANK environment.

\begin{figure}[ht!]
\centering
\begin{subfigure}[t]{.33\textwidth}
\centering
\includegraphics[width=\linewidth]{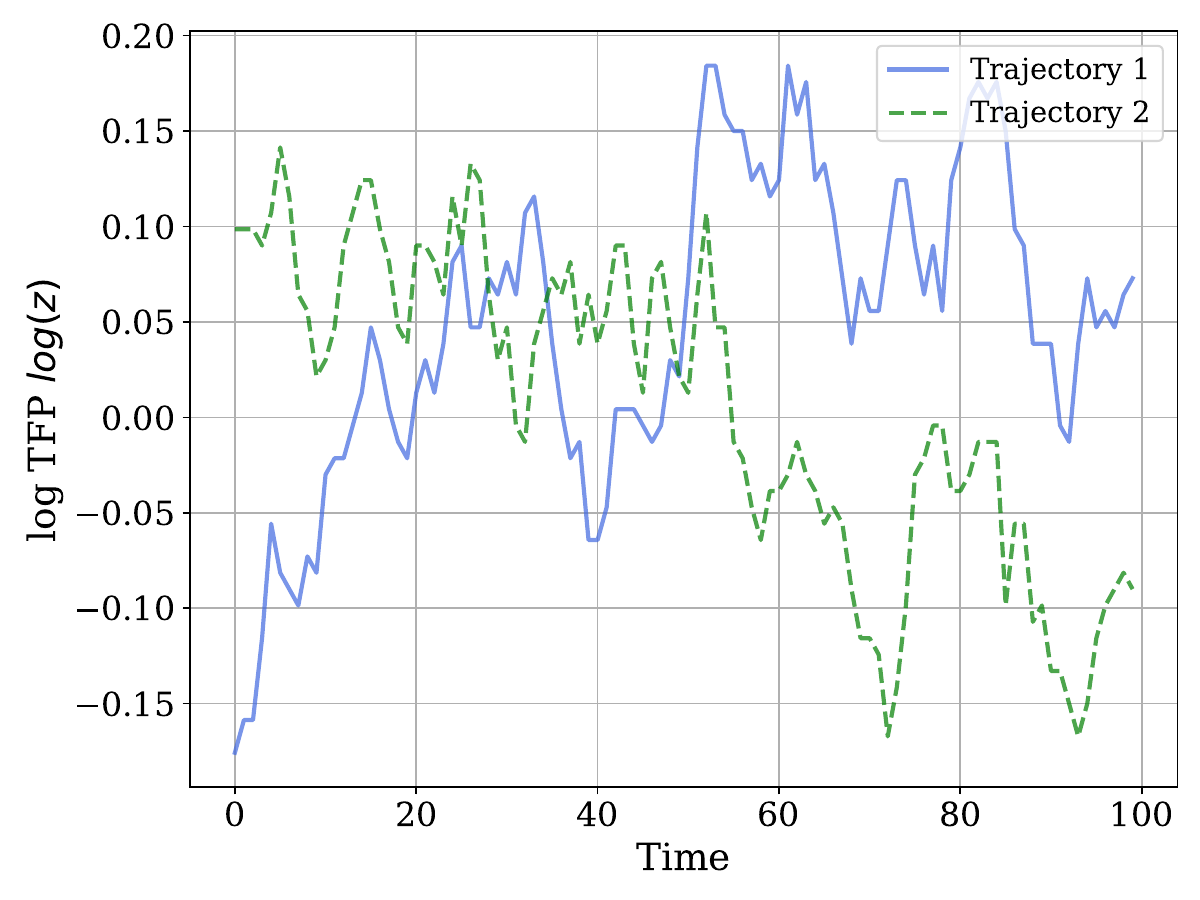}
\vspace{-7mm}
\caption{log TFP $\log(z)$}
\end{subfigure}\begin{subfigure}[t]{.33\textwidth}
\centering
\includegraphics[width=\linewidth]{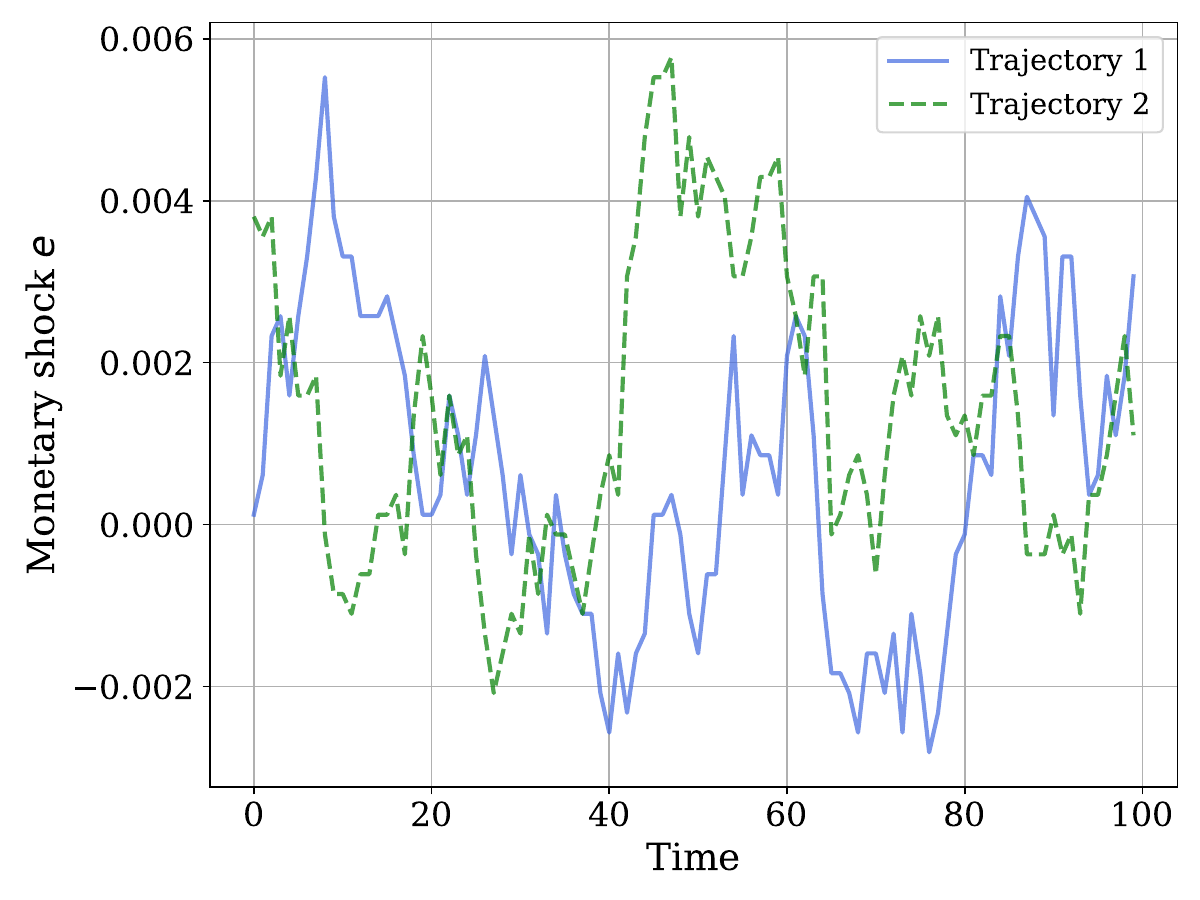}
\vspace{-7mm}
\caption{Monetary shock $\epsilon$}
\end{subfigure}
\begin{subfigure}[t]{.33\textwidth}
\centering
\includegraphics[width=\linewidth]{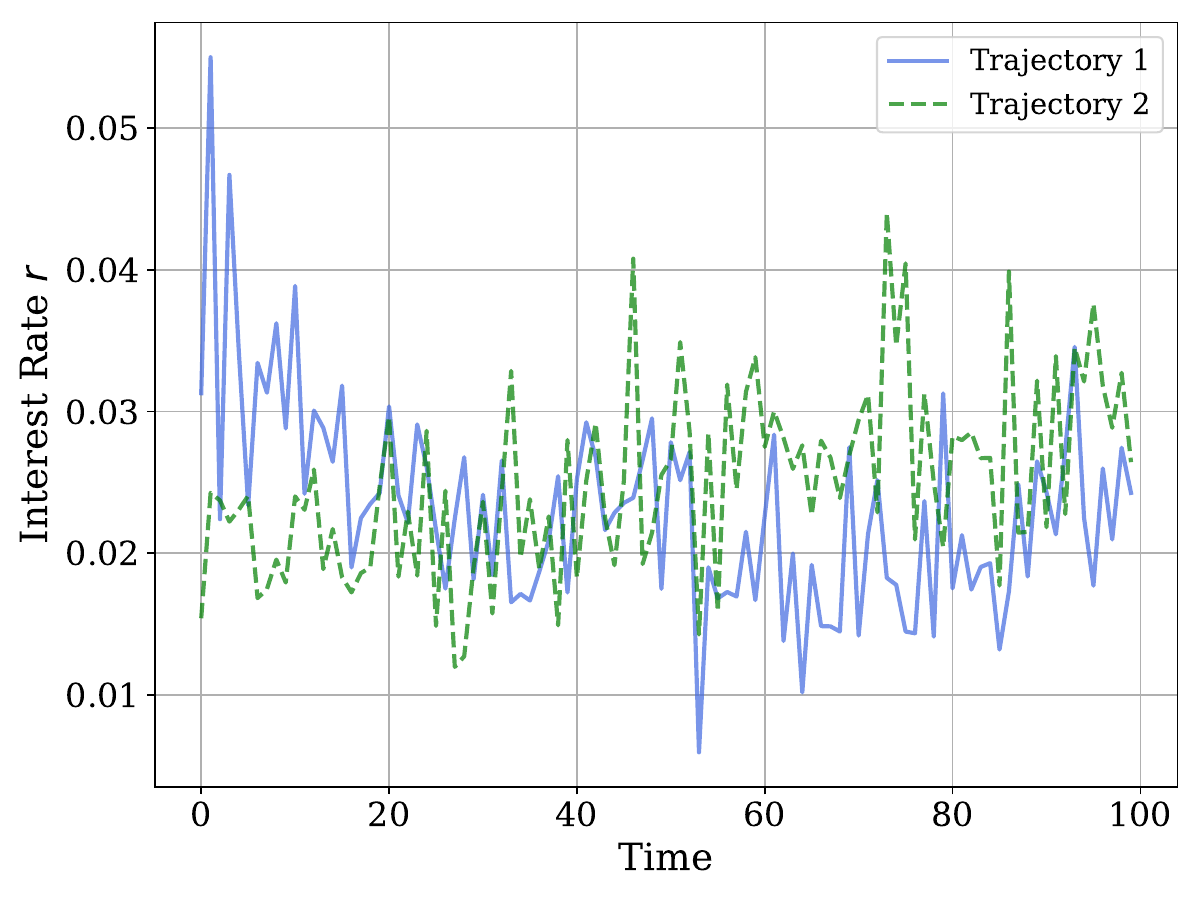}
\vspace{-7mm}
\caption{Interest rate $r$}
\end{subfigure}
\vspace{3mm}

\begin{subfigure}[t]{.33\textwidth}
\centering
\includegraphics[width=\linewidth]{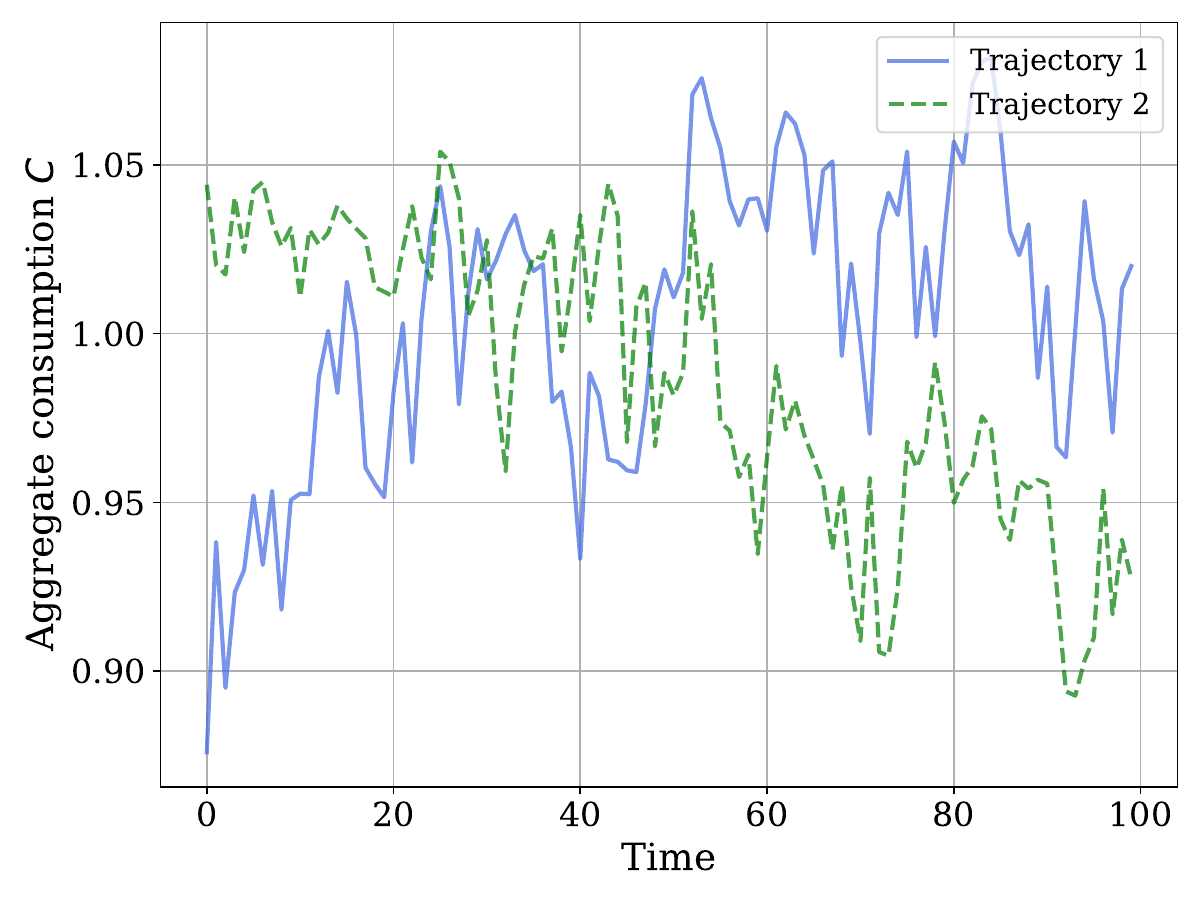}
\vspace{-7mm}
\caption{Aggregate consumption $C$}
\end{subfigure}\begin{subfigure}[t]{.33\textwidth}
\centering
\includegraphics[width=\linewidth]{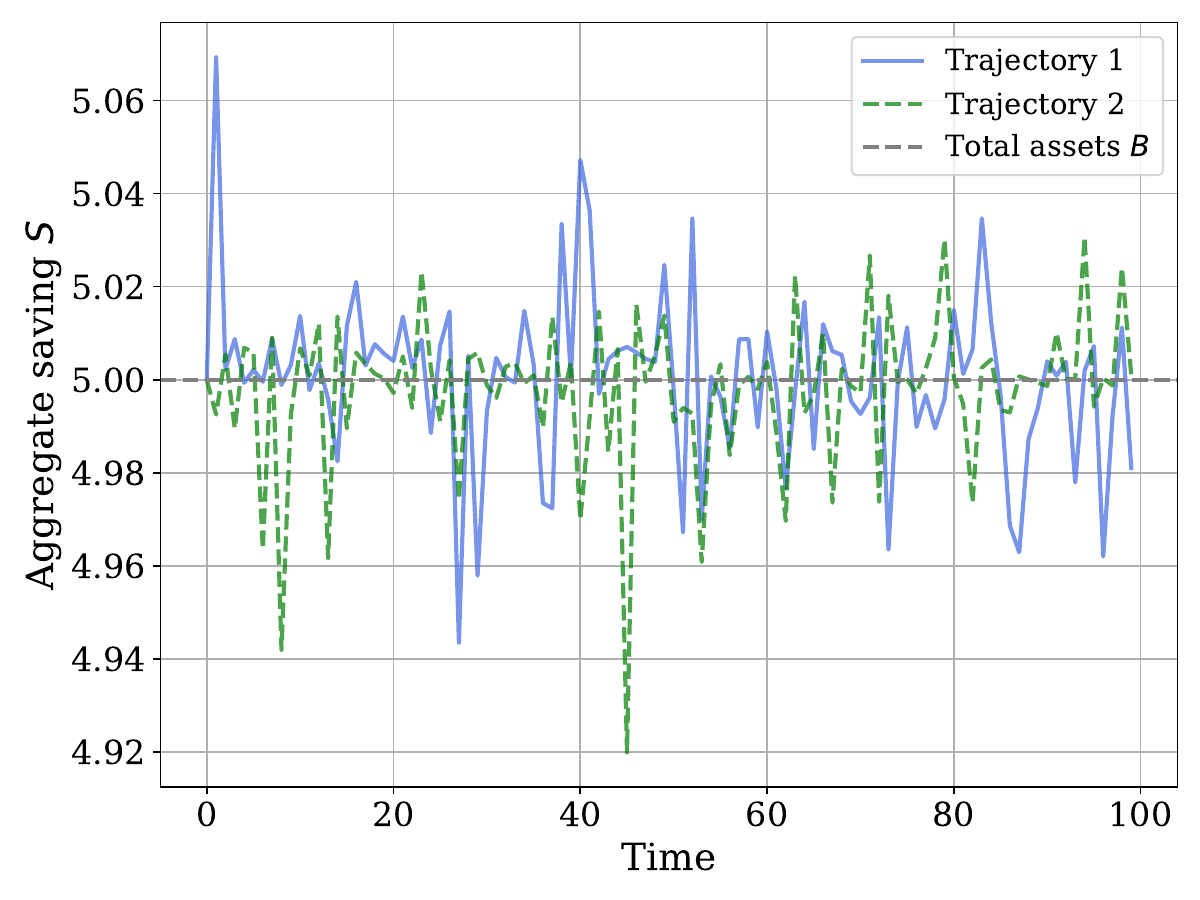}
\vspace{-7mm}
\caption{Aggregate saving $S$}
\end{subfigure}
\begin{subfigure}[t]{.33\textwidth}
\centering
\includegraphics[width=\linewidth]{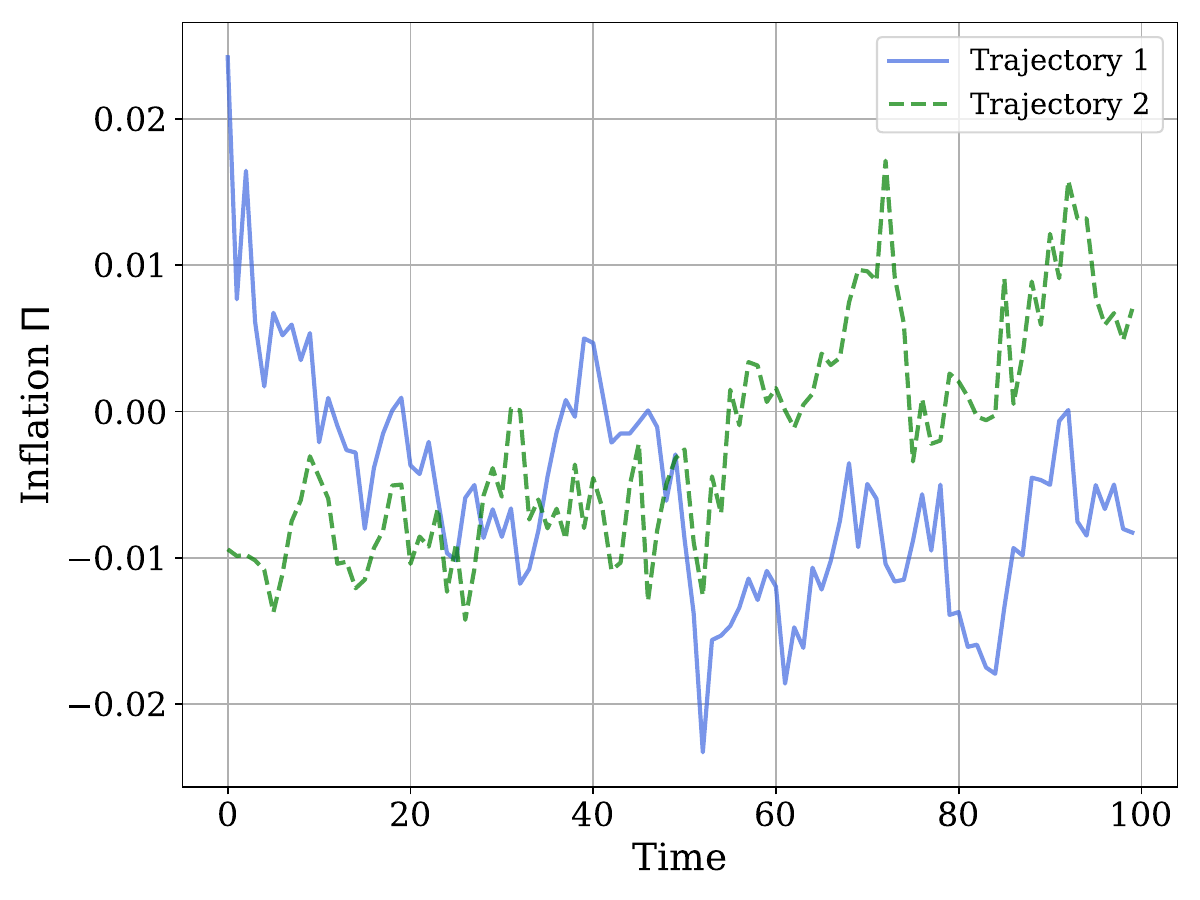}
\vspace{-7mm}
\caption{Inflation $\Pi$}
\end{subfigure}\\
\caption{Simulated Trajectories in HANK}
\label{fig:HANK_simulations}
\end{figure}

Figure \ref{fig:HANK_simulations} reports simulated time series for the HANK model.  Panels (a) and (b) display the two aggregate shocks: log TFP $z_t$ and the monetary policy shock $\epsilon_t$. Panels (c)-(f) then show the induced time series for the real interest rate, aggregate consumption, aggregate savings, and inflation. 

Aggregate consumption in Panel (d) moves procyclically with productivity: positive TFP shocks raise $Y_t$ and, through higher labor income and lower marginal costs, also increase $C_t$. Inflation in Panel (f) is countercyclical with respect to TFP, consistent with the Phillips curve \eqref{eq:NKPC}: favorable supply shocks reduce marginal costs and put downward pressure on inflation. In our calibration macro aggregates appear more sensitive to TFP innovations than to monetary shocks.

Panel (e) plots households' aggregate demand for bonds, together with the fixed supply $B = 5$. Asset demand fluctuates tightly around the supply level, and deviations from the horizontal supply line measure residuals in the bond market clearing condition. On average, the relative absolute deviation of aggregate demand from supply is $0.22\%$. These residuals primarily reflect the use of linear interpolation in solving for market-clearing prices.

%%%%%%%%%%%%%%%%%%%%%%%%%%%%%%%%%%%%%%%%%%%%%%%%%%%%%%%%%%%%
%%%%%%%%%%%%%%%%%%%%%%%%%%%%%%%%%%%%%%%%%%%%%%%%%%%%%%%%%%%%
%%%%%%%%%%%%%%%%%%%%%%%%%%%%%%%%%%%%%%%%%%%%%%%%%%%%%%%%%%%%
\section{Conclusion}\label{sec:conclusion}
We develop a new \emph{structural reinforcement learning} (SRL) approach to formulating and globally solving heterogeneous agent models with aggregate risk. We replace the cross-sectional distribution with low-dimensional prices as state variables and let agents compute price expectations directly from simulated paths. Our approach differs from standard RL in that we assume that agents have structural knowledge about the dynamics of their own individual states (e.g. their budget constraint and idiosyncratic income process). Our \emph{structural policy gradient} (SPG) algorithm sidesteps the Master equation and efficiently handles heterogeneous agent models traditional methods struggle with, like those with nontrivial market-clearing conditions. By imposing that policy functions depend only on current prices (or a short price history) we keep the state space low-dimensional so that we can work with a grid-based (tabular) approach rather than deep neural networks.

We implement our method in JAX and conduct computational experiments for three benchmark models in macroeconomics -- the \cite{huggett} model, the \cite{krusell-smith} model, and a one-asset HANK model with sticky prices.  In all three cases, the SPG algorithm converges in only a few minutes. In the Krusell-Smith model, the resulting policies are close to those obtained from alternative global solutions of the rational expectations equilibrium. Allowing agents to condition on a longer history of lagged prices hardly moves the solution, indicating that much of the relevant information for forecasting future prices is already contained in current prices. In the HANK model, we show how the same approach can be used symmetrically on the household and firm sides to globally solve the forward-looking New Keynesian Phillips curve.

The core idea of our approach -- that agents form price expectations by sampling -- could, in principle, serve as a building block for an empirically realistic theory of expectations formation in macroeconomics (which the algorithm in this paper is not). The plausibility of RL-based approaches is supported by evidence that reinforcement learning underpins a substantial share of human and animal learning.\footnote{See, for example, \citet{niv}, \citet{glimcher}, \citet{caplin-dean}, \citet{neuroeconomics-book}, \citet{gershman-daw}, and \citet{barberis-jin}.} To develop such an RL-based theory of expectations formation would require several modifications to our SRL method. First, the algorithm would need to be converted to a fully online, incremental RL algorithm, with agents updating policies and value estimates continuously while interacting with their environment, rather than only after observing $N$ price trajectories of length $T$. Second, the assumption that agents sample equilibrium prices in an unbiased way would likely need to be relaxed. Empirical evidence suggests that people disproportionately weight certain personal experiences \citep[e.g.][]{malmendier-nagel-depression-babies,malmendier-nagel-inflation}, so incorporating biased or experience-weighted sampling would be more realistic. Third, our SRL agents form price expectations in a model-free manner; this may be too simplistic and one could instead treat prices using a model-based RL approach.

%%%%%%%%%%%%%%%%%%%%%%%%%%%%%%%%%%%%%%%%%%%%%%%%%%%%%%%%%%%%
%%%%%%%%%%%%%%%%%%%%%%%%%%%%%%%%%%%%%%%%%%%%%%%%%%%%%%%%%%%%
%%%%%%%%%%%%%%%%%%%%%%%%%%%%%%%%%%%%%%%%%%%%%%%%%%%%%%%%%%%%
{\setstretch{1}
	\bibliographystyle{aer}
	\bibliography{challenge_bib}
}

%%%%%%%%%%%%%%%%%%%%%%%%%%%%%%%%%%%%%%%%%%%%%%%%%%%%%%%%%%%%
%%%%%%%%%%%%%%%%%%%%%%%%%%%%%%%%%%%%%%%%%%%%%%%%%%%%%%%%%%%%
%%%%%%%%%%%%%%%%%%%%%%%%%%%%%%%%%%%%%%%%%%%%%%%%%%%%%%%%%%%%
\newpage
\appendix
\counterwithin*{equation}{section}
\renewcommand\theequation{\thesection.\arabic{equation}}

\begin{center}
	{\Huge Online Appendix}
\end{center}
\vspace{5mm}

%%%%%%%%%%%%%%%%%%%%%%%%%%%%%%%%%%%%%%%%%%%%%%%%%%%%%%%%%%%%
%%%%%%%%%%%%%%%%%%%%%%%%%%%%%%%%%%%%%%%%%%%%%%%%%%%%%%%%%%%%
%%%%%%%%%%%%%%%%%%%%%%%%%%%%%%%%%%%%%%%%%%%%%%%%%%%%%%%%%%%%
\section{Model Calibration and Hyperparameters}\label{app:applications}

This appendix presents additional model, calibration and implementation details for the applications we solve in Section \ref{sec:applications} using our SRL approach. We start with the \cite{huggett} application in Appendix \ref{app:applications_huggett}, move to the \cite{krusell-smith} application in Appendix \ref{app:applications_krusell_smith}, and conclude with the HANK application in Appendix \ref{app:applications_HANK}.

%%%%%%%%%%%%%%%%%%%%%%%%%%%%%%%%%%%%%%%%%%%%%%%%%%%%%%%%%%%%
%%%%%%%%%%%%%%%%%%%%%%%%%%%%%%%%%%%%%%%%%%%%%%%%%%%%%%%%%%%%
\subsection{Appendix for Huggett Application}\label{app:applications_huggett}

We summarize the calibration we use in Section \ref{sec:applications_huggett} in Table \ref{tab:huggett_params}.
Table \ref{tab:huggett_hyperparams} summarizes the hyperparameters used to solve the Huggett model with our SRL algorithm. We briefly discuss the main choices and their rationale.

\paragraph{Partial equilibrium specification.}
In the general equilibrium Huggett model, the interest rate is a complicated function of the aggregate state and the cross-sectional distribution. It is not Markov. For the partial equilibrium (PE) exercise discussed in Section \ref{sec:applications_huggett}, we instead take as given an exogenous Markov law of motion for the interest rate $r_t$ and let households solve their individual problem taking as given this process.

We model the interest rate as a mean-reverting process with a square-root volatility term. %, and the real wage $p_{2, t}$ as an AR(1) in logs. 
In continuous time, this is analogous to a Cox-Ingersoll-Ross (CIR) or ``Feller square-root'' process, which ensures positivity of the interest rate. In discrete time, our PE price process is specified as
\begin{align*}
	 r_{t+1} &= (1-\rho_r) \overline r + \rho_r r_t + \nu_r \sqrt{\max\{r_t, 0\}} \; \cdot \; \varepsilon_{r, t}
	 \qquad \text{ where } \qquad
	 \varepsilon_{r, t} \sim \mathcal N(0, 1)  
	 % \log p_{2, t+1} &= (1-\rho_2) \log \overline P_2 + \rho_2 \log p_{2, t} + \nu_2 \varepsilon_{2, t}, 
	 % \qquad \text{ where } \qquad
	 % \varepsilon_{2, t} \sim \mathcal N(0, 1),
\end{align*}
where $\overline r$ is the long-run mean level of the interest rate, $\rho_r$ its autocorrelation, and $\nu_r$ the innovation volatility. The parameter values we use are reported in Table \ref{tab:huggett_params} and are chosen so that the unconditional distribution of interest rates as well as the implied aggregate bond holdings in PE are broadly consistent with general equilibrium in the Huggett calibration.

The idiosyncratic income process $y$ in PE is the same as in the GE model, a three-point discretization of a log AR(1) with persistence $\rho_y$ and volatility $\nu_y$, as reported in Table \ref{tab:huggett_params}. Thus, the only difference between PE and GE is that in PE the household takes $r_t$ as an exogenous Markov process, while in GE it is determined endogenously from bond market clearing.

For the numerical implementation, we discretize the PE interest rate process on a grid described in Table \ref{tab:huggett_hyperparams}. This grid is constructed using the CIR discretization method in \citet{farmer-toda},
which is designed for square-root processes and preserves positivity. Together with the income grid for $y$, this yields a fully specified PE environment in which we can solve the household problem using both our SPG algorithm and a conventional VFI method, as described in the main text.

\paragraph{Discretization.}
The individual state $(b, y)$ consists of bond holdings $b$ and idiosyncratic income $y$. We discretize bonds on a one-dimensional grid with $n_b = 200$ points and an upper bound $b^\texttt{max} = 50$. The income process $y$ takes $n_y = 3$ possible values. On the aggregate side, the two key state variables are the interest rate $r_t$ and aggregate income $z_t$. We approximate $r_t$ on a grid with $n_r = 20$ points covering the interval $[r_L, r_H] = [0.01, 0.06]$. This range is chosen to comfortably contain all equilibrium interest rate realizations observed in our simulations while avoiding an unnecessarily large grid. Aggregate productivity $z_t$ is discretized on a grid with $n_z = 30$ points using a standard Tauchen procedure. These choices strike a balance between accuracy and computational cost. They are fine enough to capture the relevant curvature in individual policies and the dependence of prices on the aggregate state.

\paragraph{Simulation horizon and truncation.}
We approximate lifetime utility by truncating the infinite sum in \eqref{eq:value_given_policy} at a finite horizon $T_\texttt{trunc}$. We choose $T_\texttt{trunc}$ to ensure that the tail of the discounted utility is negligible relative to a user-specified tolerance level $\epsilon_\texttt{trunc}$,
\begin{equation*}
    T_\texttt{trunc} = \min \Big\{ T \, : \, \beta^T <\epsilon_\texttt{trunc} \Big\} .
\end{equation*}
In the baseline Huggett experiment, we use $\epsilon_\texttt{trunc} = 10^{-3}$ and obtain $T_\texttt{trunc} = 170$, i.e. the contribution of periods beyond $T_\texttt{trunc}$ is bounded by $10^{-3}$ in present-value terms. To avoid numerical issues when wealth is very low, we also impose a minimal consumption floor $c_\texttt{min} = 10^{-3}$. This has no discernible effect on the economic results but prevents the utility function from being evaluated at (or extremely close to) zero.

\begin{table}[t!]
\centering
\begin{tabular}{clc}
\hline\hline
\textbf{Parameter} & \textbf{Description} & \textbf{Value} \\
\hline
$\beta$ & Discount factor & 0.96\\
$\sigma$ & Coefficient of relative risk aversion & 2 \\
$\rho_y$ & Autocorrelation of labor income & 0.6 \\
$\nu_y$ & Variance parameter of labor income & 0.2 \\
$\rho_z$ & Persistence of AR(1) for $z_t$ (log TFP)  & 0.9\\
$\nu_z$  & Volatility of AR(1) for $z_t$ (log TFP)  & 0.02 \\
$B$  & Total bond supply  & 0 \\
$\underline b$  & Borrowing constraint  & -1 \\
\hline
$\overline r$ & Mean interest rate (PE) & 0.038 \\
$\rho_r$ & Autocorrelation of interest rate (PE) & 0.8 \\
$\nu_r$ & Volatility of interest rate (PE) & 0.02 \\
\hline\hline
\end{tabular}
\caption{Huggett model calibration}
\label{tab:huggett_params}
\end{table}

\paragraph{Training schedule and learning rate.}
We train the SPG algorithm with an exponentially decaying learning rate. Let $lr_\texttt{ini}$ denote the initial learning rate and $lr_\texttt{decay} \in (0, 1)$ the decay factor. The learning rate at iteration $t$ is given by
\begin{equation*}
	lr_t = lr_\texttt{ini} \cdot lr_\texttt{decay}^{t'},
	\qquad \text{ where } \qquad
	t' = \frac{\max \{t-N_\texttt{warm-up}, 0\}}{N_\texttt{epoch}-N_\texttt{warm-up}},
\end{equation*}
so that $lr_t$ is held constant during an initial ``warm-up'' phase of length $N_\texttt{warm-up}$ and then decays smoothly to a lower value by the final epoch $N_\texttt{epoch}$. In the Huggett application, we set $N_\texttt{epoch} = 1000$, $N_\texttt{warm-up} = 50$, $lr_\texttt{ini} = 10^{-3}$, and $lr_\texttt{decay} = 0.5$, and we use an exponential scheduler (denoted by $lr_\texttt{sche}$ in Table \ref{tab:huggett_hyperparams}). We declare convergence when the change in the policy parameters across epochs falls below the threshold $\epsilon_\text{converge} = 3\times 10^{-4}$.

\paragraph{Sampling, batching, and memory constraints.}
Due to GPU memory constraints, we do not use all simulated data to update the policy in each iteration. Instead, we sample data in mini-batches. In each update, the effective data size is $N_\texttt{sample} \times N_\texttt{update}$, where $N_\texttt{sample}$ denotes the number of simulated trajectories per batch (we set $N_\texttt{sample}=512$ in the baseline Huggett experiment) and $N_\texttt{update}$ the number of time steps used from each trajectory for the gradient update. This mini-batching keeps memory requirements manageable while preserving enough variation in the data to obtain stable gradient estimates.

\begin{table}[t!]
\centering
\begin{tabular}{clc}
\hline\hline
\textbf{Parameter} & \textbf{Description} & \textbf{Value} \\
\hline
$n_b$ & Number of $b$ grid points & 200 \\
$b^{\texttt{max}}$ & Upper bound of $b$ grid & 50 \\
$n_y$ & Number of $y$ grid points & 3 \\
$n_r$ & Number of $r$ grid points & 20 \\
$r_L$ & Lower bound of $r$ grid & 0.01 \\
$r_H$ & Upper bound of $r$ grid & 0.06 \\
$n_z$ & Number of $z$ grid points & 30 \\
\hline
$c_\texttt{min}$ & Minimum consumption & $10^{-3}$ \\
$T_\texttt{trunc}$ & Truncation horizon for simulations & 170 \\
$\epsilon_\texttt{trunc}$ & Truncation threshold & $10^{-3}$ \\
\hline
$N_{\texttt{epoch}}$ & Maximum number of parameter updates & 1000 \\
$N_{\texttt{warm-up}}$ & Number of warm-up epochs & 50 \\
$lr_{\texttt{ini}}$ & Initial learning rate & $10^{-3}$ \\
$lr_{\texttt{decay}}$ & Learning rate decay rate & 0.5 \\
$lr_{\texttt{sche}}$ & Learning-rate scheduler & exponential \\
$N_{\texttt{sample}}$ & Batch size (trajectories per update) & 512 \\
$\epsilon_{\text{converge}}$ & Convergence threshold & $3\times 10^{-4}$ \\
\hline\hline
\end{tabular}
\caption{Hyperparameters for solving the Huggett model}
\label{tab:huggett_hyperparams}
\end{table}

\paragraph{Initialization and warm-up.}
We initialize the policy as described in Footnote \ref{footnote:guess} to guarantee that the initial aggregate savings schedule is at least weakly responsive to the interest rate. The training process is then split into two phases. During the warm-up phase of length $N_\texttt{warm-up}$, we fix the cross-sectional distribution of agents at some simple initial guess $\bm g_0$ and do not update it. In this phase, the sole objective is to move the policy away from its crude initial guess and toward a reasonable neighborhood of the eventual solution. Keeping $\bm g$ fixed prevents the badly informed initial policy from ``polluting'' the distribution.

After warm-up, we switch to an adaptive phase in which the distribution is updated endogenously, and which may last up to $N_\texttt{epoch} - N_\texttt{warm-up}$ epochs ( though convergence typically occurs earlier). 
We use the simulated distribution implied by the most recent policy as the initial distribution for each trajectory. In other words, after warm-up, each new batch of trajectories starts from a cross-section that is itself an equilibrium object. This iterative updating of the initial distribution ensures that the policy is trained on data drawn from its own induced stationary distribution, which is important for the accuracy of the final solution.

%%%%%%%%%%%%%%%%%%%%%%%%%%%%%%%%%%%%%%%%%%%%%%%%%%%%%%%%%%%%
%%%%%%%%%%%%%%%%%%%%%%%%%%%%%%%%%%%%%%%%%%%%%%%%%%%%%%%%%%%%
\subsection{Appendix for Krusell-Smith Application}\label{app:applications_krusell_smith}

\paragraph{Calibration.}
Table \ref{tab:KS_params} summarizes the calibration for the Krusell-Smith model used in Section \ref{sec:applications_krusell_smith}. We use a discount factor of $\beta = 0.95$. The utility function is CRRA with a coefficient of relative risk aversion $\sigma = 3$, and the borrowing constraint is set at $\underline b = 0$. 

The idiosyncratic income process is modeled as a log AR(1) with autocorrelation $\rho_y = 0.6$ and innovation volatility $\nu_y = 0.2$, the same specification as in the Huggett model. On the production side, we follow the standard Krusell-Smith calibration: the capital share is $\alpha = 0.36$, the depreciation rate is $\delta = 0.08$, and aggregate productivity $z_t$ follows a log AR(1) with persistence $\rho_z = 0.9$ and volatility $\nu_z = 0.03$.

\begin{table}[t!]
\centering
\begin{tabular}{clc}
\hline\hline
\textbf{Parameter} & \textbf{Description} & \textbf{Value} \\
\hline
$\beta$ & Discount factor & 0.95\\
$\sigma$ & Utility parameter & 3 \\
$\underline b$ & Borrowing constraint & 0 \\
$\rho_y$ & Autocorrelation of idiosyncratic shock & 0.6 \\
$\nu_y$ & Volatility of idiosyncratic shock & 0.2 \\
$a$ & Capital share & 0.36\\
$\delta$ & Capital depreciation rate & 0.08 \\
$\rho_z$ & Persistence of AR(1) for $z_t$ (log TFP)  & 0.9\\
$\nu_z$  & Volatility of AR(1) for $z_t$ (log TFP)  & 0.03 \\
\hline\hline
\end{tabular}
\caption{Krusell--Smith model calibration}
\label{tab:KS_params}
\end{table}

\paragraph{Discretization and grids.}
Table \ref{tab:KS_hyperparams} reports the hyperparameters used for the SPG solution of the Krusell-Smith model. The individual capital state $b$ is discretized on a grid with $n_b = 200$ points and an upper bound $b^\texttt{max} = 100$, which is higher than in the Huggett experiment. The larger upper bound reflects the fact that, in Krusell-Smith, agents accumulate capital rather than unproductive bonds, and the equilibrium wealth distribution is more dispersed. The idiosyncratic income state $y$ again takes $n_y = 3$ values.

The aggregate price vector consists of the interest rate $r_t$ and the real wage $w_t$. We approximate the price space on two separate grids: the interest rate is discretized with $n_r = 30$ points on $[r_L, r_H] = [0.02, 0.07]$, and the wage with $n_w = 50$ points on $[w_L, w_H] = [0.9, 1.5]$. These ranges comfortably contain the realizations observed in our simulations and allow the policy to respond flexibly to movements in both prices without requiring a prohibitive number of grid points.

\begin{table}[t!]
\centering
\begin{tabular}{clc}
\hline\hline
\textbf{Parameter} & \textbf{Description} & \textbf{Value} \\
\hline
$n_b$ & Number of $b$ grid points & 200 \\
$b^{\texttt{max}}$ & Upper bound of $b$ grid & 100 \\
$n_y$ & Number of $y$ grid points & 3 \\
$n_r$ & Number of $r$ grid points & 30 \\
$r_L$ & Lower bound of $r$ grid & 0.02 \\
$r_H$ & Upper bound of $r$ grid & 0.07 \\
$n_w$ & Number of $p_2$ grid points & 50 \\
$w_L$ & Lower bound of $w$ grid & 0.9 \\
$w_H$ & Upper bound of $w$ grid & 1.5 \\
\hline
$c_{\texttt{min}}$ & Minimum consumption & $10^{-3}$ \\
$c_{\texttt{init}}$ & Initial guess for consumption share & 0.5 \\
$T_{\texttt{trunc}}$ & Truncation horizon for simulations & 90 \\
$\epsilon_{\texttt{trunc}}$ & Truncation threshold & $10^{-2}$ \\
\hline
$N_{\texttt{epoch}}$ & Maximum number of parameter updates & 1000 \\
$N_{\texttt{warm-up}}$ & Number of warm-up epochs & 50 \\
$lr_{\texttt{ini}}$ & Initial learning rate & $5\times 10^{-4}$ \\
$lr_{\texttt{decay}}$ & Learning-rate decay rate & 0.5 \\
$lr_{\texttt{sche}}$ & Learning-rate scheduler & exponential \\
$N_{\texttt{sample}}$ & Batch size (trajectories per update) & 512 \\
$\epsilon_{\text{converge}}$ & Convergence threshold & $2\times 10^{-4}$ \\
\hline\hline
\end{tabular}
\caption{Hyperparameters for solving the Krusell--Smith model}
\label{tab:KS_hyperparams}
\end{table}

\paragraph{Simulation horizon and truncation.}
As in the Huggett case, lifetime utility is computed by truncating the infinite sum at a finite horizon $T_\texttt{trunc}$. For Krusell-Smith we set $T_\texttt{trunc}=90$ and use a truncation tolerance $\epsilon_\texttt{trunc}=10^{-2}$, i.e.
\begin{equation*}
	\beta^{T_\texttt{trunc}} < \epsilon_\texttt{trunc},
\end{equation*}
so that the tail of the discounted utility stream is negligible at the scale of our numerical accuracy. We also impose a minimal consumption level $c_\texttt{min} = 10^{-3}$ to avoid evaluating utility at zero or extremely small consumption levels.

\paragraph{Training schedule and convergence.}
We use the same general training structure as in the Huggett exercise but with slightly different numerical values. The maximum number of epochs is $N_\texttt{epoch} = 1000$, with $N_\texttt{warm-up}=50$ warm-up epochs during which the learning rate is kept constant and the initial distribution is fixed. The learning rate starts at $lr_\texttt{ini} = 5 \times 10^{-4}$ and decays exponentially at rate $lr_\texttt{decay} = 0.5$ according to the scheduler denoted by $lr_\texttt{sche}$ in Table \ref{tab:KS_hyperparams}. As in the Huggett case, the decay is only activated after the warm-up phase. We declare convergence when the change in parameters between successive epochs falls below $\epsilon_\text{converge} = 2 \times 10^{-4}$; in practice, the algorithm typically converges well before hitting the hard cap of $N_\texttt{epoch}$.

Mini-batching is again used to handle memory constraints and stabilize gradient estimates. Each update uses $N_\texttt{sample}=512$ simulated trajectories, and a fixed number of time steps per trajectory, to form the stochastic gradient. This yields a total data size per update of $N_\texttt{sample} \times N_\texttt{update}$, which we choose to fully utilize the available GPU memory without inducing excessive variance in the gradient.

\paragraph{Initialization and warm-up.}
The warm-up logic mirrors that used in the Huggett application but is adapted to the two-price environment. During the first $N_\texttt{warm-up}$ epochs, all trajectories are initialized from a cross-sectional distribution that is held fixed, while the policy is being updated. After warm-up, we update the cross-sectional distribution based on the most recent policy and allow the learning rate to adjust. The initial conditions for each new set of trajectories are drawn from the simulated distribution generated by the most recent policy.

%%%%%%%%%%%%%%%%%%%%%%%%%%%%%%%%%%%%%%%%%%%%%%%%%%%%%%%%%%%%
%%%%%%%%%%%%%%%%%%%%%%%%%%%%%%%%%%%%%%%%%%%%%%%%%%%%%%%%%%%%
\subsection{Appendix for HANK Application}\label{app:applications_HANK}

\paragraph{Calibration.}
Table \ref{tab:HANK_calibration} reports the calibration of the HANK model used in Section \ref{sec:applications_HANK}. One model period corresponds to a year, and the discount factor is set such that the annual real interest rate is in a plausible range; in the baseline we use a discount factor of $0.975$. Preferences over consumption and labor are CRRA and separable, with coefficient of relative risk aversion equal to $1$ and inverse Frisch elasticity of labor supply $\eta = 1$. 

We follow a standard New Keynesian calibration. Price-setting firms face Rotemberg adjustment costs with parameter $\theta = 100$ and an elasticity of substitution across intermediate goods of $\varepsilon = 10$. Monetary policy follows the Taylor rule specified in the main text, with coefficient $\phi = 1.5$ on inflation and a gross steady-state real interest rate target $\bar R = 1.025$.

Aggregate risk is two-dimensional. TFP $z_t$ follows an AR(1) process with persistence $\rho_z = 0.9$ and innovation volatility $\nu_z = 0.07$. The monetary policy shock $\epsilon_t$ is also AR(1) with the same persistence $\rho_\epsilon = 0.9$ and volatility $\nu_\epsilon = 0.002$. These values are summarized in Table \ref{tab:HANK_calibration} and are chosen so that both real and nominal variables display non-trivial but stable dynamics in the simulations.

\begin{table}[t!]
\centering
\begin{tabular}{clc}
\hline\hline
\textbf{Parameter} & \textbf{Description} & \textbf{Value} \\
\hline
$\beta$ & Discount factor & 0.975 \\
$\sigma$ & Coefficient of relative risk aversion & 1 \\
$\eta$ & Inverse of Frisch elasticity & 1 \\
$\phi$ & Coefficient of Taylor rule & 1.5 \\
$\theta$ & Price adjustment cost & 100 \\
$\epsilon$ & Elasticity of substitution & 10 \\
$\bar R$ & Target for gross interest rate & 1.025 \\
$\rho_z$ & Autocorrelation of aggregate TFP shock & 0.9 \\
$\nu_z$ & Volatility of aggregate TFP shock & 0.07 \\
$\rho_\epsilon$ & Autocorrelation of monetary policy shock & 0.9 \\
$\nu_\epsilon$ & Volatility of monetary policy shock & 0.002 \\
\hline\hline
\end{tabular}
\caption{Calibration for the HANK model}
\label{tab:HANK_calibration}
\end{table}

\paragraph{Discretization and grids.}
Table \ref{tab:hank_hyperparams} lists the hyperparameters and grid choices for our solution of the HANK model. The individual asset state $b$ is discretized on a grid with $n_b = 200$ points and an upper bound $b^\texttt{max} = 100$. The idiosyncratic income state $y$ again takes $n_y = 3$ values, using the same discretization as in the Huggett and Krusell-Smith applications for ease of comparison.

The aggregate state combines a two-dimensional price vector $(r_t, w_t)$ with the two exogenous shocks $z_t$ and $\epsilon_t$. We approximate the real interest rate on a grid with $n_r = 30$ points over the interval $[r_L, r_H] = [0.01, 0.04]$. The real wage is discretized with $n_w = 30$ points on $[w_L, w_H] = [0.7, 1.0]$. The ranges are chosen to cover comfortably the realizations observed in equilibrium simulations, while keeping the price grids small enough for efficient training.

For the aggregate shocks, we use $n_z = 50$ grid points for log TFP $z_t$ and $n_\epsilon = 50$ points for the monetary shock $\epsilon_t$. Both grids are obtained by discretizing the respective AR(1) processes with a standard Tauchen method. The relatively fine grids for $(z_t, \epsilon_t)$ help the algorithm capture the interaction between real and nominal disturbances in the HANK model.

\begin{table}[t!]
\centering
\begin{tabular}{clc}
\hline\hline
\textbf{Parameter} & \textbf{Description} & \textbf{Value} \\
\hline
$n_b$ & Number of $b$ grid points & 200 \\
$b^\texttt{max}$ & Upper bound of $b$ grid & 100 \\
$y$ & Number of $y$ grid points & 3 \\
$n_r$ & Number of $r$ grid points & 30 \\
$r_L$ & Lower bound of $r$ grid & 0.01 \\
$r_H$ & Upper bound of $r$ grid & 0.04 \\
$n_w$ & Number of $p_2$ grid points & 30 \\
$w_L$ & Lower bound of $w$ grid & 0.7 \\
$w_H$ & Upper bound of $w$ grid & 1.0 \\
$n_z$ & Number of $z$ grid points & 50 \\
$n_\epsilon$ & Number of $e$ grid points & 50 \\
\hline
$c_\texttt{min}$ & Minimum consumption & $10^{-3}$ \\
$c_\texttt{init}$ & Initial guess of consumption share & 0.5 \\
$n_\texttt{init}$ & Initial guess of labor supply & 1.5 \\
$\Pi_\texttt{init}$ & Initial guess of inflation & 0 \\
$T_\texttt{trunc}$ & Truncation horizon for simulations & 182 \\
$\epsilon_\texttt{trunc}$ & Truncation threshold & $10^{-2}$ \\
$N_{\texttt{epoch}}$ & Maximum number of parameter updates & 1000 \\
$N_{\texttt{warm-up}}$ & Number of warm-up epochs & 50 \\
$lr_{\texttt{ini}}$ & Initial learning rate & $5\times 10^{-3}$ \\
$lr_{\texttt{decay}}$ & Learning-rate decay rate & 0.5 \\
$N_\texttt{sample}$ & Baseline sampling size & 512 \\
$\epsilon_\text{converge}$ & Convergence threshold & $2\times 10^{-3}$ \\
\hline\hline
\end{tabular}
\caption{Hyperparameters for solving the HANK model}
\label{tab:hank_hyperparams}
\end{table}

\paragraph{Simulation horizon, truncation, and initial guesses.}
The HANK model combines persistence in both TFP and monetary shocks with sluggish price adjustment. To accommodate this accurately, we use a longer truncation horizon $T_\texttt{trunc} = 182$ periods and a truncation tolerance $\epsilon_\texttt{trunc} = 10^{-2}$, which implies
\begin{equation*}
	\beta^{T_\texttt{trunc}} < \epsilon_\texttt{trunc},
\end{equation*}
so that the contribution of periods beyond $T_\texttt{trunc}$ is negligible at the scale of our numerical accuracy. As in the other applications, we impose a minimal consumption level $c_\texttt{min} = 10^{-3}$ to avoid evaluating the utility function at zero consumption.

The HANK environment includes both consumption-saving and labor-supply decisions, as well as firm price-setting. We initialize these policy functions using simple guess rules; for $c_\texttt{init}$, we set an initial constant consumption share of total cash-in-hand, a constant $n_\texttt{init}$ sets an initial level of hours worked, and $\Pi_\texttt{init}=0$ initializes inflation. These initial values are deliberately crude and their sole purpose is to place the policy in a reasonable region of the parameter space before learning from simulated data. In practice, the final solution is insensitive to these initial guesses once training has converged.

\paragraph{Training and convergence.}
The remaining training hyperparameters follow the logic of the Huggett and Krusell-Smith applications. We use a baseline batch size of $N_\texttt{sample} = 512$ simulated trajectories per update, chosen to saturate GPU memory without generating excessive variance in the policy gradient. The convergence threshold is set at $\epsilon_\text{converge} = 2 \times 10^{-3}$ for the HANK model, which reflects the greater complexity of the joint household-firm problem and the fact that small changes in the policy parameters can translate into larger differences in aggregate dynamics.

Other aspects of the training schedule --- notably the total number of epochs, the length of the warm-up phase, and the learning-rate schedule --- are chosen in line with the Krusell-Smith specification discussed in Appendix \ref{app:applications_krusell_smith} and are not repeated here. In practice, the HANK model converges somewhat more slowly than Huggett and Krusell-Smith but still within a few minutes on a single GPU, as reported in Table \ref{tab:Table_runtime_all} in the main text.

\iffalse
\section{Comparison of SRL solution to the \cite{krusell-smith} solution.}
For the the Krusell–Smith model, before we have deep learning based rational expectation solutions, the original \citet{krusell-smith} algorithm has emerged as a de facto benchmark in the literature. This makes it also a natural testing ground for our SRL approach in a full general equilibrium setting.

In this Appendix, we solve the same Krusell–Smith economy using two methods: the original Krusell–Smith procedure and our SRL algorithm, and compare the resulting objects. Figure \ref{fig:KS_comparison_KS_method} reports this comparison. Each panel plots the optimal consumption policy as a function of wealth $b$ for different fixed values of individual income and aggregate productivity. Across all panels, the two solution methods deliver almost indistinguishable policies; differences are small relative to the overall variability generated by the aggregate shocks.

Taken together with the PE results and our comparison exercises against the deep learning based RE solution, this comparison indicates that, at least in the Krusell–Smith environment, our SRL approach can replicate the benchmark global solution while retaining the flexibility and scalability of a reinforcement-learning implementation.

\begin{figure}[ht!]
\centering
\includegraphics[scale=0.3]{figures/KS1998/KS_RE_comparison.pdf}
\caption{Comparison to Krusell-Smith (1998) Solution Method}
\label{fig:KS_comparison_KS_method}
\end{figure}
\fi

\end{document}